\documentclass[11pt,a4paper]{article}

\usepackage{jheppub}
\usepackage{booktabs}

\newcommand{\ycut}{y_\text{cut}}
\newcommand{\zcut}{z_\text{cut}}
\newcommand{\rhofat}{\rho_\text{fat}}
\newcommand{\zfat}{z_\text{fat}}
\newcommand{\Rfact}{R_\text{fact}}
\newcommand{\Rprune}{R_\text{prune}}
\newcommand{\Rsub}{R_\text{sub}}

\newcommand{\asbar}{{\bar \alpha}_s}
\newcommand{\muNP}{\mu_\text{NP}}
\newcommand{\nfilt}{n_\text{filt}}
\newcommand{\Rfilt}{R_\text{filt}}
\newcommand{\NP}{\text{NP}}
\newcommand{\PT}{\text{PT}}

\newcommand{\GeV}{\,\mathrm{GeV}}
\newcommand{\TeV}{\,\mathrm{TeV}}
\newcommand{\Erf}{\mathrm{Erf}}

\newcommand{\jet}{\mathrm{jet}}

\newcommand{\as}{\alpha_s}
\newcommand{\order}[1]{{\cal O}\left(#1\right)}
\newcommand{\subjet}[1]{$j_{#1}$}

\newcommand{\cN}{{\cal N}}

\definecolor{darkgreen}{rgb}{0,0.5,0}
\definecolor{darkblue}{rgb}{0,0,0.5}
\definecolor{darkred}{rgb}{0.5,0,0.0}

\newcommand{\sanepruning}{\text{Y-pruning}}
\newcommand{\Sanepruning}{\text{Y-pruning}}
\newcommand{\saneprune}{\text{(Y-prune)}}
\newcommand{\sanepruneNLO}{\text{(Y-prune, NLO)}}
\newcommand{\sane}{\text{Y}}
\newcommand{\anomalouspruning}{\text{I-pruning}}
\newcommand{\Anomalouspruning}{\text{I-pruning}}
\newcommand{\anomprune}{\text{(I-prune)}}
\newcommand{\anompruneNLO}{\text{(I-prune, NLO)}}
\newcommand{\anomalous}{\text{I}}

\newcommand{\beq}{\begin{equation}}
\newcommand{\eeq}{\end{equation}}
\newcommand{\bea}{\begin{eqnarray}}
\newcommand{\eea}{\end{eqnarray}}
\newcommand{\bdm}{\begin{displaymath}}
\newcommand{\edm}{\end{displaymath}}
\def\as{\alpha_s}

\def\ord{{\cal O}}

\def\d{\partial}

\def \d{{\rm d} }
\def \d0 {D\O \;}

\title{Towards an understanding of jet substructure}

\author[a,b]{Mrinal Dasgupta,}
\author[b]{Alessandro Fregoso,}
\author[c]{Simone Marzani}
\author[d,e,1]{and Gavin P.~Salam\note{On leave from Department of Physics, Princeton University, Princeton, NJ 08544, USA.}}
\affiliation[a]{Consortium for
  Fundamental Physics, School of Physics \& Astronomy, University
  of Manchester, Oxford Road, Manchester M13 9PL, United Kingdom}
\affiliation[b]{School of Physics \& Astronomy, University
  of Manchester, Oxford Road, Manchester M13 9PL, United Kingdom}
\affiliation[c]{Institute for Particle Physics Phenomenology, Durham University, South Road, Durham DH1 3LE, United Kingdom}
\affiliation[d]{CERN, PH-TH, CH-1211 Geneva 23, Switzerland}
\affiliation[e]{LPTHE; CNRS UMR 7589; UPMC Univ. Paris 6; Place Jussieu, Paris 75252, France}

\emailAdd{mrinal.dasgupta@manchester.ac.uk}
\emailAdd{alessandro.fregoso@hep.manchester.ac.uk}
\emailAdd{simone.marzani@durham.ac.uk}
\emailAdd{gavin.salam@cern.ch}

\preprint{
\begin{flushright}
  CERN-PH-TH/2013-145\\
  DCPT/13/86\\
  IPPP/13/43 \\
  LPN13-036\\
  MAN/HEP/2013/12
\end{flushright}
}


\abstract{ 
  We present first analytic, resummed calculations of the rates at
  which widespread jet substructure tools tag QCD jets.
  As well as considering trimming, pruning and the mass-drop tagger,
  we introduce modified tools with improved analytical and
  phenomenological behaviours.
  Most taggers have double logarithmic resummed structures.
  The modified mass-drop tagger is special in that it involves only
  single logarithms, and is free from a complex class of terms known
  as non-global logarithms.
  The modification of pruning brings an improved ability to
  discriminate between the different colour structures that
  characterise signal and background.
  As we outline in an extensive phenomenological discussion, these
  results provide valuable insight into the performance of existing
  tools and help lay robust foundations for future substructure
  studies.
  }

\begin{document}
\maketitle
\section{Introduction}

The Large Hadron Collider (LHC) at CERN is increasingly exploring
phenomena at energies far above the electroweak scale.
One of the features of this exploration is that analysis techniques
developed for earlier colliders, in which electroweak-scale particles
could be considered ``heavy'', i.e.\ slow-moving, have to be
fundamentally reconsidered at the LHC. 
In particular, in the context of jet-related studies, the large boost
of electroweak bosons and top quarks causes their hadronic decays to
become collimated inside a single jet.
Consequently a vibrant research field has emerged in recent years,
investigating how best to identify the characteristic substructure that
appears inside the single ``fat'' jets from electroweak scale objects,
as reviewed in Refs.~\cite{Boost2010,Boost2011,PlehnSpannowsky}.
In parallel, the ``tagging'' and ``grooming'' methods that have been developed have started to be
tested and applied in numerous experimental analyses
(e.g. Refs~\cite{ATLAS:2012am,Aad:2012meb,Aad:2013gja,CMS-substructure-studies} for
studies on QCD jets and Refs~ \cite{Aad:2012raa,Aad:2012dpa, ATLAS:2012dp, ATLAS:2012ds, Chatrchyan:2012ku,Chatrchyan:2012cx, Chatrchyan:2012yxa} for searches).

The taggers' and groomers' action is twofold: they aim to suppress or reshape
backgrounds, while retaining signal jets and enhancing their
characteristic jet-mass peak at the $W/Z/$Higgs/top/etc.\ mass.
Nearly all the theoretical discussion of these aspects has taken place
in the 
context of Monte Carlo simulation studies (see for instance
Ref.~\cite{Boost2011} and references therein), with 
tools such as 
Herwig~\cite{Herwig6,Herwig++}, Pythia~\cite{Pythia6,Pythia8} and
Sherpa~\cite{Sherpa}.
While Monte Carlo simulation is a powerful tool, its
intrinsic numerical nature can make it difficult to extract the key
characteristics of individual substructure methods and understand the relations between
them.
As an example of the kind of statements that exist about them in the
literature, we quote from the Boost 2010 proceedings: 
\begin{quotation}
  The [Monte Carlo] findings discussed above indicate that while
  [pruning, trimming and filtering] have qualitatively similar
  effects, there are important differences. For our choice of
  parameters, pruning acts most aggressively on the signal and
  background followed by trimming and filtering.
\end{quotation}
While true, this brings no insight about whether the differences are
due to intrinsic properties of the substructure methods analysed or instead due to the
particular parameters that were chosen; nor does it allow one to
understand whether any differences are generic, or restricted to some
specific kinematic range, e.g.\ in jet transverse momentum.
Furthermore there can be significant differences between Monte Carlo
simulation tools and among tunes (see
e.g. \cite{Boost2011,Richardson:2012bn,ATLAS:2012am,CMS-substructure-studies}),
which may be  
hard to diagnose experimentally, because of the many kinds of physics
effects that contribute to the jet structure (final-state showering,
initial-state showering, underlying event, hadronisation, etc.).
Overall, this points to a need to carry out analytical
calculations to understand the interplay between tagging/grooming techniques and the
quantum chromodynamical (QCD) showering that occurs in both signal and
background jets.

So far there have been three main investigations into the analytical
features that emerge from substructure techniques.
Refs.~\cite{Rubin:2010fc,Rubin:2010boa} investigated the mass
resolution that can be obtained on signal jets and how to optimize the
parameters of a method known as filtering~\cite{Butterworth:2008iy}.
Ref.~\cite{Walsh:2011fz} discussed constraints that might arise if one
is to apply Soft Collinear Effective Theory (SCET) to jet substructure
calculations. 
Ref.~\cite{Feige:2012vc} observed that for narrow jets the
distribution of the $N$-subjettiness shape variable~\cite{Thaler:2010tr} for 2-body signal
decays can be resummed to high accuracy insofar as it is related to
the thrust distribution in
$e^+e^-$~\cite{Catani:1992ua,Becher:2008cf,GehrmannDeRidder:2007bj,Weinzierl:2008iv},
though for phenomenological purposes this still needs to be supplemented with
a calculation of the interplay with practical cuts on the jet mass.
Other calculations that relate to the field of jet substructure
include those of planar flow~\cite{Field:2012rw}, energy-energy correlations~\cite{Larkoski:2013eya} and jet multiplicities in
the small-jet-radius limit~\cite{Gerwick:2012fw}.
Additionally Ref.~\cite{QuirogaArias:2012nj} has examined the extent
to which simple approximations about the kinematics involved in tagging and grooming can bring insight into different methods.

Here we embark on a comparative, analytical study of multiple
commonly-used taggers and groomers.
Ideally we would include all existing methods for both background
(QCD jet) and signal-induced jets, however given the many techniques
that have been proposed, this would be a gargantuan task.
In practice we find that a background-only study, for just a handful
of substructure techniques, already brings significant insight into the way the
taggers function.

The three commonly used methods that we concentrate on are: the
mass-drop tagger (MDT)~\cite{Butterworth:2008iy},
pruning~\cite{Ellis:2009me,Ellis:2009su} and
trimming~\cite{Krohn:2009th}.
They all involve the identification of subjets within an original jet,
and share the characteristic that they attempt to remove subjets
carrying less than some (small) fraction of the original jet's
momentum.

\begin{figure}
  \centering
  \includegraphics[width=0.48\textwidth]{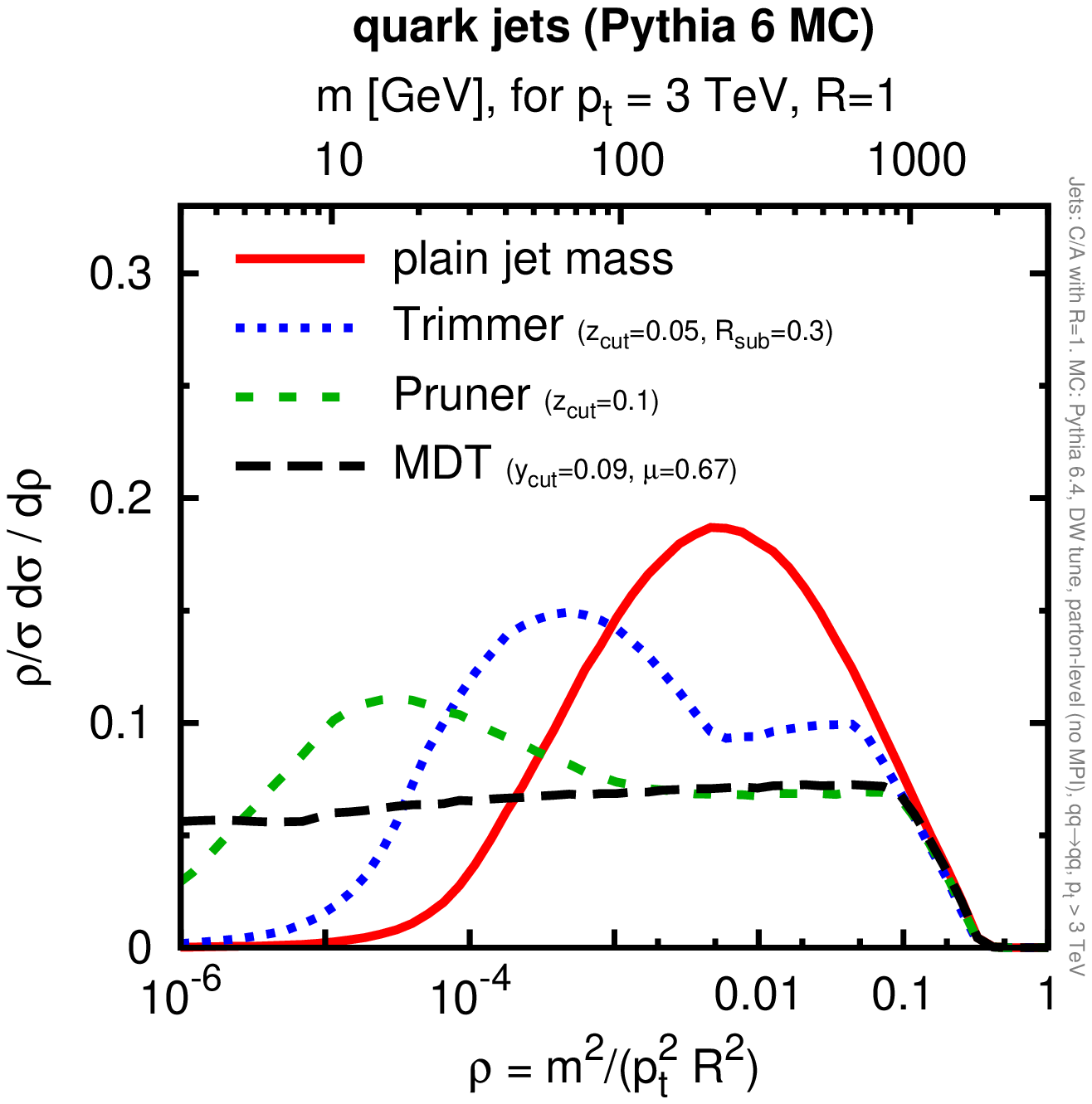}\hfill
  \includegraphics[width=0.48\textwidth]{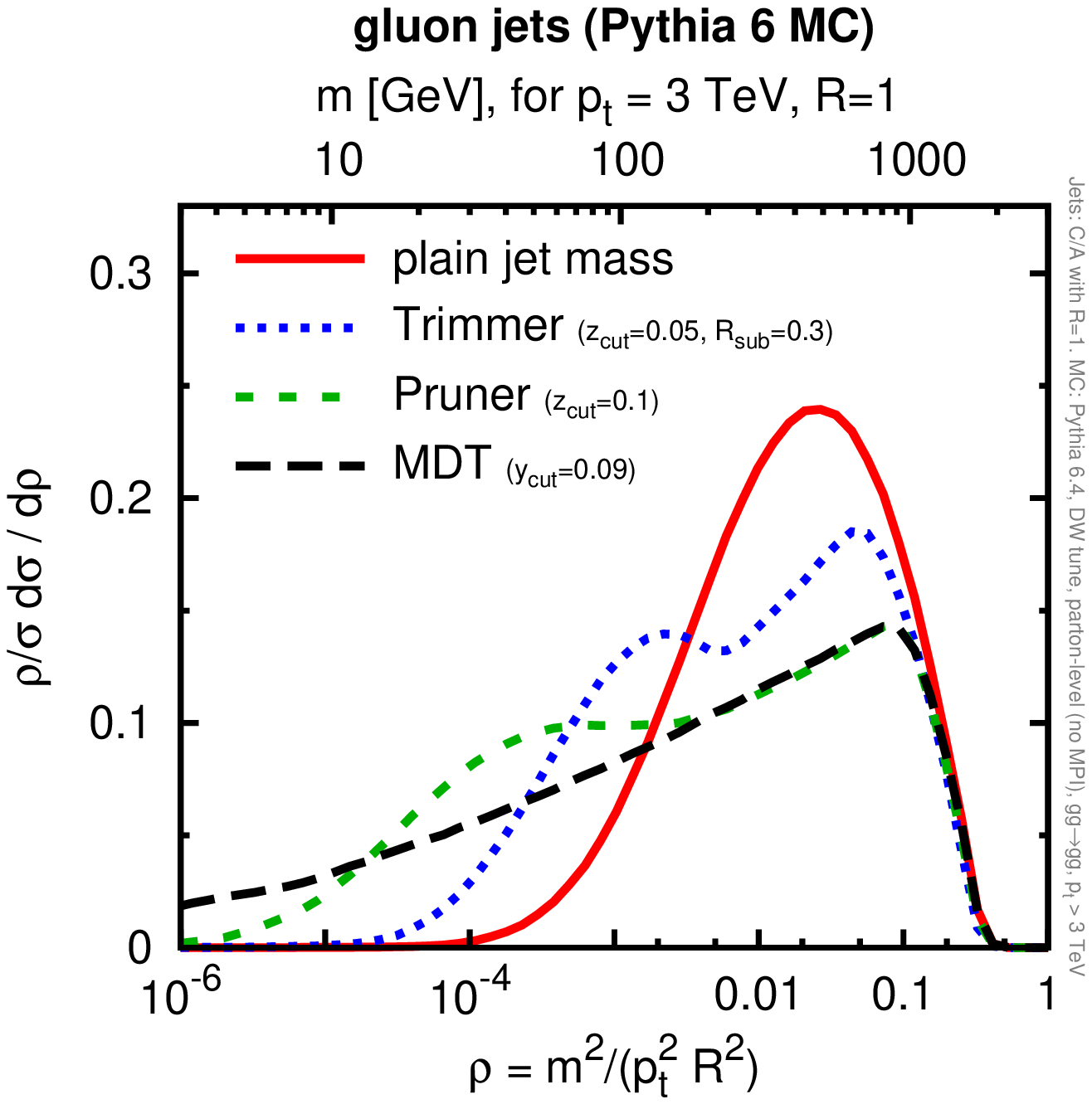}
  \caption{The distribution of $\rho = m^2/(p_t^2 R^2)$ for tagged jets, with
    three taggers/groomers: trimming, pruning and the mass-drop tagger (MDT). 
    The results have been  obtained from Monte Carlo simulation
    with Pythia~6.425~\cite{Pythia6} in the DW tune~\cite{DW}
    (virtuality-ordered shower), with a minimum $p_t$ cut in the
    generation of $3\TeV$, for $14\TeV$ $pp$ collisions, at parton
    level, including initial and final-state showering, but without
    the underlying event (multiple interactions). 
    The left-hand plot shows  $qq \to qq$ scattering, 
    the right-hand plot $gg \to gg$ scattering.
    In all cases, the taggers have been applied to the two leading
    Cambridge/Aachen~\cite{Dokshitzer:1997in,Wobisch:1998wt} jets ($R=1.0$). 
    The parameters chosen for mass-drop ($\ycut=0.09$, $\mu=0.67$),
    pruning ($\zcut=0.1$, $R_\text{fact}=0.5$) and trimming
    ($\zcut=0.05$, $\Rsub=0.3$) all correspond to widely-used
    choices. 
  }
  \label{fig:tagged-mass-MC}
\end{figure}

To provide a starting point for our discussion, consider
Fig.~\ref{fig:tagged-mass-MC}, which shows Monte Carlo simulation for
the mass distribution of tagged/groomed jets with the three substructure methods considered here (and also
for the plain jet mass), plotted as a function of a variable $\rho$,
\begin{equation}
  \rho \equiv \frac{m^2}{p_t^2 R^2}\,,
\end{equation}
where $m$ is the jet's mass, $p_t$ its transverse momentum and $R$ the
radius for the jet definition;
the upper axis gives the correspondence in terms of jet mass for jets
with $p_t = 3\TeV$.
The left-hand plot is for quark-induced jets, the right-hand plot for
gluon-induced jets.
A first observation is that all three methods are identical to the
plain jet mass for $\rho \gtrsim 0.1$.
At that point, pruning and MDT have a kink, and in the quark-jet case
exhibit a flat distribution below the kink.
Trimming has a kink at a lower mass value, and also then becomes flat.
For gluon jets, the kinks appear in the same location, but below the kink
there is no flat region.
Pruning and trimming then each have an additional transition point,
at somewhat smaller $\rho$ values, below which they develop
peaks that are reminiscent (but at lower $\rho$) of that of the
plain jet mass.
Knowing about such features can be crucial, for example in data-driven
background estimates, where there is often an implicit assumption of
smoothness of background shapes.
In this context one observes that for the upper-range of $p_t$'s that
the LHC will eventually cover, $p_t \gtrsim 3 \TeV$, the lower
transition points of pruning and trimming occur precisely in the
region of electroweak-scale masses.\footnote{At this point, a question
  arises of whether the LHC experiments are able to accurately measure
  EW-scale masses for TeV-scale jets. Challenges can arise, for
  example in terms of the angular resolution of the hadronic
  calorimeter, which may be relevant with current experimental
  reconstruction methods. Work in Ref.~\cite{Katz:2010mr}, however,
  suggests that with full use of information from tracking and
  electromagnetic calorimetry, which have higher angular resolution,
  good mass resolution for multi-TeV scale jets may well be possible.}

To our knowledge the similarities and differences observed in
Fig.~\ref{fig:tagged-mass-MC} have not been systematically commented
on before, let alone understood.
Questions that one can ask include: why do the taggers/groomers have these
characteristic shapes for the mass distributions? 
Is there any significance to the fact that pruning and MDT appear very
similar over some extended range of masses?
How do the positions of the kinks and transition-points depend on the substructure methods' parameters?
Good taggers and groomers should probably not generate such rich
structures for the background shapes and, as we shall see, a deeper
understanding can point to desirable modifications of these methods.
Finally, what classes of perturbative terms are associated with the substructure techniques, specifically what kinds of logarithms of jet mass arise at each
order in the strong coupling $\as$ and what are the implications for
the likely reliability of fixed-order, resummed and Monte Carlo predictions?
These are the types of question that we shall address here.
A companion paper~\cite{taggersNLO} discusses the first two orders of
log-enhanced terms in substantially more depth and includes
comparisons to fixed-order results for jets in $e^+e^-$ collisions.

\section{Definitions and approximations} \label{sec:def}


Let us start with a question of nomenclature: tagging v.\ grooming,
for which there is no generalised agreement. 
One definition of grooming that is in widespread use is that,
given an input jet, a groomer is a procedure that always returns
an output jet, although possibly with a different mass.
A tagger could then instead be construed as a procedure that might
sometimes not return an output jet (so pruning and trimming are
groomers, while the mass-drop method is a tagger).

An alternative definition of grooming comes from the 2010 Boost
report~\cite{Boost2010}, and is more restricted: grooming is
``elimination of uncorrelated UE/PU radiation from a target jet''.
With this definition, consider a signal jet, say from $W$ or top
decay: in the absence of showering, hadronisation, underlying-event or
pileup, the groomed version of the jet should be identical to the
original, ungroomed jet, because there is no radiation to groom away.
A tagger would instead be a procedure that, through a
combination of cuts (e.g.\ on an invariant mass, but also internal jet
variables), rejects background jets more often than it rejects signal
jets. 
In this definition even a simple cut on plain jet mass is to be
considered a tagging step and all the procedures that we consider here
involve both tagging and grooming elements when they are used in
conjunction with a mass cut.\footnote{The only pure groomer would be
  plain filtering~\cite{Butterworth:2008iy}.}
For simplicity we will just refer to them as taggers.

The techniques that we will be investigating have, in general, quite
complicated dynamics.
To help make their analysis tractable, we shall focus on their
behaviour for small values of the $\rho=m^2/p_t^2 R^2$ ratio, considering the
differential distribution
 $ \frac{\rho}{\sigma} \frac{d\sigma}{d\rho}\,,$
 or its integral up to some value $\rho$, $\Sigma(\rho) = \int^{\rho} d{\rho'}
 \frac{d\sigma}{d{\rho'}}$, 
 which we shall call the integrated distribution.

We will work with jet algorithms in the limit of small jet radius $R$.
This enables us to consider only the radiation from the parton that
initiated the jet, and to ignore considerations such as large-angle
radiation from other final-state partons and from the initial-state
partons. 
In practice the small-$R$ approximation is known to be reasonable even
up to quite large values of angle $\sim
1$~\cite{Dasgupta:2007wa,Dasgupta:2012hg}.

When considering multiple emissions, we will assume that they are
ordered either in angle or in energy.
This kind of approximation, together with an appropriate treatment of
the running coupling, is generally sufficient to obtain what is known
as single-logarithmic accuracy, i.e.\ terms $\as^n \ln^n m/p_t$ in
the integrated distribution.
Note that we will not always aim for single-logarithmic accuracy, and
the specific accuracy we reach will be different for each tagger, in
part because the complications that one encounters differ
substantially for each one.
In terms of choosing what accuracy to aim for, our guiding principle
will be to capture the key features of each tagger.
In many cases we will supplement our full results with versions in a
fixed-coupling approximation, often easier to assimilate, while
nevertheless encoding the essence of the results.
When examining fixed-order expansions of the results, we will label
our results with ``LO'' (leading-order) and ``NLO'' (next-to leading
order). It is understood that these expressions are not the full
fixed-order results but, rather, their logarithmic-enhanced parts. 

All of the taggers that we consider involve a parameter called $\ycut$
or $\zcut$ that effectively cuts on the energy fraction of soft
radiation. 
Since the taggers tend to be used with values of these parameters in
the range $0.05-0.15$, it will be legitimate to assume that terms
suppressed by powers of $\ycut$ or $\zcut$ can be neglected.
However, given that $\ycut$ or $\zcut$ are not usually taken
parametrically small, we shall not systematically resum logarithms of
$\ycut$ or $\zcut$, even if such a resummation could conceivably be
carried out.

Our results will apply to jets produced both at hadron colliders and
at $e^+e^-$ colliders.
We will imagine the hadron-collider jets to be produced at rapidity $y
= \frac12 \ln \frac{E+p_z}{E-p_z} = 0$, as a result of which $E = p_t$
and the boost-invariant angular separations $\Delta_{ij} = [(\phi_i -
\phi_j)^2 + (y_i - y_j)^2]^{\frac12}$ are equal to angular separations
$\theta_{ij}$ for small $\theta_{ij}$.
Thus results will be identical whether we use hadron-collider ($p_t$
and $\Delta$ based) or $e^+e^-$ ($E$ and $\theta$ based) formulations
of the jet algorithms.
For simplicity of notation we will use energies and angles as our main
variables. 

In the introduction we already defined the variable $\rho = m^2
/(p_t^2 R^2)$ (or equivalently $\rho = m^2 /(E^2 R^2)$). 
In the small-angle approximation, $\rho$ is invariant under boosts
along the jet direction, since they scale the jet $p_t$ up by some
factor (say $\gamma$) and scale its opening angle by the inverse
factor ($1/\gamma$) while leaving the mass unchanged.
Because of this invariance, the analytical results are often simplest
when expressed in terms of $\rho$, rather than separately in terms of
$m$, $p_t$ and $R$.

All jets will be assumed to have been found with the Cambridge/Aachen
(C/A) algorithm~\cite{Dokshitzer:1997in,Wobisch:1998wt}, which is the
algorithm of choice for both the mass-drop tagger and pruning. 
In its hadron-collider version, the algorithm successively recombines
the pair of particles with the smallest $\Delta_{ij}$, until no pairs
are left with $\Delta_{ij} < R$.
All objects that remain at this stage are called jets. 
The $e^+e^-$ version of the algorithm simply replaces $\Delta_{ij}$
with $\theta_{ij}$.

Finally, we will explicitly derive results only for quark-initiated jets. 
This is for reasons of brevity: gluon-initiated jets are no more
complicated to consider, usually involving just trivial modifications
of the results that we give. Results for gluon jets are collected in
appendix~\ref{sec:gluon-jets}.

The companion paper~\cite{taggersNLO}, limited to the first two
perturbative orders in $e^+e^-$ collisions, lifts the small-$R$ and
small-$\ycut$ (or $\zcut$) approximations.

\section{Recap of plain jet mass}
\label{sec:plain-jet-mass}

For concreteness, and subsequent reference, it is perhaps worthwhile
writing the integrated jet-mass distribution (for quark-initiated
jets) with the approximations mentioned above.
Let us define
\begin{subequations}
  \label{eq:jet-mass-Delta}
  \begin{align}
    D(\rho) &= \int_\rho^1 \frac{d\rho'}{\rho'} \int_{\rho'}^1 dz\,\,
    p_{gq}(z)\,\, \frac{\as(z \rho' R^2 p_t^2 ) C_F}{\pi}\,,\\[3pt]
    &\simeq \frac{\as C_F}{\pi} \left[ \frac12 \ln^2 \frac1\rho \;-\;
      \frac34 \ln \frac1\rho + \order{1}\right], \qquad \text{(fixed coupling approx.)}\,,
    \label{eq:jet-mass-Delta-b}
  \end{align}
\end{subequations}
where $p_{gq} = \frac{1+(1-z)^2}{2z}$ is the quark-gluon splitting
function, stripped of its colour factor, and the fixed-coupling
approximation in the second line helps visualise the
double-logarithmic structure of $D(\rho)$.

\begin{figure}[t]
  \centering
  \begin{minipage}[b]{0.45\linewidth}
  \caption{Lund diagrams~\cite{Andersson:1988gp} represent
    emission kinematics in terms of two variables: vertically, the
    logarithm of an emission's 
    transverse momentum $k_t$ with respect to the jet axis,
    and horizontally, the logarithm of the inverse of the emission's
    angle $\theta$ 
    with respect to the jet axis, i.e.\ its rapidity with respect to
    the jet axis.
    Here the diagram shows a line of constant jet mass, together with a shaded
    region corresponding to the part of the kinematic plane where
    emissions are vetoed, leading to a Sudakov form
    factor.} 
  \label{fig:plain-jet-mass-lund}
  \end{minipage}\hspace{0.05\linewidth}
  \begin{minipage}[b]{0.4\linewidth}
    \mbox{ }\\
    \includegraphics[width=\textwidth]{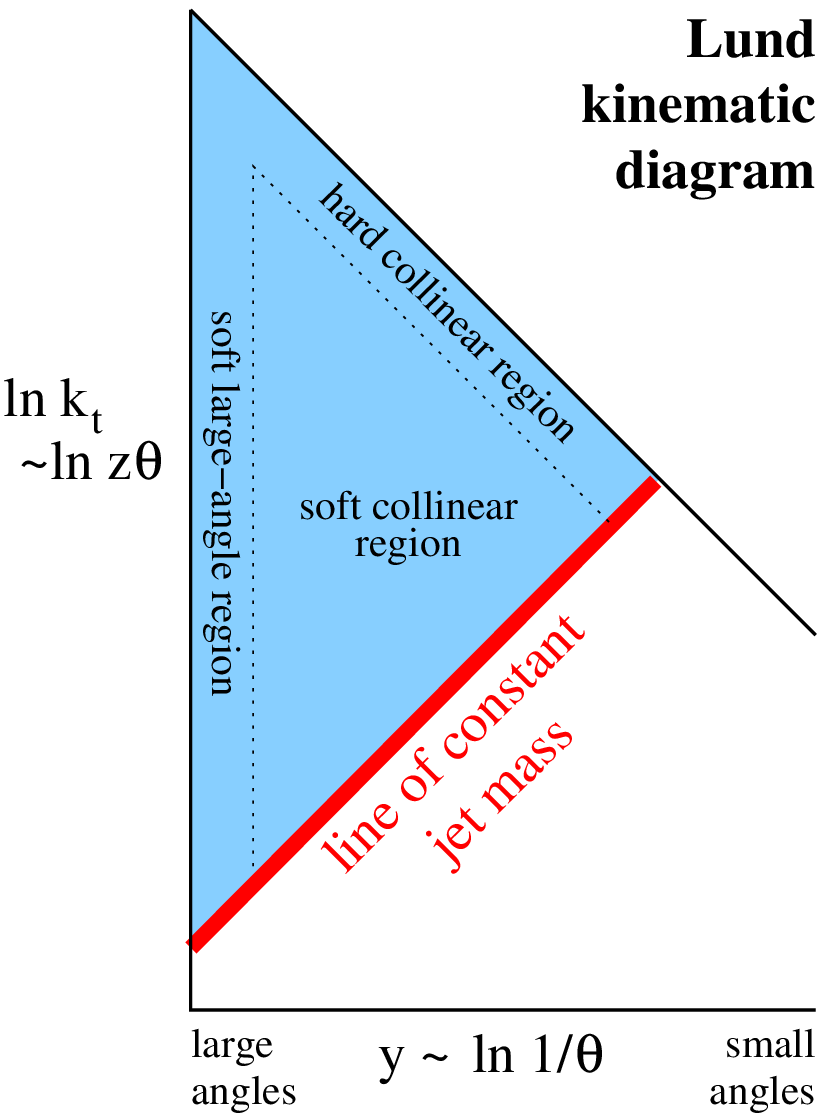}
  \end{minipage}
\end{figure}

To NLL accuracy,\footnote{Which requires the coupling in
  Eq.~(\ref{eq:jet-mass-Delta}) to run with a two-loop
  $\beta$-function, and to be evaluated in the CMW
  scheme~\cite{Catani:1990rr}, or equivalently taking into account the
  two-loop cusp anomalous dimension.} i.e.\ control of terms $\as^n
L^{n+1}$ and $\as^n L^n$ in $\ln \Sigma(\rho)$, where $L \equiv \ln
\frac1{\rho}$, the integrated jet mass distribution is given by
\begin{equation}
  \label{eq:Sigma-plain-jet-mass}
  \Sigma(\rho) = e^{-D(\rho)} \cdot \frac{e^{-\gamma_E
     D'(\rho)}}{\Gamma(1 +D'(\rho))} \cdot \cN(\rho)\,.
\end{equation}
The first factor, which is double logarithmic, accounts for the
Sudakov suppression of emissions that would induce a (squared,
normalised) jet mass greater than $\rho$.
In terms of the ``Lund'' representation of the kinematic
plane~\cite{Andersson:1988gp}, Fig.~\ref{fig:plain-jet-mass-lund}, it
accounts for the probability of there being no emissions in the shaded
region, with the $\frac12 \ln^2 1/\rho$ term in
Eq.~(\ref{eq:jet-mass-Delta-b}) for $D(\rho)$ coming from the bulk of the area (soft
divergence of $p_{gq}$), while the $-\frac34 \ln 1/\rho$ term comes
from the hard collinear region (finite $z$).
The second factor in Eq.~(\ref{eq:Sigma-plain-jet-mass}), defined in
terms of $D'(\rho) \equiv \partial_L D$, encodes the
single-logarithmic corrections associated with the fact that the
effects of multiple emissions add together to give the jet's overall
mass.
These emissions tend to be close to the constant-jet-mass boundary in
Fig.~\ref{fig:plain-jet-mass-lund}.
The third factor, also single logarithmic, accounts for modifications
of the radiation pattern in the jet (non-global
logarithms~\cite{Dasgupta:2001sh}) and boundaries of the jet
(clustering logarithms~\cite{Appleby:2002ke,Delenda:2006nf,Kelley:2012kj}) induced
by soft radiation near the jet's edge, i.e.\ near the left-hand,
vertical edge of the shaded region.
Had we been working with the anti-$k_t$ jet
algorithm~\cite{Cacciari:2008gp}, only the non-global logarithms
would have been present, which could then be parametrised (in the
large-$N_C$ limit) as a function ${\cal S}(t)$ of a variable $t(\rho)
=\frac1{2\pi} \int_\rho^1 \frac{d\rho'}{\rho'} \as({\rho'}^2 p_t^2
R^2)$~\cite{Dasgupta:2001sh}.
Note that non-global logarithms are moderately problematic, because
their
resummation~\cite{Dasgupta:2001sh,Dasgupta:2002bw,Banfi:2002hw,Hatta:2009nd,Rubin:2010fc}
has until very recently always been restricted to the large-$N_C$
limit.\footnote{
A resummation at finite $N_C$ has been performed in
Ref.~\cite{Hatta:2013iba}, using an approach initially developed in
Ref.~\cite{Weigert:2003mm}. 
  Some of the complications that occur beyond leading $N_C$ have also
  been explored in~\cite{Forshaw:2006fk}, finding terms enhanced by
  additional logarithms that are associated with emissions collinear
  to the beam directions.}
In effect, non-global logarithms are the main reason why there does
not exist a full resummation of the standard jet mass beyond NLL accuracy (for
work towards higher accuracy, see Refs.~\cite{Chien:2012ur,Jouttenus:2013hs})
and why even the NLL calculations have to neglect some of the terms
suppressed by powers of $1/N_C^2$, as done in
Ref.~\cite{Dasgupta:2012hg}.

To visualize the expected behaviour of the jet mass distribution, we
can resort to a fixed-coupling approximation, ignoring all but the
first factor in Eq.~(\ref{eq:Sigma-plain-jet-mass}), leading to the
following differential jet mass distribution
\begin{equation}
  \label{eq:jet-mass-fixed-coupling}
  \frac{\rho}{\sigma} \frac{d\sigma}{d\rho} 
  \simeq \frac{\as C_F}{\pi}\left(\ln \frac1\rho - \frac34\right)
    e^{-\frac{\as C_F}{2\pi} \left(\ln^2 \frac1\rho - \frac32
        \ln\frac1\rho + \order{1}\right)}\,.
\end{equation}
This shows a characteristic initial
growth linear in $\ln \frac1\rho$ as $\rho$ decreases, cut off by a
Sudakov suppression (the exponent) as $\rho$ decreases further. Both
of those features are visible in Fig.~\ref{fig:tagged-mass-MC}.
It is also simple to use Eq.~(\ref{eq:jet-mass-fixed-coupling}) to
analytically estimate the position of the peak in $\rho
d\sigma/d\rho$.
It is given by $L_{\text{peak}} = 1/\sqrt{\asbar} + \order{1}$, where
$\asbar = \as C_F/\pi$ for quark-jets and $\asbar = \as C_A/\pi$ for gluon-jets .
Substituting $\as=0.12$ gives a reasonable degree of agreement with the Monte Carlo
peak positions.

\section{Trimming}
\label{sec:trimming}

Trimming~\cite{Krohn:2009th}, in the variant that is most widely used
today, takes all the particles in a jet of radius $R$ and reclusters
them into subjets with a jet definition with radius $\Rsub < R$.
All resulting subjets that satisfy the condition $p_t^\text{(subjet)}
> \zcut p_t^\text{(jet)}$ are kept and merged to form the trimmed
jet.\footnote{In usual formulations
  of trimming, the parameter that we refer to as $\zcut$ is called
  $f_\text{cut}$. We use $\zcut$ in order to emphasize the connection
  with the parameters used in other taggers.}
The other subjets are discarded.
While our Monte Carlo results are obtained using the Cambridge/Aachen
algorithm (for both the original jet finding and the reclustering), at
the accuracy that we shall consider here, our analytical results will
hold independently of the jet algorithm used, at least for any member
of the generalised-$k_t$
family~\cite{Kt,KtHH,Dokshitzer:1997in,Wobisch:1998wt,Cacciari:2008gp}. 

\subsection{Leading-order calculation}

Let us first consider the situation at leading order.
If a gluon is emitted at an angle $\theta > \Rsub$ it will be
included in the final trimmed jet only if it carries an energy
fraction $z > \zcut$.
On the other hand, if it is emitted at an angle $\theta <
\Rsub$, it will be included in the same subjet as the leading
parton and will automatically pass the trimming condition.
In this case it will contribute to the jet mass independently of its
energy fraction $z$.

The above understanding leads to the following integral for the
trimmed-mass distribution,
\begin{multline}
  \label{eq:trimming-LO-full-start}
 \frac{1}{\sigma} \frac{d\sigma}{d m^2}^\text{(trim, LO)} 
  = \frac{\alpha_s C_F}{\pi} 
  \int_0^1 dz\, p_{gq}(z) \int \frac{d\theta^2}{\theta^2} \,
  \delta\!\left(m^2 - z(1-z) p_t^2 \theta^2 \right) 
  \times \\ \times
  \Big[
    \Theta\left(z - \zcut \right) \Theta\left(1-z - \zcut \right)
    \Theta(\theta^2- \Rsub^2) 
    + \Theta(\Rsub^2 - \theta^2)
  \Big]
  \Theta \left (R^2 - \theta^2 \right).
\end{multline}
It is straightforward to evaluate this for any value of
$\zcut$~\cite{taggersNLO}, but the expressions that we obtain and the
subsequent resummation will be much simpler if we assume that $\zcut$
is small (as it usually is in practice), so that we can neglect terms
suppressed by powers of $\zcut$.
Working furthermore in the approximation $m^2 \ll p_t^2 R^2$,
i.e. $\rho \ll 1 $, and making use of the fact that $p_{gq}(z)$ is
finite for $z \to 1$, we can then discard the middle $\Theta$-function
in the first term in square brackets and ignore the $(1-z)$ factors in
the $\delta$-function.
One may then reorganise the contents of the second line so as to
obtain
\begin{multline}
  \label{eq:trimming-start}
 \frac{1}{\sigma} \frac{d\sigma}{d m^2}^\text{(trim, LO)} 
  = \frac{\alpha_s C_F}{\pi} 
  \int_0^1 dz\, p_{gq}(z) \int \frac{d\theta^2}{\theta^2} \,
  \delta\!\left(m^2 - z p_t^2 \theta^2 \right) 
  \times \\ \times
  \Big[
    \Theta\left(z - \zcut \right) \Theta \left (R^2 - \theta^2 \right)
    + \Theta\left(\zcut -z \right)  \Theta(\Rsub^2 - \theta^2)
  \Big]
  \,.
\end{multline}
Carrying out the integration over $\theta$, and expressing the result
in terms of $\rho$ and $r \equiv \Rsub/R$ gives
\begin{equation}
  \label{eq:trimming-next}
 \frac{\rho}{\sigma} \frac{d\sigma}{d \rho}^\text{(trim, LO)} 
  = \frac{\alpha_s C_F}{\pi}  
  \int_0^1 dz \, p_{gq}(z) \,
  \Big[
    \Theta\left(z - \zcut \right) \Theta \left (z - \rho \right)
    + \Theta\left(\zcut -z \right)  \Theta(z r^2 - \rho)
  \Big]
  \,.
\end{equation}
The remaining $z$ integral is straightforward to evaluate and leads to
the following result:
\begin{multline}
  \label{eq:trimming-LO}
  \frac{\rho}{\sigma} \frac{d\sigma}{d\rho}^\text{(trim, LO)} 
  = \frac{\as C_F}{\pi} 
  \left[ 
    \Theta\left( \rho-\zcut  \right) \ln \frac{1}{\rho} +
    \Theta\left(\zcut- \rho \right) \ln \frac{1}{\zcut} 
    - \frac34  
  \right. + \\ + \left. 
     \Theta\left(\zcut r^2 - \rho \right)  \ln \frac{\zcut r^2}{\rho}
  \right].
\end{multline}
For $\rho > \zcut$ this is simply the same as the leading-order jet
mass distribution, with a linear growth of the distribution as $\ln
1/\rho$. In the integrated distribution $\Sigma(\rho)$, this
corresponds to an $\as L^2$ growth, with the two powers of $L$
associated with simultaneous soft and collinear divergences.
For $\rho < \zcut$ but $\rho > r^2 \zcut$ still, the $\zcut$ condition
tames the soft divergence: the integrated distribution then goes as
$\as L \ln \frac{1}{\zcut}$, dominated by just the collinear
divergence.
However, because the $\zcut$ condition is applied only to subjets
separated by at least $\Rsub$ from the main jet, this
taming is short-lived: small jet masses with arbitrarily small $z$
values can come from angular regions $\theta < \Rsub$.
As a result, for $\rho < r^2 \zcut$, the structure of the result
reverts to that for a standard jet mass,
\begin{equation}
  \label{eq:trimming-LO-small-mass}
  \frac{\rho}{\sigma} \frac{d\sigma}{d\rho}^\text{(trim, LO)} 
  = \frac{\as C_F}{\pi} \left[  \ln \frac{r^2}{\rho} - \frac34 \right],
  \qquad\quad \rho < \zcut r^2\,,
\end{equation}
albeit with a reduced radius, $\Rsub \equiv r R$.

\begin{figure}
  \centering
  \includegraphics[width=0.32\textwidth]{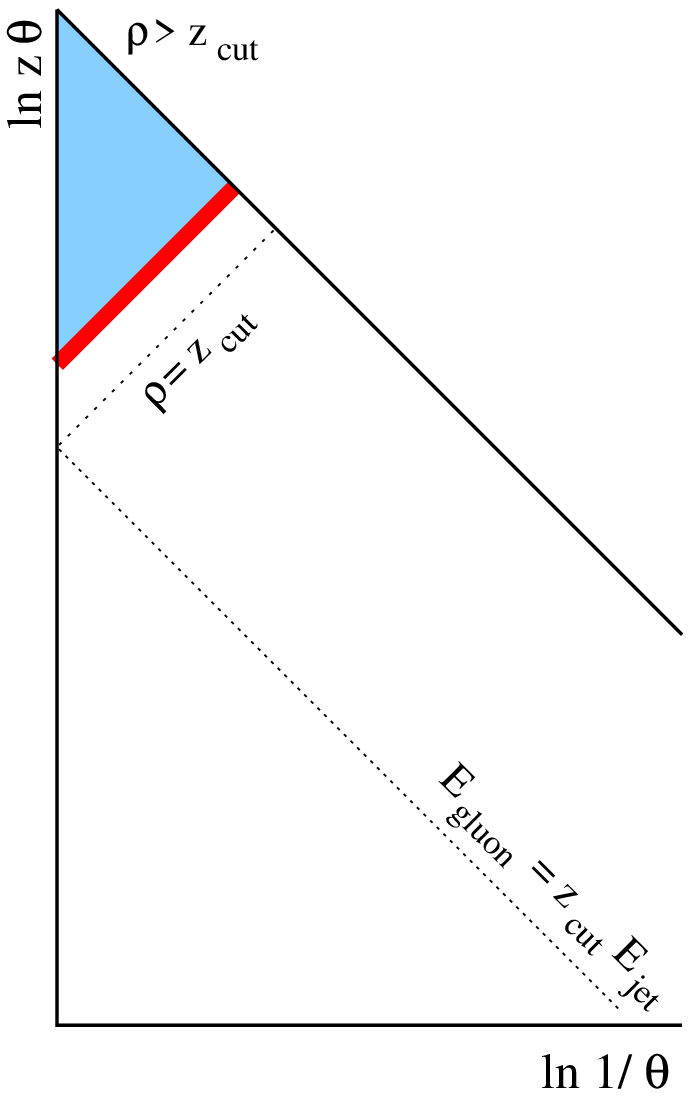}\hfill
  \includegraphics[width=0.32\textwidth]{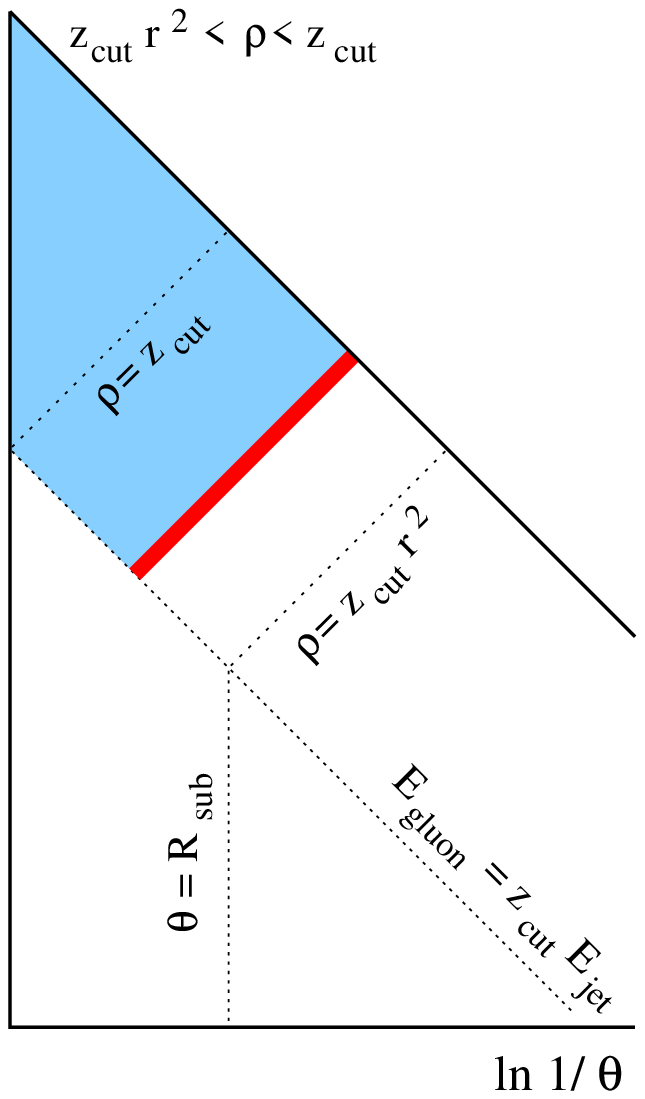}\hfill
  \includegraphics[width=0.32\textwidth]{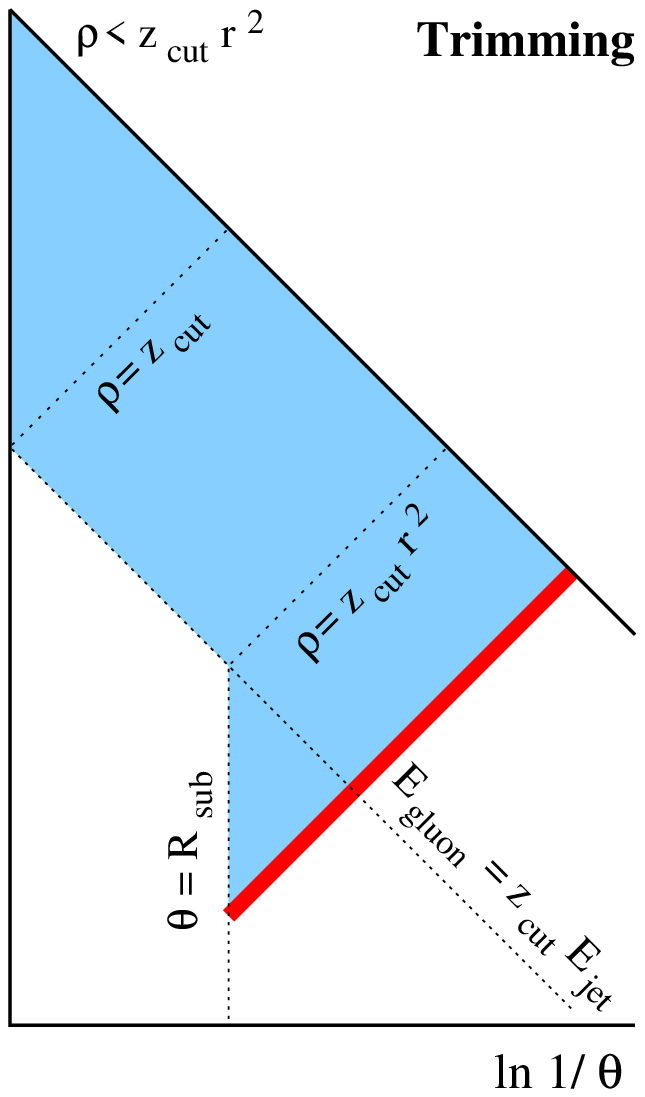}\hfill
  \caption{Lund kinematic diagrams for trimming, considering three
    different possible values of $\rho$. 
    In each case, to obtain the given value of $\rho$, there must be
    an emission somewhere along the thick (red) line, and there must be
    no emissions in the shaded region.
    Emissions in the unshaded regions have no impact on the trimmed
    jet mass.
    Dotted lines serve to indicate transition regions in the kinematic
    plane and their relation to the parameters of the trimmer.}  
  \label{fig:trimmed-jet-mass-lund}
\end{figure}

The three situations for the trimmed jet mass can be visualised in
Fig.~\ref{fig:trimmed-jet-mass-lund} with the help of appropriate Lund
kinematic diagrams.
The LO integrated cross section $\Sigma(\rho)$ is proportional to the
area of the shaded regions, and the differential cross section
proportional to the length of the thick (red) line.
For $\rho > \zcut$ the integrated cross section corresponds to a
triangular region, hence a dependence on $L^2$.
For $\rho < \zcut$ but $\rho > r^2 \zcut$, the extra contribution to
the integrated cross section comes from a rectangular region, with one
side growing with $L$ and the other of fixed length $\simeq \ln
1/\zcut$. 
This gives an integrated cross section that grows as $L \ln 1/\zcut$,
i.e.\ with only one power of $L$.
Finally for $\rho < r^2 \zcut$ there is once more a triangular region,
and so a dependence on $L^2$.

\subsection{Resummed calculation}

Thanks to the above considerations it is relatively straightforward to
obtain an understanding of the all-order trimmed jet-mass
distribution.
The key result that we use from the extensive literature on
event-shape and jet-mass resummations (see
e.g.~Ref.\cite{Catani:1992ua,Dokshitzer:1998kz}) is that one can
effectively use an independent-emission approximation, ignoring
subsequent splittings of those emissions, other than in the treatment
of the running coupling.
This can be understood as a consequence of angular ordering and is
sufficient to derive all of Eq.~(\ref{eq:Sigma-plain-jet-mass}) except
for the non-global terms.
This approach is not necessarily appropriate for all taggers, however
it will be suitable for most of the cases in this paper where we give
a final resummed answer.
The resummation is most easily written for the integrated cross
section, involving a sum over an arbitrary number of independent
emissions and corresponding virtual corrections.
We parametrise each emission in terms of its momentum fraction $z_i =
E_i/E_\text{jet}$\footnote{There is a potential subtlety as to whether
  the denominator should be the jet energy or the energy that remains
  after all emissions $1 \ldots (i-1)$. At our accuray the difference is
  irrelevant, as discussed in the context of the mass-drop tagger in
  appendix~\ref{sec:finite-ycut-modMD}. } and its individual contribution $\rho_i = z_i
\theta_i^2/R^2$ to the squared, normalised jet mass:
\begin{multline}
  \label{eq:all-orders-trimming-start}
  \Sigma^{\text{(trim)}}(\rho) = 
  \sum_{n=0}^\infty 
  \frac{1}{n!}
  \prod_{i=1}^{n}
    \int
    dz_i p_{gq}(z_i)\frac{d\rho_i}{\rho_i} 
    \frac{\alpha_s(\rho_i z_i p_t^2 R^2) \, C_F}{\pi} 
    \bigg[
      \Theta(\rho - \rho_i) 
      \,+ \\
      + \Theta(\rho_i - \rho) \Theta(\zcut - z_i) \Theta(\rho_i / r^2
      - z_i)
     -1
    \bigg]
    \Theta(z_i - \rho_i)\,,
\end{multline}
There are three terms in the square brackets: the last one
corresponds to virtual corrections, while the first two correspond to
different regions of real phase-space: the first states that we can
sum over any emission whose individual contribution is $\rho_i <
\rho$;
the second states that we can sum over emissions with $\rho_i > \rho$,
{\it if} they are trimmed away, i.e.\ have $z < \zcut$ and $\theta_i >
\Rsub$ (which is straightforward to express as a condition on
$z_i < \rho_i/r^2$).
The total contents within the square brackets equal $-1$ in the shaded
kinematic regions of Fig.~\ref{fig:trimmed-jet-mass-lund} and $0$
elsewhere. 

The sum over $n$ in Eq.~(\ref{eq:all-orders-trimming-start}) simply
leads to an exponential and we can write the final result as
\begin{multline}
  \label{eq:trimming-LL-small-mass}
  \Sigma^\text{(trim)} (\rho)
  =
  \exp\Bigg[ 
    - D(\max(\zcut,\rho)) 
    - S(\zcut,\rho) \Theta(\zcut - \rho)
    \\ \left.
    - \Theta(\zcut r^2 - \rho) \int_\rho^{\zcut r^2} \frac{d\rho'}{\rho'}
    \int^{\zcut}_{\rho'/r^2} \frac{dz}{z} \frac{C_F}{\pi}\, \as(\rho' z p_t^2 R^2)
  \right]\,.
\end{multline}
where $D$ was defined in Eq.~(\ref{eq:jet-mass-Delta}) and the function $S$ is given by
\begin{subequations}
  \label{eq:Sab}
  \begin{align}
    S(a,b) &\equiv \frac{C_F}{\pi} \int^a_b \frac{d\rho'}{\rho'} \int_{\zcut}^1
    dz \,p_{gq}(z) \, \as(\rho' z \, p_t^2 R^2)\,,\\[5pt]
    &\simeq \frac{\as C_F}{\pi} \left[ \ln \frac{1}{\zcut} - \frac34 +
      \order{\zcut} \right] \ln \frac{a}{b}, \qquad \text{(fixed coupling approx.)}
  \end{align}
\end{subequations}
and contains only single logarithms, $\as^n \ln^n \frac{a}{b}$
(treating powers of $\ln \frac1{\zcut}$ as finite coefficients).
To help better visualise structure of
Eq.~(\ref{eq:trimming-LL-small-mass}), one may prefer 
to examine its closed form for fixed coupling:
\begin{multline}
  \label{eq:trimming-LL-small-mass-FC}
  \Sigma^\text{(trim)} (\rho)
  \simeq
  \exp\left[ -\frac{\as C_F}{2\pi} \left( - \frac32 \ln \frac1{\rho} \;+ \;
      \Theta(\rho - \zcut) \ln^2 \frac{1}{\rho} + 
      \right.\right. \\ \left.\left.
      + \Theta(\zcut -\rho) \left(\ln^2 \frac{1}{\zcut} + 2\ln
        \frac{\zcut}{\rho} \ln \frac{1}{\zcut}\right) 
      + \Theta(\zcut r^2 -\rho) \ln^2 \frac{\zcut r^2}{\rho}
    \right)
  \right]\,.
\end{multline}

Eq.~(\ref{eq:trimming-LL-small-mass}) resums terms $\as^n L^{2n}$ and
$\as^n L^{2n-1}$ in $\Sigma(\rho)$ (neglecting finite $\zcut$ effects
and terms enhanced by powers of $\ln \zcut$).
It also resums all terms $\as^n L^{n+1}$ in $\ln \Sigma(\rho)$.
To obtain what is commonly referred to as NLL accuracy, i.e.\ all
terms $\as^n L^{n}$ in $\ln \Sigma(\rho)$, would require a treatment
of several additional effects: the two-loop $\beta$-function and cusp
anomalous dimension,  
non-global logarithms involving resummation of terms $\ln (\zcut^2
r^2/\rho)$, related clustering logarithms, and multiple-emission
effects on the observable.
The clustering logarithms will depend on the jet algorithm used for
the trimming, but the rest of the structure will be independent of
this (as long as the algorithm belongs to the generalised-$k_t$ family).
These terms are all relatively straightforward to include, since they
follow the structure of the plain jet-mass distribution.
However, we leave their study to future work.
Analogous results can be also derived for gluon-induced jets. Explicit expressions are collected in appendix~\ref{sec:gluon-jets}.

\subsection{Comparison with Monte Carlo results}
\label{sec:trim-MC}
\begin{figure}
  \centering
      \includegraphics[width=0.49\textwidth]{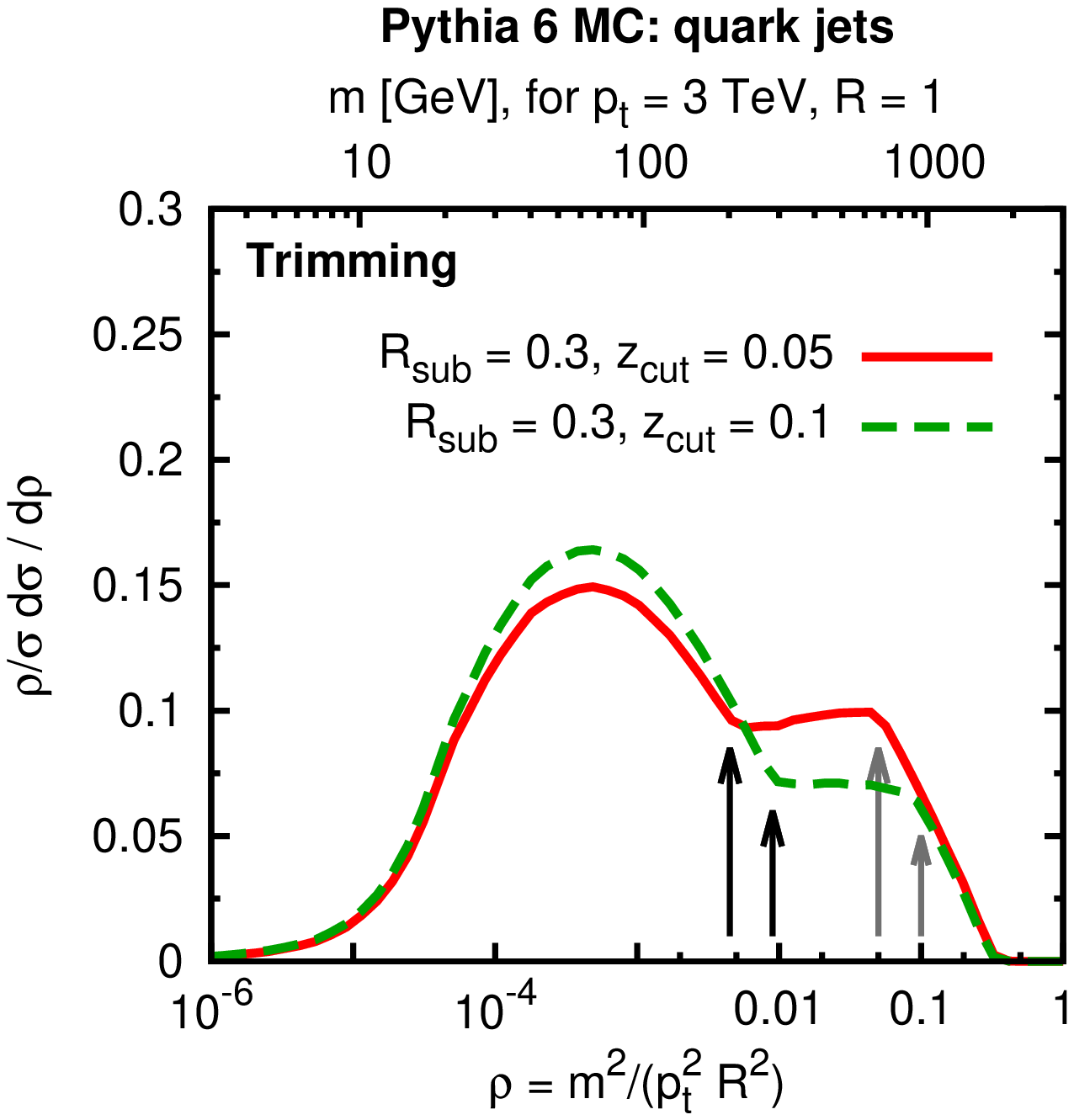}\hfill
      \includegraphics[width=0.49\textwidth]{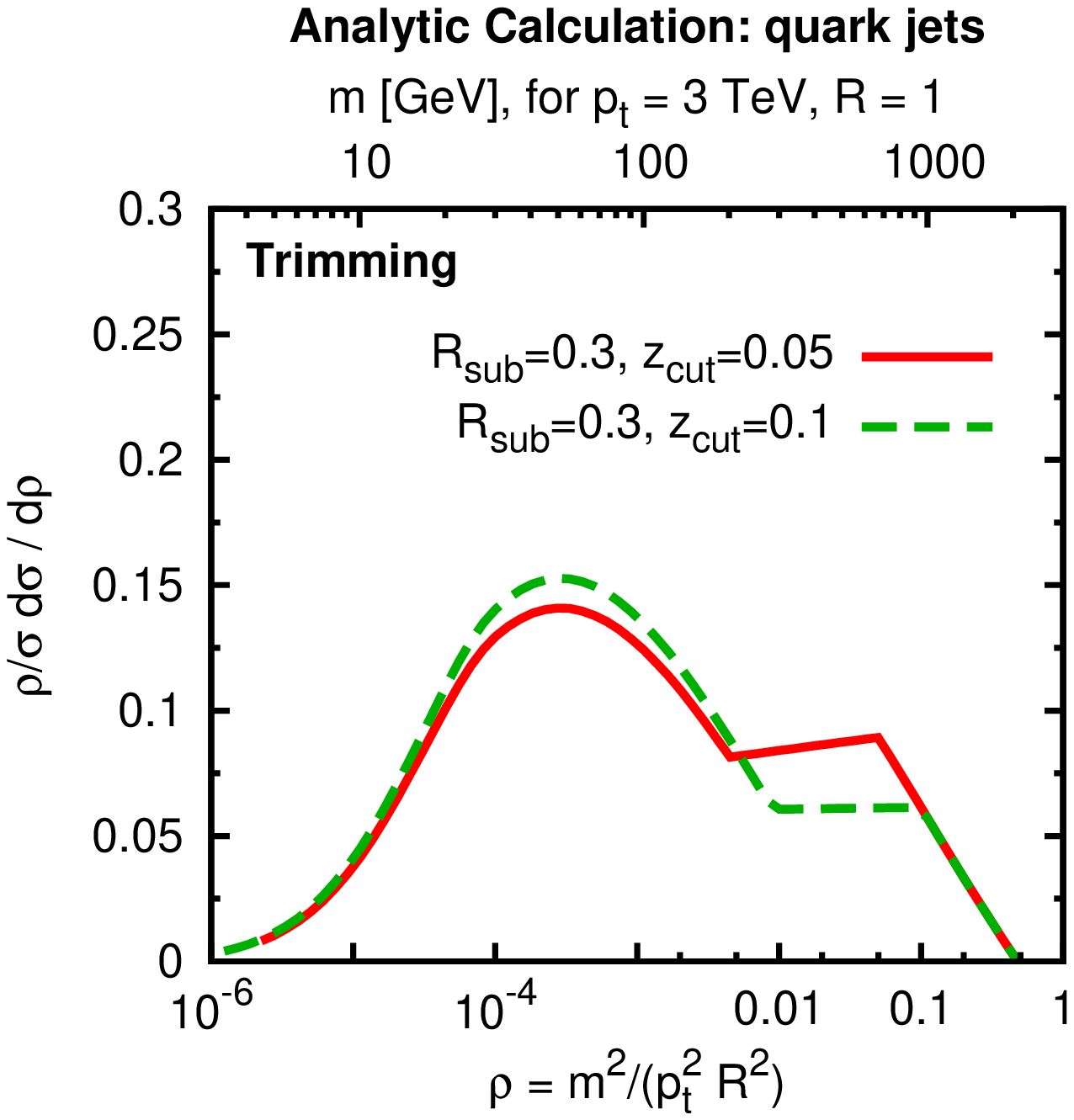}\\[5pt]
      \includegraphics[width=0.49\textwidth]{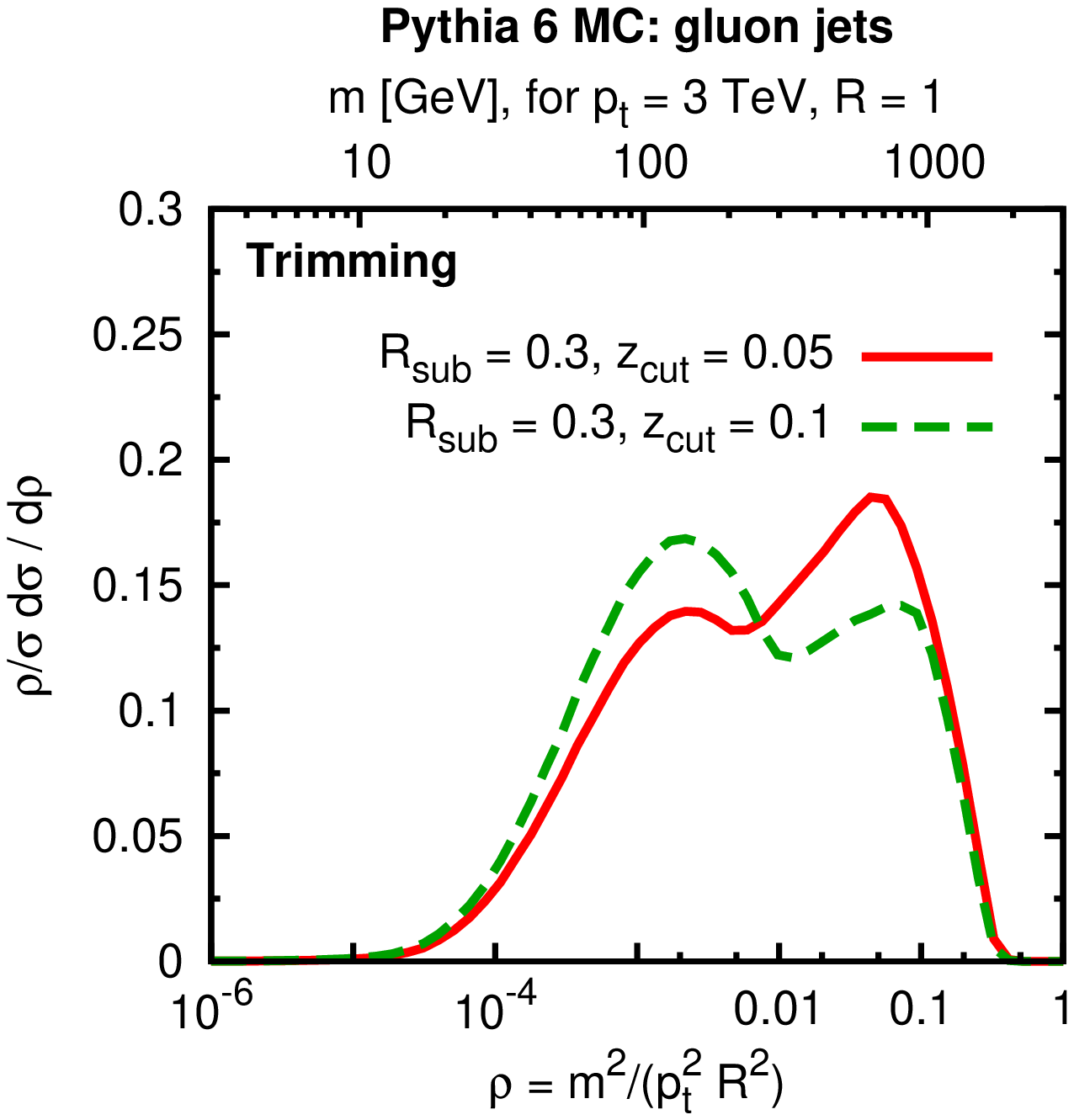}\hfill
      \includegraphics[width=0.49\textwidth]{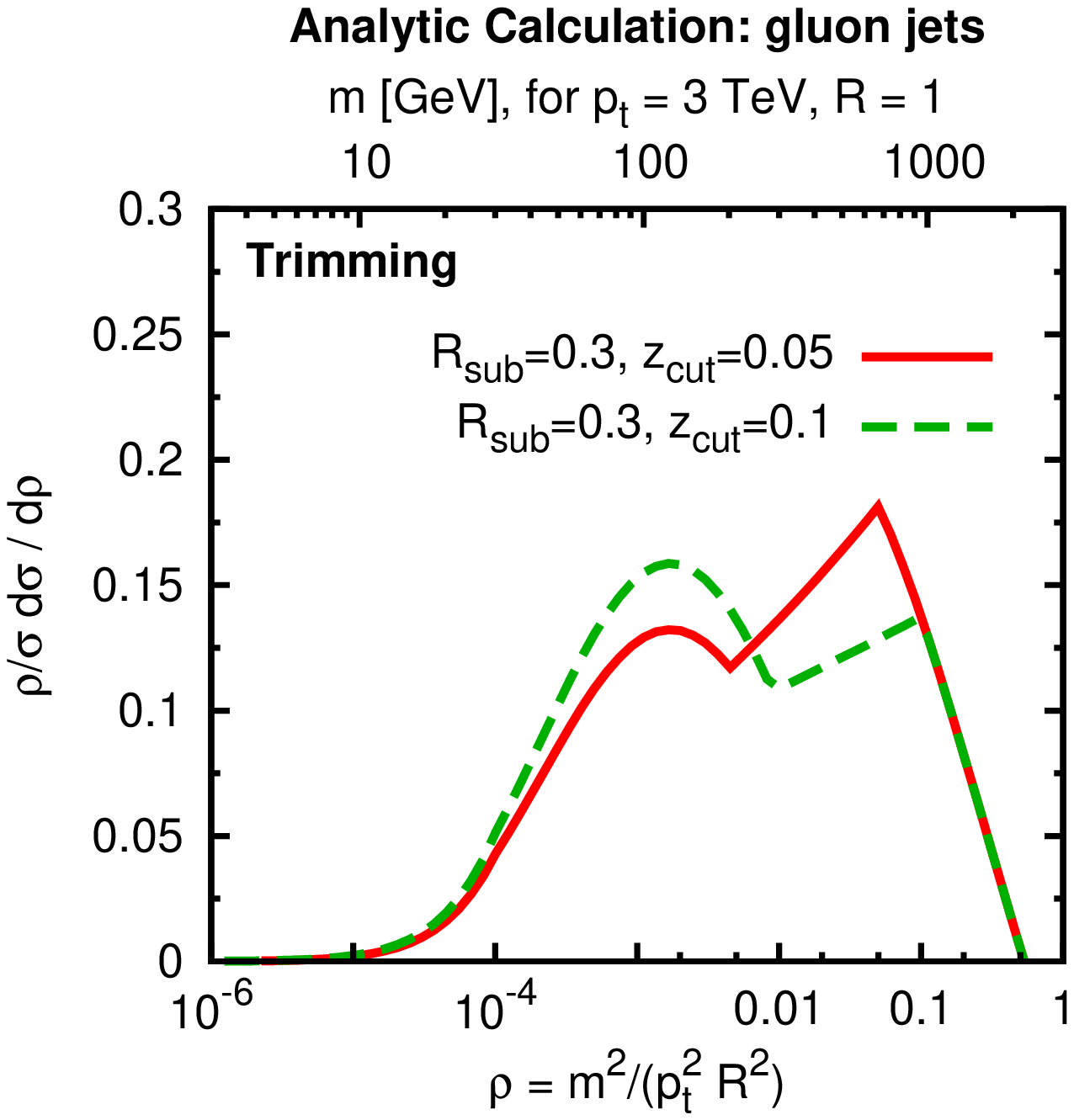}
  \caption{%
    Comparison of Monte Carlo (left panels) and analytic
    results (right panels) for trimming. The upper panels are for
    quark jets, the lower panels for gluon jets. 
    Two sets of trimming parameters are illustrated. 
    In the upper left panel, arrows indicate the expected transition
    points, at $\rho = r^2 \zcut$ (in black) and $\rho = \zcut$ (in
    grey), where $r = \Rsub/R$.  
    The details of the MC event generation are as for
    Fig.~\ref{fig:tagged-mass-MC}.  
  }
  \label{fig:trimming-MC-analytic}
\end{figure}

One test of Eq.~(\ref{eq:trimming-LL-small-mass}) is to compare it
to the Monte Carlo results. 
We do this in Fig.~\ref{fig:trimming-MC-analytic} where the left-hand
plots show the trimmed-mass distribution as obtained with Monte Carlo simulation
and the right-hand plots shows the corresponding analytical results.\footnote{ 
Resummed expressions for the various taggers (as well as for the plain jet mass) contain integrals of the strong coupling $\as(k_t^2)$. In order to evaluate these integrals down to low scales, we must introduce a prescription to deal with the non-perturbative region. We decide to freeze the coupling below a non-perturbative scale $\mu_\text{NP}$:
\begin{equation*}\label{eq:coupling-freezing}
\as (k_t^2)= \as^\text{1-loop}(k_t^2)\Theta\left (k_t^2-\mu_\text{NP}^2\right)+\as^\text{1-loop}(\mu_\text{NP}^2)\Theta\left (\mu_\text{NP}^2-k_t^2\right)\,,
\end{equation*}
where $\as^\text{1-loop}(k_t^2)$ is the usual one-loop expression for
the strong coupling, i.e.\ its running is evaluated with $\beta_0$
only.  We use $\as(m_Z)=0.118$, $n_f=5$ and $\mu_\text{NP}=1$~GeV
throughout this paper.} 
The upper row is for quark-initiated jets, while the lower one is for
gluon-initiated jets.
Two sets of trimming parameters are shown, to help visualize the
dependence on them.

The three regions of $\rho$ are clearly distinguishable in each plot,
with a close correspondence of the Monte Carlo and analytic shapes and
transition points, as well as their dependence on the trimming
parameters.
Specifically, in the case of quark jets, for $\rho > \zcut$, one sees a linear rise with $\ln
1/\rho$. 
For $\rho < \zcut$, down to $\rho = r^2
\zcut$ there is an approximate plateau, whose height increases for
smaller $\zcut$, as expected from the $\ln 1/\zcut$ term for this
region in the LO formula, Eq.~(\ref{eq:trimming-LO}).
For $\rho < r^2 \zcut$, the linear rise starts again, but is quickly
suppressed by a Sudakov form factor, giving the usual jet-mass type
peak. The case of gluon-initiated jets is similar, although the
single-logarithmic region is not flat, because of the specific choices
of $\zcut$. 

Insofar as $\zcut$ and $\Rsub$ are not too small, the peak position
is essentially given by the peak position for the mass of a jet of
size $\Rsub$ rather than $R$,
\begin{equation}
  \label{eq:trimming-peak}
  L_{\text{peak}}^\text{trim} = \frac1{\sqrt{\asbar}} - 2 \ln r +
  \order{1}\,. 
\end{equation}
i.e.\ at a $\rho$ value that is a factor $r^2$ smaller than for the
plain jet mass.
This is consistent with what is observed comparing the Monte Carlo
results for the plain and trimmed jet masses.
A final comment is that while the peak position is independent of
$\zcut$, its height is not: the smaller the value of $\zcut$, the
greater the Sudakov suppression associated with vetoing emissions in
the range $\zcut r^2 < \rho < \zcut$, and so the smaller the peak
height, again in accord with the Monte Carlo results.

\section{Pruning}
\label{sec:pruning}

Pruning~\cite{Ellis:2009me,Ellis:2009su} takes an initial jet, and
from its mass deduces a pruning 
radius $R_\text{prune} = \Rfact \cdot \frac{2m}{p_t}$, where $\Rfact$
is a parameter of the tagger.
It then reclusters the jet and for every clustering step, involving
objects $a$ and $b$, it checks whether $\Delta_{ab} > R_\text{prune}$
and $\min(p_{ta},p_{tb}) < \zcut p_{t,(a+b)}$, where $\zcut$ is a second
parameter of the tagger.
If so, then the softer of the $a$ and $b$ is discarded.
Otherwise $a$ and $b$ are recombined as usual.
Clustering then proceeds with the remaining objects, applying the
pruning check at each stage.

In analysing pruning, we will take $\Rfact = \frac12$, i.e.\ its
default suggested value~\cite{Ellis:2009su}.
In analogy with our approach for trimming, we will work in the limit
of small $\zcut$ (but $\ln \zcut$ not too large).
We will assume that the reclustering is performed with the
Cambridge/Aachen algorithm, the most common choice, and that adopted
by CMS~\cite{CMS-substructure-studies}.
ATLAS~\cite{Aad:2013gja} have instead performed the reclustering with
the $k_t$ algorithm~\cite{Kt,KtHH}). 
Similar methods could be used to study that case, but we leave such an
investigation to future work.

\subsection{Leading-order calculation}

At leading order, i.e.\ a jet involving a single $1\to 2$ splitting, $R_\text{prune} = \frac{m}{p_t} = \Delta_{ab} \sqrt{z(1-z)}$,
which guarantees that $\Delta_{ab}$ is always larger than
$R_\text{prune}$.
To establish the pruned jet mass, one then needs to examine the second
part of the pruning condition: if $\min(z,1-z) > \zcut$ then the
clustering is accepted and the pruned jet has a finite mass.
Otherwise the pruned jet mass is zero. 
This pattern is true independently of the angle between the two
prongs.
This leads to the following result for the mass distribution:
\begin{align}
  \label{eq:LO-pruning-start}
  \frac{1}{\sigma} \frac{d\sigma}{d m^2}^\text{(prune, LO)} 
  &= \frac{\alpha_s C_F}{\pi} 
  \int dz \,p_{gq}(z) \, \frac{d\theta^2}{\theta^2} \,
  \delta\!\left(m^2 - z(1-z) p_t^2 \theta^2 \right) 
  \times \\ &\qquad \qquad \qquad\nonumber \times
  \Theta\left(z - \zcut \right) 
  \Theta\left((1-z) - \zcut \right) 
  \Theta \left (R^2 - \theta^2 \right),
  \\
  \label{eq:LO-pruning-thetadone}
  &= \frac{\alpha_s C_F}{\pi} 
  \int dz\, p_{gq}(z) \, \frac{1}{m^2} \,
  \Theta\left(z - \zcut \right) 
  \Theta \left (z - \frac{m^2}{p_t^2 R^2} \right),
\end{align}
where to obtain  the last line we have made use of the fact that $\zcut$ is
small and that the integral is dominated by the region $z\ll 1$.
The final $z$-integration is straightforward to perform and gives
\begin{equation}
  \label{eq:pruning-LO}
  \frac{\rho}{\sigma} \frac{d\sigma}{d\rho}^\text{(prune, LO)} 
  = \frac{\as C_F}{\pi} 
  \left[ 
    \Theta(\rho - \zcut) \ln \frac{1}{\rho} +
    \Theta(\zcut-\rho) \ln \frac1\zcut - \frac34
  \right].
\end{equation}
This has the structure of a rise linear in $\ln \rho$ for $\rho$ down
to $\zcut$, and then it is constant below. 
For small $\rho$, the corresponding integrated cross section has the
remarkable property that it contains no double-logarithmic terms,
i.e.\ no $\as L^2$ contribution.
This is, in a certain sense, what pruning was, in our understanding,
intended to 
achieve: the double-log contribution comes from the region of
arbitrarily soft gluon emission, and pruning removes such soft
emissions. 

\subsection{3-particle configurations: $\sanepruning$ and
  $\anomalouspruning$} 

\begin{figure}
  \centering
  \includegraphics[height=0.25\textwidth]{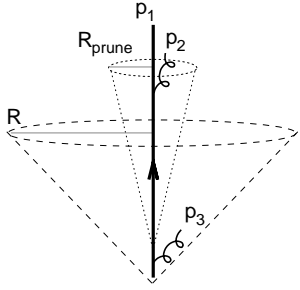}
  \caption{Configuration that illustrates generation of double logs
    in pruning at $\order{\as^2}$.
    Soft gluon $p_3$ dominates the jet mass, thus determining the pruning
    radius. 
    However, because of $p_3$'s softness, it is then pruned away,
    leaving only the central core of the jet, which has a usual
    double-logarithmic type mass distribution.  
  }
  \label{fig:pruning-double-log-config}
\end{figure}

When we consider 3-particle configurations the behaviour of pruning
develops a certain degree of complexity.
Fig.~\ref{fig:pruning-double-log-config} illustrates the type of
configuration that is responsible: there is a soft parton ($p_3$) that
dominates the total jet mass and so sets the pruning radius,
but it does not pass the pruning $\zcut$ threshold, meaning that it does not
contribute to the pruned mass; meanwhile there is another parton
($p_2$), within the pruning radius, that contributes to the pruned jet
mass independently of how soft it is.
We call this ``$\anomalouspruning$'', because at the angular scale
$\Rprune$, the final pruned jet consists of a single prong.
It is to be contrasted with the type of configuration that contributed
to the leading order result Eq.~(\ref{eq:pruning-LO}), for which at an
angular scale $\Rprune$, the pruned jet always consisted of two prongs. 
That we call ``$\sanepruning$''.\footnote{In preliminary
  presentations given about this work, the working names that had been
  used for $\sanepruning$ and $\anomalouspruning$ were, respectively,
  ``sane'' and ``anomalous'' pruning.} 
  
Let us work through $\anomalouspruning$ quantitatively. 
For gluon $3$ to be discarded by pruning it must have $z_3 < \zcut \ll
1$, i.e.\ it must be soft.
Then the pruning radius is given by $R_\text{prune}^2 = z_3
\theta_3^2$ and for $p_2$ to be within the pruning core we have
$\theta_2 < R_\text{prune}$.
This implies $\theta_2 \ll \theta_3$, which allows us to treat $p_2$
and $p_3$ as being emitted independently (i.e.\ due to angular
ordering) and also means that the C/A algorithm will first cluster
$1+2$ and then $(1+2)+3$.
The leading-logarithmic contribution that one then obtains at
$\order{\as^2}$ is 
\begin{subequations}
\begin{align}
  \label{eq:prtdl}
  \frac{\rho}{\sigma} \frac{d\sigma^\anompruneNLO}{d\rho}
  &\simeq \left(\frac{C_F \alpha_s}{\pi} \right) ^2 \int_0^{\zcut}\frac{d z_3}{z_3} 
  \int^{R^2}\frac{d\theta_3^2}{\theta_3^2} \int_{0}^{1} 
  \frac{dz_2}{z_2} \int_0^{z_3 \theta_3^2} \frac{d\theta_2^2}{\theta_2^2}\; 
  \rho \,\delta \left(\rho-z_2 \frac{\theta_2^2}{R^2} \right)
  \\ 
  &= \left(\frac{C_F \alpha_s}{\pi}\right)^2 \frac{1}{6} \ln^3
  \frac{\zcut}{\rho}+ \ord\left(\as^2 \ln^2 \frac{1}{\rho} \right),
  \qquad\text{(valid for $\rho < \zcut$)},
\end{align}
\end{subequations}
where we have directly taken the soft limits of the relevant splitting
functions.

The $\ln^3 \rho$ contribution that one observes here in the
differential distribution corresponds to a double logarithmic
($\alpha_s^2 \ln^4 \rho$) behaviour of the integrated cross-section,
i.e.\ it has as many logs as the raw jet mass, with both soft and
collinear origins.
This term is the first of a whole tower of terms $\as^n \ln^{2n}
\rho$, all associated with configurations where the emission(s) that
set the total jet mass are discarded during pruning, leaving just the
mass of the core of the jet (at angles smaller than
$R_{\text{prune}}$).

In general, substructure taggers aim to eliminate contributions from
soft emission.
What we see here is that this is not entirely the case for pruning.
However, in an experimental analysis, it is easy to diagnose whether
configurations such as that in
Fig.~\ref{fig:pruning-double-log-config} have arisen.
Accordingly, we introduce explicit operative definitions for $\anomalouspruning$
and its converse, $\sanepruning$:
\begin{itemize}
\item[ ] \textbf{$\Sanepruning$:} if at any stage during the sequential
  recombination there was a clustering that satisfied the $\Delta_{ab}
  > R_\text{prune}$ condition and the requirement
 $\frac{\min(p_{t,a},p_{t,b})}{p_{t,(a+b)}} > \zcut$, the jet is deemed to
  pass the $\sanepruning$ (i.e.\ two-prong) requirement. The jet mass was dominated by
  (semi)-hard radiation and it is likely that the pruning radius was set
  appropriately for that radiation.\footnote{It is equally
    possible to define ``Y-trimming'', which supplements
    trimming with the requirement that at least two subjets must pass
    the trimming cuts. Because Y-trimming involves a fixed
    subjet radius, it is of more limited 
    phenomenological interest than \sanepruning, and we leave its
    discussion to appendix~\ref{sec:Ytrimming}. }
\item[ ] \textbf{$\Anomalouspruning$:} if during the sequential
  recombination there was never a clustering satisfying the
  $\Delta_{ab} > R_\text{prune}$ condition and the requirement
  $\frac{\min(p_{t,a},p_{t,b})}{p_{t,(a+b)}} > \zcut$, the jet is deemed to
  belong to the $\anomalous$- (i.e.\ one-prong) pruned
  class. Typically, for this class of jets, the jet mass was
  dominated by soft emissions, leading to a pruning radius
  that had no relation to any hard substructure potentially
  present in the jet. 
\end{itemize}

According to our first definition of grooming and tagging in
section~\ref{sec:def}, generic pruning is a grooming procedure:
given an initial jet, 
there is always a corresponding pruned jet, though often with a different
mass.
In contrast, according to that same definition, $\sanepruning$ is a
tagger: i.e.\ given some initial jet, there will not always be a
corresponding $\sane$-pruned jet. 
In the Monte Carlo results that we will discuss below in
section~\ref{sec:pruning-analytic-VS-MC}, for our default choice of
pruning parameters, $\sanepruning$ tags about $40\%$ of QCD
jets. 

Let us examine the $\as^2$ contribution for $\sanepruning$.
Physically, the key addition relative to the LO result (for which we exclusively have $\sanepruning$) is the requirement that there should have been
no radiation $p_3$ that would set a pruning radius larger than
$\theta_2$, i.e.\ no radiation with $\rho_3 \equiv z_3 \theta_3^2 >
\theta_2^2$.
Insofar as we neglect logarithms of $\zcut$, we can replace this with
the condition $\rho_3 > \rho_2\equiv \rho$, resulting in a structure
up to $\as^2$ of
\begin{equation}
  \label{eq:sane-prune}
  \frac{\rho}{\sigma} \frac{d\sigma^\sanepruneNLO}{d\rho}  
   \simeq \frac{\as C_F}{\pi} \left(\ln \frac{1}{\zcut} -
     \frac34\right)
   \times \left[1 -\frac{\as C_F}{2\pi} \ln^2 \frac{1}{\rho}
   \right]\,,\qquad
 \rho < \zcut^2\,.
\end{equation}
where the round bracket comes (as at LO), from the integral over
allowed $z_2$ values, and we have used a double-logarithmic
approximation for the contents of the square brackets.
Translating to the integrated distribution, Eq.~(\ref{eq:sane-prune})
implies the presence of a term of the form $\as^2 \ln^3 1/\rho$, i.e.\
with one logarithm fewer than the $\anomalouspruning$ contribution.
As we shall see below, this difference will be related to highly
distinct resummation structures for the two types of contribution.

\subsection{Resummed results}

\begin{figure}
  \centering
  \includegraphics[width=0.32\textwidth]{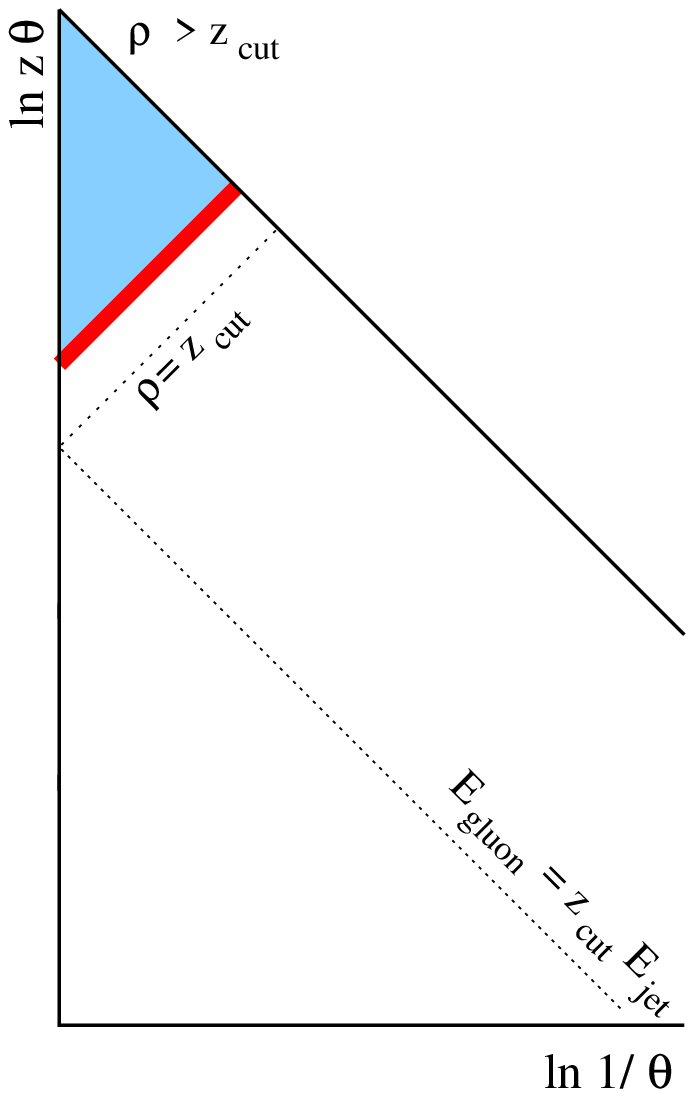}\hfill
  \includegraphics[width=0.32\textwidth]{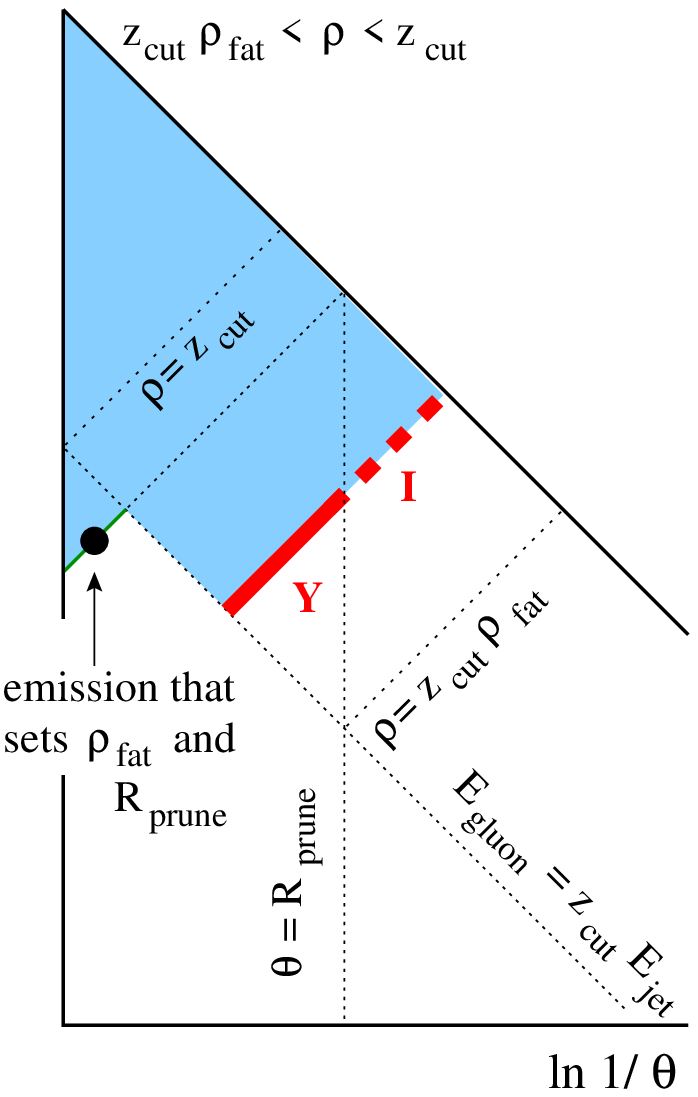}\hfill
  \includegraphics[width=0.32\textwidth]{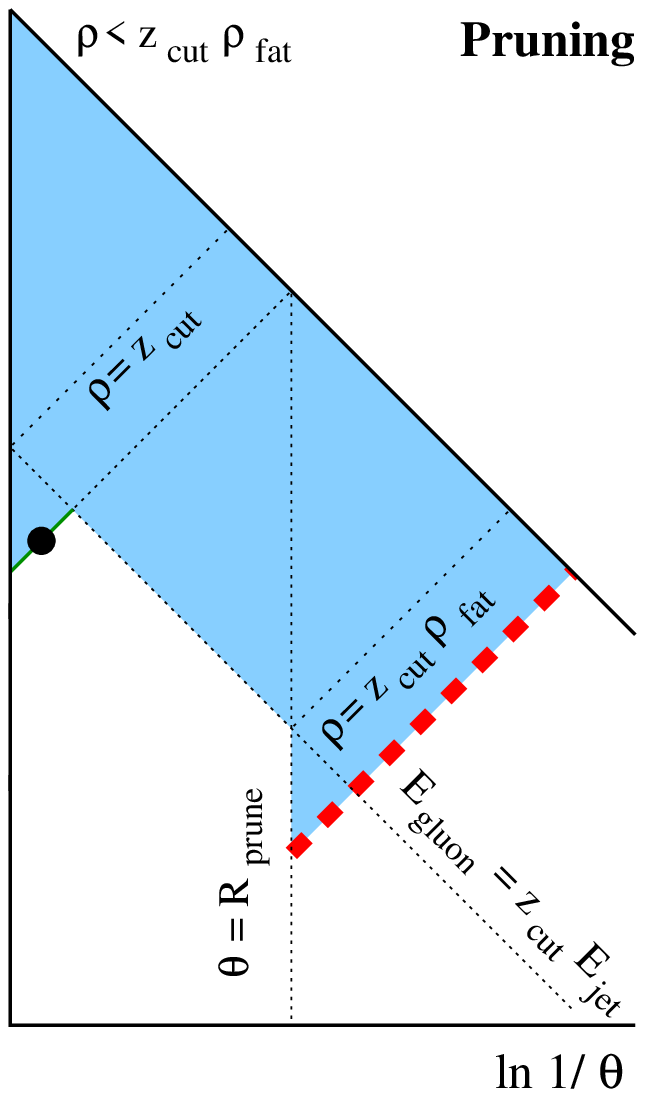}\hfill
  \caption{Lund kinematic diagrams for pruning, considering three
    different possible values of $\rho$. 
    In each case, to obtain the given value of $\rho$, there must be
    an emission somewhere along the thick (red) line, and there must be
    no emissions in the shaded region.
    The solid part of the thick line corresponds to $\sanepruning$,
    while the dashed part gives $\anomalouspruning$.
    Emissions in the unshaded regions have no impact on the pruned
    jet mass.
    The behaviour of the pruner can be affected by the presence of an
    emission that dominates $\rhofat$ (and so sets the pruning
    radius), but is discarded because it is below the pruning energy cut.  
    The dotted line that shows the pruning energy cut is parametrised
    in terms of the jet energy; this is a simplification, insofar as
    pruning uses the local subjet to provide its reference energy.
  }
  \label{fig:pruned-jet-mass-lund}
\end{figure}
 
To understand how to resum the pruned jet mass, for both the $\sane$ and
$\anomalous$ components, it is useful to refer to
Fig.~\ref{fig:pruned-jet-mass-lund}. 
The left-most figure corresponds to the the region $\rho > \zcut$ and
is essentially identical to the plain jet mass (as for trimming in
this region). 
In this region we only have $\sanepruning$.

The middle and right-hand plots illustrate two of the main
configurations that are relevant when $\rho < \zcut$.
Both show an emission (small black disk) that dominates the total jet
mass ($\rhofat$) and so sets the pruning radius 
\begin{equation} \label{eq:pruning-radius}
R_\text{prune}^2 =
\rhofat \, R^2\,.
\end{equation}
It will always be at an angle larger than $\Rprune$, and for the
discussion here it will be interesting to consider the cases where it
has a momentum fraction $\zfat < \zcut$, so that it is pruned away. 
We then need to consider a second emission, somewhere along the thick
(red) solid and dashed lines, with momentum fraction $z$ and angle
$\theta$, that sets 
the final pruned mass $\rho$.
The two possible situations are:
\begin{subequations}
  \label{eq:pruning-separation}
  \begin{align}
    \text{$\sanepruning$:}&\quad \theta>\Rprune,\; z > \zcut \quad\longrightarrow
    \quad \zcut < z < \frac{\rho}{\rhofat}\,,\\
    \text{$\anomalouspruning$:}&\quad \theta<\Rprune,\phantom{\; z > \zcut} \quad\longrightarrow
    \quad z > \frac{\rho}{\rhofat}\,,
  \end{align}
\end{subequations}
where the conditions on $z$ have been derived by combining the
relation $\theta^2 = \rho R^2/z$ with Eq.~(\ref{eq:pruning-radius}). 

In the middle panel of Fig.~\ref{fig:pruned-jet-mass-lund}, the
$\sanepruning$ region is represented by a thick (red) solid line, while
the $\anomalouspruning$ region is represented by a thick (red) dashed
line.

In the rightmost panel, with $\rho / \rhofat < \zcut$, there can be no
$\sanepruning$, because emissions with $\theta > \Rprune$ necessarily
have $z < \rho / \rhofat <\zcut$.
There is then only $\anomalouspruning$, and because there is no direct
constraint on the momentum fraction of emissions with $\theta <
\Rprune$, any $z > \rho/\rhofat$ contributes to the $\anomalouspruning$,
even if $z < \zcut$.
Given that $\rhofat < \zcut$, $\anomalouspruning$ with $z < \zcut$
starts to appear only for $\rho < \zcut^2$.

To determine the distributions for $\sane$- and $\anomalouspruning$, we will
work, as for trimming, in an independent emission picture. 
However, for brevity, we will not explicitly write the independent
emissions here, but instead make use of the result that when one
forbids emissions (i.e. the shaded regions of
Fig.~\ref{fig:pruned-jet-mass-lund}), one simply includes a factor
corresponding to the exponential of (minus) the integral of the
coupling times the splitting function over the forbidden
region.
%

\subsubsection{$\Sanepruning$}
\label{sec:sane-pruning}

For $\sanepruning$, one way of writing the result is as an integral over
the momentum fraction $z$ of the emission that gives the final pruned
mass. 
For a given $z$ to contribute it must obviously satisfy $z> \zcut$.
In addition the fat jet mass must be smaller than $\rho/z$.
From the considerations of the previous section, this then gives us,
for $\rho < \zcut$,
\begin{equation}
  \label{eq:sane-prune-result}
  \frac{\rho}{\sigma} \frac{d\sigma^\saneprune}{d\rho} =   
  \int_{\zcut}^1 dz \; p_{gq}(z)\; 
  e^{-D\left(\min(\zcut,\frac{\rho}{z})\right) 
      - S\left(\min(\zcut,\frac{\rho}{z}),\rho \right)}
  \frac{\as (\rho z\, p_t^2 R^2) \, C_F}{\pi} \,.
\end{equation}
The $D\left(\min(\zcut,\frac{\rho}{z})\right)$ terms accounts for the
suppression of all emissions that would produce a $\rhofat > \rho/z$
(or $\rhofat > \zcut$).
The term $S\left(\min(\zcut,\frac{\rho}{z}),\rho \right)$ accounts for
the further required suppression of emissions with $z > \zcut$
contributing a mass between $\rho/z$ and $\rho$.

Another, equivalent way of writing the result makes the $\rhofat$
integral more explicit:
\begin{multline}
  \label{eq:sane-prune-result-v2}
  \frac{\rho}{\sigma} \frac{d\sigma^\saneprune}{d\rho} = 
  e^{-D\left(\rho\right)} 
  \int_{\zcut}^1 dz \; p_{gq}(z) \frac{\as (\rho z\, p_t^2 R^2) \, C_F}{\pi}
  \;+
  \\
  + 
  \int_\rho^{\min(\zcut,\rho/\zcut)}
  \frac{d\rhofat}{\rhofat}
  \left(e^{-D(\rhofat)} \int_{\rhofat}^{\zcut} \frac{dz'}{z'}
    \frac{\as (\rhofat z'\, 
    p_t^2 R^2) \, C_F}{\pi} \right)
  \times
  \\
  \times
  e^{- S\left(\rhofat,\rho \right)}
  \int_{\zcut}^{\rho/\rhofat} dz \; p_{gq}(z) \frac{\as (\rho z\,
    p_t^2 R^2) \, C_F}{\pi}\,.
\end{multline}
The term on the first line corresponds to configurations in which the
emission that dominates the pruned mass also dominates the overall
fat-jet mass.
The term on the second and third lines corresponds to situations where
there is an explicit emission with momentum fraction $z' < \zcut$ that
gets pruned away.\footnote{In integrating over $z'$ we have replaced
  $p_{gq}(z') \to 1/z'$, because $z' < \zcut \ll 1$.} 
It sets a fat-jet mass substantially larger than the
final pruned mass, $\rhofat \gg \rho$, while the emission that
dominates the pruned mass still has $\theta > \Rprune$.

The above two expressions should capture terms $\as^n L^{2n-1}$ and
$\as^n L^{2n-2}$ in $\Sigma^\saneprune(\rho)$.
It is less straightforward to discuss the accuracy for $\ln
\Sigma^\saneprune(\rho)$: this is because unlike the cases of
plain jet mass and trimming, pruning does not lead to a simple
exponentiated structure.
Analogous results for gluon-initiated jets are given in appendix~\ref{sec:gluon-jets-prune}.

To help understand the structure of Eqs.~(\ref{eq:sane-prune-result})
and (\ref{eq:sane-prune-result-v2}), it is useful to evaluate them
in a fixed-coupling approximation, neglecting terms $\sim \as \ln^2
\zcut$, which for $\rho < \zcut^2$ yields
\begin{subequations}
  \begin{align}
    \label{eq:sane-prune-fixed-coupling}
    \frac{\rho}{\sigma} \frac{d\sigma^\saneprune}{d\rho}
    &\simeq e^{-D(\rho)} \left[ \left( e^{\frac{\as C_F}{\pi} \ln
          \frac1{\zcut} \ln \frac{1}{\rho}} - 1\right)
      \frac{1}{\ln \frac{1}{\rho}} - \frac{3}{4} \frac{\as
        C_F}{\pi} \right]
    \\
    &\simeq e^{-D(\rho)}\, \frac{\as C_F}{\pi} \left[\ln \frac{1}{\zcut}
      - \frac{3}{4} \right]\,,
    \qquad \as \ln \frac1{\zcut} \ln \frac{1}{\rho} \ll 1\,,
  \end{align}
\end{subequations}
where 
the second line provides a
further simplification for situations where $\rho$ is not too small
and illustrates the consistency with Eq.~(\ref{eq:sane-prune}).

\subsubsection{$\Anomalouspruning$}
\label{sec:anom-pruning}

The resummed result for $\anomalouspruning$ reads for $\rho < \zcut$
\begin{multline}
  \label{eq:anom-prune-result}
  \frac{\rho}{\sigma} \frac{d\sigma^\anomprune}{d\rho} = 
  \int_\rho^{\zcut}
  \frac{d\rhofat}{\rhofat}
  \left(e^{-D(\rhofat)} \int_{\rhofat}^{\zcut} \frac{dz'}{z'}
    \frac{\as (\rhofat z'\, 
    p_t^2 R^2) \, C_F}{\pi} \right)
  \times
  \\
  \times
  e^{- S\left(\rhofat,\rho \right)}
  \int_{\rho/\rhofat}^1 dz \; p_{gq}(z) \frac{\as (\rho z\,
    p_t^2 R^2) \, C_F}{\pi}
  \left[
    \Theta\left(\frac{\rho}{\rho_{\text{fat}}} - \zcut\right)
    + \right.
    \\
    \left. +
    \Theta\left(\zcut - \frac{\rho}{\rho_{\text{fat}}}\right)
    \exp\left(-\int_\rho^{\zcut\rho_{\text{fat}}} \frac{d\rho'}{\rho'}
    \int_{\rho'/\rhofat}^{\zcut} \frac{dz'}{z'}
    \frac{C_F}{\pi} \as(\rho' z' p_t^2 R^2)
    \right)
  \right]\,.
\end{multline}
In order to have $\anomalouspruning$, there must be an emission that
sets the fat-jet mass and pruning radius such that that first emission
gets pruned away and a second emission falls within the pruning
radius. 
The first line of Eq.~(\ref{eq:anom-prune-result}) gives the
distribution for the fat-jet mass, assuming that the corresponding
emission has $z < \zcut$, i.e.\ gets pruned away.
The second line includes a Sudakov suppression $e^{-S(\rhofat,\rho)}$
for forbidding emissions with $z > \zcut$ between the scales of
$\rhofat$ and $\rho$, and also includes an integral over the allowed
$z$ values for emissions that fall within the pruning radius.
This multiplies a square bracket containing two terms: the first
corresponds to the middle diagram of
Fig.~\ref{fig:pruned-jet-mass-lund}, while the second corresponds to
the right-hand diagram, and accounts for the
required additional Sudakov suppression of emissions with $z < \zcut$
and $\theta < \Rprune$. 
In this factor, we have directly replaced $dz \, p_{gq}(z)$ with $dz/z$,
neglecting corrections suppressed by powers of $\zcut$.

Eq.~(\ref{eq:anom-prune-result}) should account for terms $\as^n
L^{2n}$ and $\as^n L^{2n-1}$ in $\Sigma^\anomprune$, i.e.\ the
first two towers of logarithms. 
Note that overall we have one power of $L$ more than for $\sanepruning$.
As for the case of $\sanepruning$, it is less straightforward to discuss
the accuracy for $\ln \Sigma^\anomprune$. Analogous results for gluon-initiated jets are given in appendix~\ref{sec:gluon-jets-prune}.
A calculation beyond the small-$\zcut$ limit reveals that there are flavour-changing contributions that mix quark-initiated and gluon-initiated jets. They give rise to terms $\sim \zcut \as^n L^{2n-1}$~\cite{taggersNLO}, and they are neglected here because they vanish as $\zcut \to 0$.

The structure of Eq.~(\ref{eq:anom-prune-result}) is relatively
complicated. 
Accordingly, to gain some insight into it we will make a double
logarithmic approximation, considering just terms $\as^n L^{2n}$ in
$\Sigma^\anomprune(\rho)$.
Within this approximation we can replace $p_{gq}(z)$ with $1/z$,
assume $\zcut$ to be of order $1$ and take $\as$ fixed.
This then gives
\begin{equation}
  \label{eq:anom-prune-DL-resummed}
    \frac{\rho}{\sigma} \frac{d\sigma^{\anomprune}}{d\rho}  
   \simeq
     \left(\frac{\as C_F}{\pi}\right)^2 \int_\rho^1  
    \frac{d\rhofat}{\rhofat} \ln \rhofat e^{- \frac12 \frac{\as
        C_F}{\pi} \ln^2 \frac{1}{\rhofat}}
    \,
    \ln \frac{\rho}{\rhofat} e^{- \frac12 \frac{\as
        C_F}{\pi} \ln^2 \frac{\rhofat}{\rho}}\,,
\end{equation}
which integrates to
\begin{equation}
  \label{eq:anom-prune-DL-errf}
  \frac{\rho}{\sigma} \frac{d\sigma^{\anomprune}}{d\rho}  
  \simeq 
  \frac{\asbar L}{2}  e^{-\frac12 \asbar L^2}
  + 
  \frac{\sqrt{\asbar \pi}}{4} e^{-\frac14 \asbar L^2} (-2 + \asbar L^2
  ) \, \Erf\left(\frac{\sqrt{\asbar} L}{2}\right)\,.
\end{equation}
It is straightforward to verify that this has no $\as$ term and is equivalent to
Eq.~(\ref{eq:prtdl}) at order $\as^2$.
The structure involving the factor $L^2 e^{-\frac14 \asbar L^2}$ can be
seen to arise from the point where the integrand in
Eq.~(\ref{eq:anom-prune-DL-resummed}) is maximal. 
Insofar as it is legitimate to consider just this structure, one might
expect the $\anomalous$-pruned mass distribution to have a maximum
situated near $L = 2/\sqrt{\asbar}$.
Using the full form of Eq.~(\ref{eq:anom-prune-DL-errf}), the maximum
is at $L \simeq 2.284/\sqrt{\asbar}$, which is to be compared to the maximum of the plain jet-mass
distribution, situated at $L = 1/\sqrt{\asbar}$.
We will return to these observations when we discuss comparisons with
Monte Carlo below. 

\subsubsection{Sum of $\sane$ and $\anomalous$ components}
\label{sec:total-pruning}

Finally let us add together $\sane$- and $\anomalouspruning$ in the region
$\zcut^2 < \rho < \zcut$, working in a fixed-coupling approximation
for simplicity.
In this region, the upper limit of the $\rhofat$ integrals in
Eqs.~(\ref{eq:sane-prune-result-v2}) and (\ref{eq:anom-prune-result})
becomes $\zcut$.
In the square brackets of Eq.~(\ref{eq:anom-prune-result}), it is the
first of the 
$\Theta$-functions that is relevant (because we have $\rhofat <
\zcut$ and $\rho > \zcut^2$).
The $z$ integrals in Eqs.~(\ref{eq:sane-prune-result-v2}) and
(\ref{eq:anom-prune-result}) are associated with the same prefactors
and $\rhofat$ integration, and have complementary limits in $z$,
$\zcut < z < \rho/\rhofat$ and $\rho/\rhofat < z < 1$ respectively and
so add together to give an integral over $z$ from $\zcut$ to $1$.
We can therefore write the sum as
\begin{equation}
  \label{eq:sum-prunings-start}
  \frac{\rho}{\sigma} \frac{d\sigma^\text{(prune)}}{d\rho} = 
  \int_{\zcut}^1 dz \; p_{gq}(z) \asbar
  \left(
    e^{-D\left(\rho\right)} 
    + 
    \int_\rho^{\zcut} d\rhofat
    \left(e^{-D(\rhofat)- S\left(\rhofat,\rho \right)} \int_{\rhofat}^{\zcut} \frac{dz'}{z'}
      \asbar \right)
  \right)\,.
\end{equation}
Using a fixed-coupling approximation for simplicity, and making use of
the fact that  
\begin{equation}
  \label{eq:D-manip}
  D(\rho) = D(\zcut) + S(\zcut,\rho)  + \frac{\asbar}{2} \ln^2
  \frac{\zcut}{\rho} + \order{\asbar \zcut}\,,
\end{equation}
we then obtain the simple result
\begin{equation}
  \label{eq:sum-prunings}
  \frac{\rho}{\sigma} \frac{d\sigma^\text{(prune)}}{d\rho} = 
    e^{-D(\zcut) + S(\zcut,\rho)} \asbar\left(\ln\frac{1}{\zcut} -
      \frac34 \right)\,, \qquad
  \zcut^2 <\rho < \zcut\,,
\end{equation}
which corresponds to the following integrated cross section:
\begin{equation}
  \label{eq:sum-prunings-integrated}
  \Sigma^\text{(prune)}(\rho) = e^{-D(\zcut) + S(\zcut,\rho)}\,, \qquad
  \zcut^2 <\rho < \zcut\,.
\end{equation}
This second form holds also with running coupling effects included.

Several comments can be made about
Eq.~(\ref{eq:sum-prunings-integrated}).
Relative to the middle panel of Fig.~\ref{fig:pruned-jet-mass-lund},
the key point is that for $\zcut^2 < \rho < \zcut$, the presence or
not of a distinct ``fat-jet'' emission (one with $z' < \zcut$) only
modifies the separation between $\anomalous$ and $\sanepruning$, but not
their sum. 
As a result, $-\ln \Sigma(\rho)$ is effectively just the integral of
the leading order distribution, Eq.~(\ref{eq:pruning-LO}).
This is the pattern that is seen also for trimming and the plain jet
mass (at NLL accuracy in $\Sigma$), but with the difference that in
the case of pruning the pattern breaks down for $\rho < \zcut^2$,
whereas for trimming and plain jet mass it holds for all $\rho$
values.

Another point of interest is that
Eq.~(\ref{eq:sum-prunings-integrated}) is identical to the result for
trimming, Eq.~(\ref{eq:trimming-LL-small-mass-FC}), in the
corresponding region $r^2 \zcut < \rho < \zcut$. Trimming and pruning
are also identical, at our accuracy, for $\rho > \zcut$.
We will return to this point later when we discuss the comparisons
between taggers in section \ref{sec:perturbative-comparisons}.

Finally, as in the case of trimming, to go beyond the accuracy aimed for in this paper for pruning 
would require the treatment
of several additional effects: non-global logarithms and related
clustering logarithms, multiple-emission 
effects on the observable and the two-loop cusp anomalous dimension.

Non-global logarithms  enter in a number of ways: in particular,
from the boundary at $\theta \sim R$, they affect the fat-jet mass,
and through it the distribution of the pruning radius.
This has implications for both the $\sane$ and $\anomalous$ components
starting, in the small-$\zcut$ limit, from order $\as^3$.
Moreover, at finite $\zcut$,  $\anomalouspruning$ receives non-global
contributions already at order $\as^2$~\cite{taggersNLO}. 
We leave a full resummation of pruning to single-logarithmic accuracy
to future work.

\subsection{Comparison with Monte Carlo results}
\label{sec:pruning-analytic-VS-MC}

\begin{figure}
  \centering
    \includegraphics[width=0.49\textwidth]{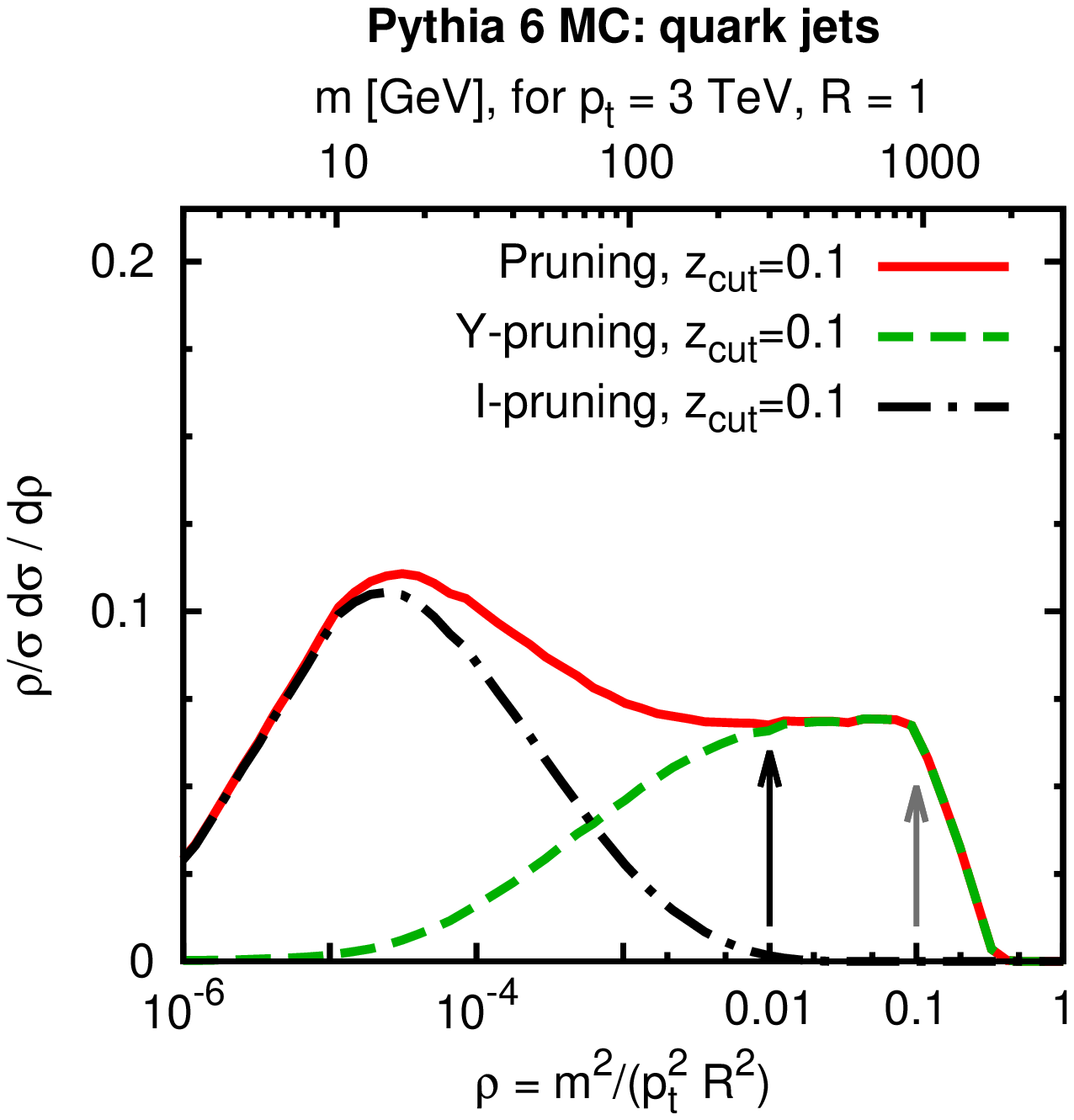}\hfill
    \includegraphics[width=0.49\textwidth]{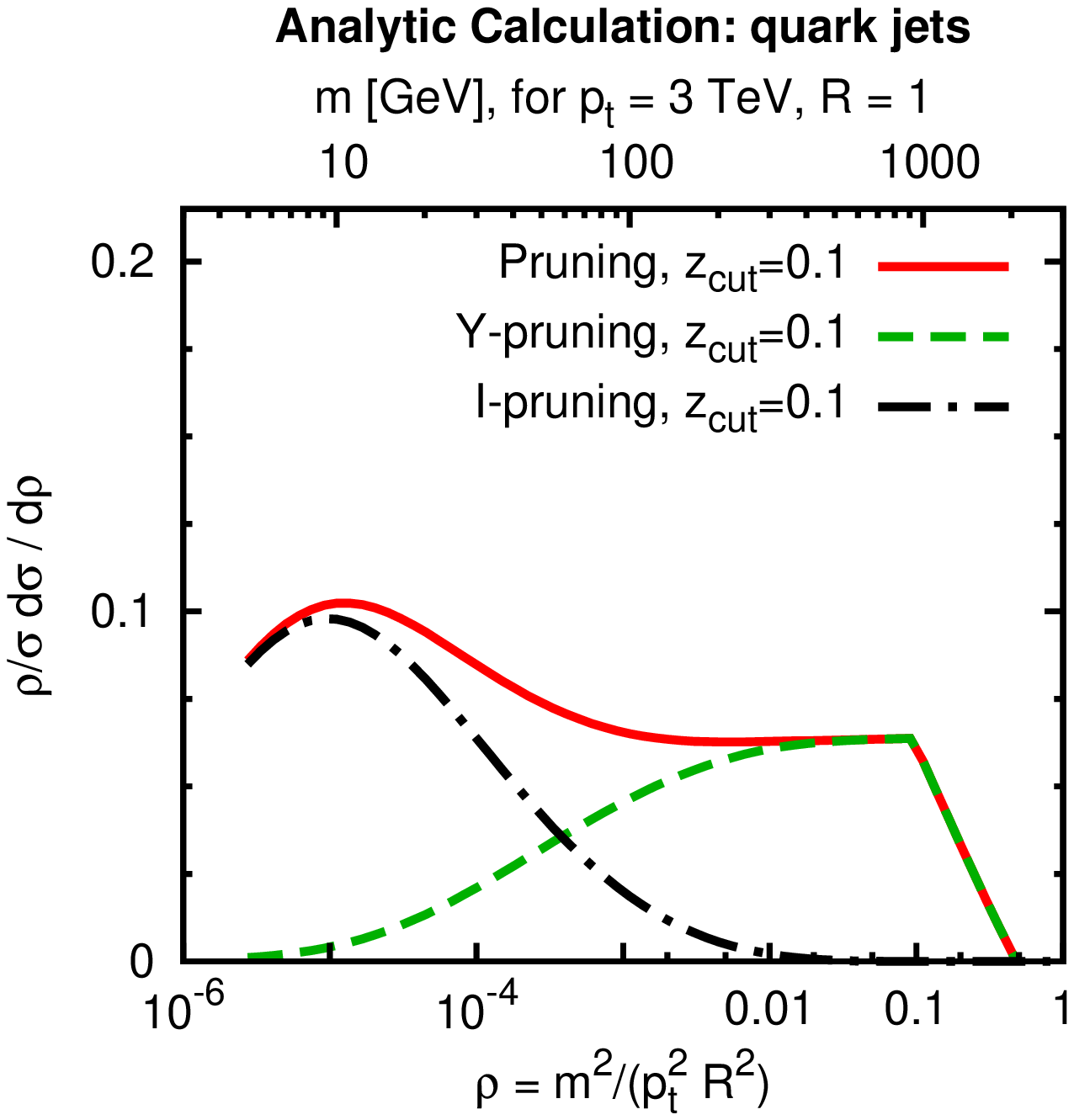}\\[5pt]
    \includegraphics[width=0.49\textwidth]{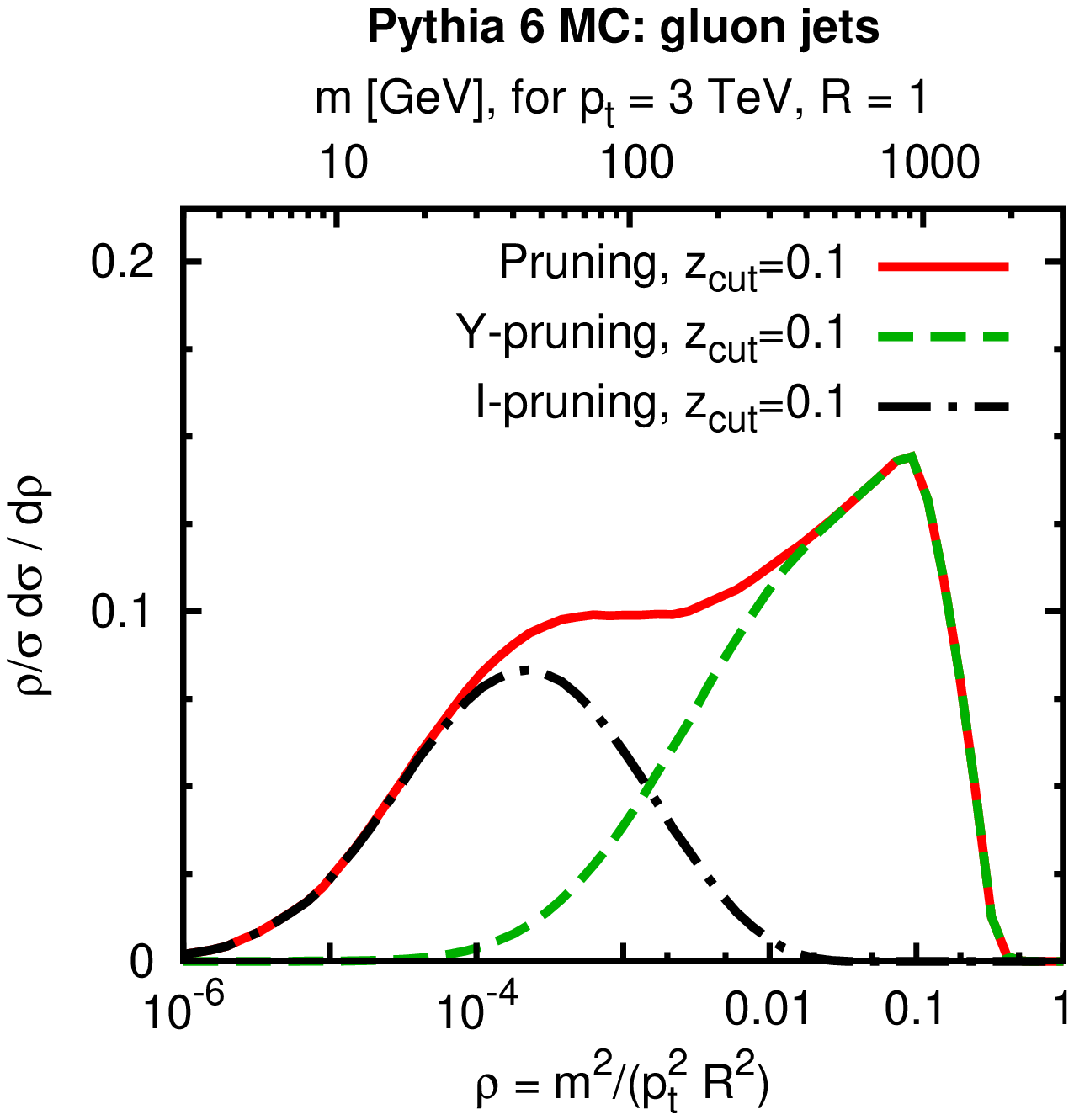}\hfill
    \includegraphics[width=0.49\textwidth]{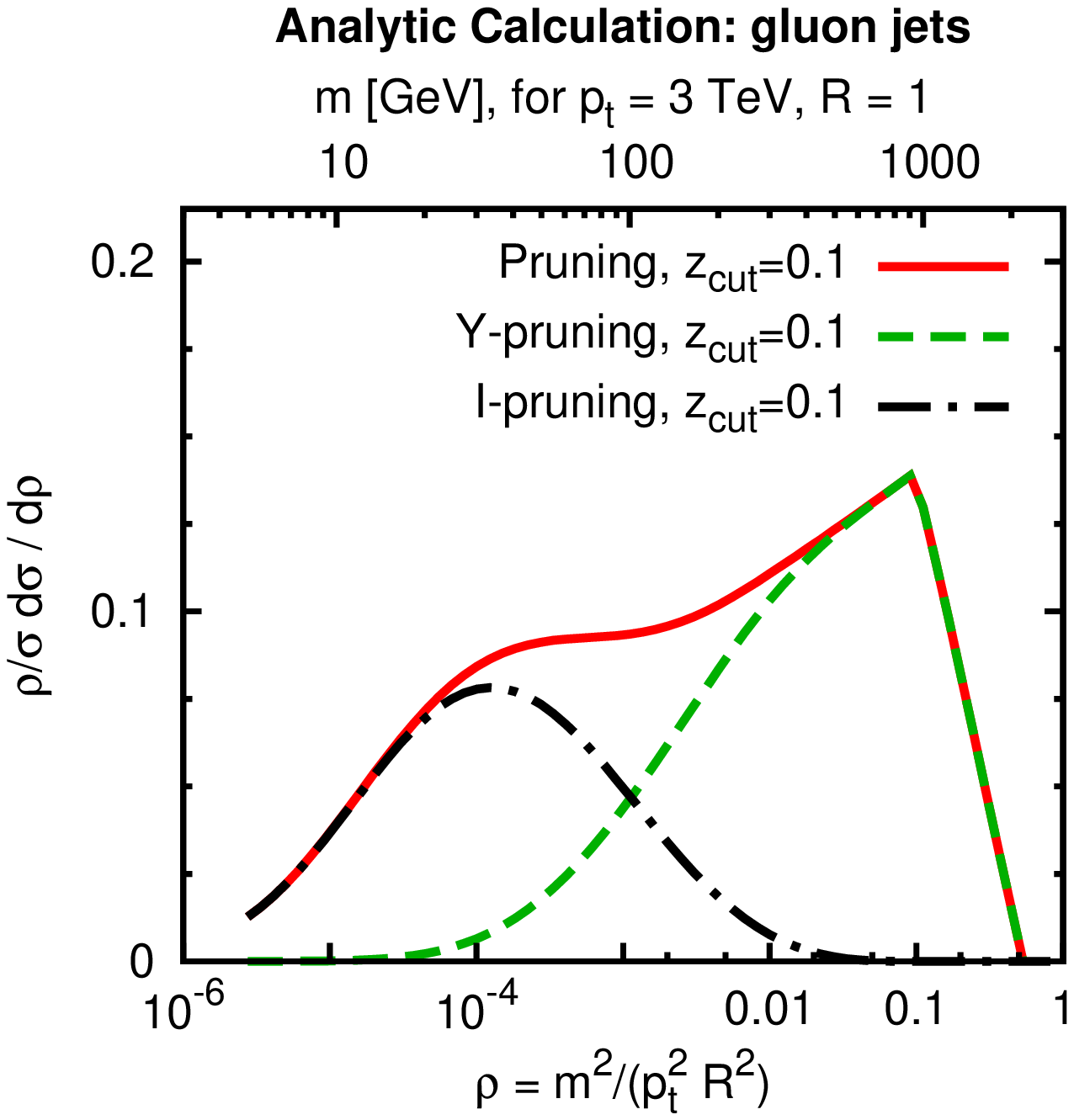}
  \label{fig:pruning_and_sane_pruning}
  \caption{%
    Comparison of Monte Carlo (left panels) and analytic
    results (right panels) for pruning.
    The upper panels are for
    quark jets, the lower panels for gluon jets. 
    The plots show full pruning as well as its breakdown into $\sane$ and
    $\anomalous$ components.
    In the upper left panel, arrows indicate the expected transition
    points, at $\rho = \zcut^2$ (in black) and $\rho = \zcut$ (in
    grey).  
    The details of the MC event generation are as for
    Fig.~\ref{fig:tagged-mass-MC}.  }
\end{figure}

Figure~\ref{fig:pruning_and_sane_pruning} shows predictions for the
pruned mass distribution from Pythia in the left-hand panels and from
our analytical calculation in the right-hand panels. Upper and lower
rows correspond to quark jets and gluon jets respectively.
As was the case with trimming, the agreement between the MC and
analytical results is reasonable.
The expected transition points at $\rho = \zcut$ and $\zcut^2$ are labelled
with arrows in the upper MC plot. 
Above $\rho = \zcut$ we see a similar behaviour as for the plain jet
mass.
For $\zcut^2 < \rho < \zcut$, we see a flat region in the quark case,
akin to the leading-order result, however in the gluon case that
flatness is strongly modified by higher orders
(the exact impact of these higher orders depends strongly on $\zcut$).
The transition at $\rho = \zcut^2$ is much smoother than that
at $\zcut$.
Recall that the transition occurs because phase space opens for
emissions with $z < \zcut$ to dominate the pruned jet mass. 
As one can verify analytically, that phase space initially opens up
slowly (cf.\ also Fig.~\ref{fig:pruned-jet-mass-lund}) and the most singular contribution for pruning ($\sane$+$\anomalous$ components) goes as $\as^2 \ln^3
  \zcut^2/\rho$. The transition is therefore gradual. 

Going substantially below $\rho = \zcut^2$, for quark jets, one sees a
clear peak in total pruning, which results from the $\anomalous$
component.
In the gluon case, while that peak is similarly visible in the
$\anomalous$ component, in the sum with $\sanepruning$ it manifests itself
as a shoulder, because the peak occurs in a region where the $\sanepruning$
component is not entirely suppressed.
As before, this precise picture holds for our specific choice of $\zcut$.

The position of the peak for the $\anomalous$ component, in 
the case of quark-initiated jets, is in reasonable agreement with the one determined 
by the fixed-coupling approximation,
Eq.~(\ref{eq:anom-prune-DL-errf}), though the agreement is poorer for
gluon jets:
for a reliable quantitative treatment of the peak region it is
important to include subleading terms.

\section{Mass Drop Tagger}
\label{sec:MDT}

The mass-drop tagger~\cite{Butterworth:2008iy} was designed to be used
with jets found by the Cambridge/Aachen
algorithm~\cite{Dokshitzer:1997in,Wobisch:1998wt}. 
It involves two parameters $\ycut$ and $\mu$ and, for an initial jet
labelled $j$, proceeds as follows:
\begin{enumerate}
\item Break the jet $j$ into two subjets by undoing its last stage of
  clustering. Label the two subjets $j_1, j_2$ such that $m_{j_1} >
  m_{j_2}$.
\item \label{item:MDT-ycut-step} If there was a significant mass drop, $m_{j_1} < \mu m_{j}$, and
  the splitting is not too asymmetric, $y = \min(p_{tj_1}^2,
  p_{tj_2}^2) \Delta R_{j_1 j_2}^2/ m_j^2 > \ycut$, then deem $j$ to
  be the tagged jet. 
  
\item \label{item:MDT-bad-step} Otherwise redefine $j$ to be equal to
  $j_1$ and go back to step 1 (unless $j$ consists of just a single
  particle, in which case the original jet is deemed untagged).
\end{enumerate}
Typical parameter choices are for example $\mu = 2/3$ and $\ycut$ in
the range $0.09 - 0.15$.
While the $\ycut$ parameter will appear explicitly in our results,
$\mu$ will not, and indeed we shall see that its exact value is not
critical as long as it is not parametrically small.


\subsection{Leading order calculation}
\label{sec:MDT-LO}

As usual, it is useful to start with a leading-order configuration,
for which the jet consists of just two partons.
When the jet is declustered, each of the prongs is massless, so that
the mass-drop condition is automatically satisfied, rendering the
$\mu$ parameter irrelevant.
There are then two possibilities: if the asymmetry condition is
satisfied the jet is tagged, with the tagged mass equal to the
original jet mass.
Otherwise the jet does not contribute to the tagged jet mass
distribution.

Considering a quark that splits into a quark with momentum fraction
$1-z$ and a gluon with momentum fraction $z$, we have $m^2_j = z(1-z) E^2$.
The asymmetry condition then becomes $\frac{z}{1-z} > \ycut$ and
$\frac{1-z}{z} > \ycut$. 

We may now write the differential cross section for the jet to have
a given tagged mass:
\begin{multline}
  \frac{1}{\sigma} \frac{d\sigma}{d m^2}^\text{(MDT, LO)} 
  = C_F \frac{\alpha_s}{\pi} 
  \int dz p_{gq}(z) \frac{d\theta^2}{\theta^2} \,
  \delta\!\left(m^2 - z(1-z) p_t^2 \theta^2 \right) 
  \times \\ \times
  \Theta\left(\frac{z}{1-z} - \ycut \right) 
  \Theta\left(\frac{1-z}{z} - \ycut \right) 
  \Theta \left (R^2 - \theta^2 \right).
\end{multline}
Proceeding as with our other LO calculations, including a requirement
$\ycut \ll 1$, leads us to the following result
\begin{equation}
  \label{eq:MDT-LO}
  \frac{\rho}{\sigma} \frac{d\sigma}{d\rho}^\text{(MDT, LO)} 
  = \frac{\as C_F}{\pi} 
  \left[ 
    \Theta(\rho - \ycut) \ln \frac{1}{\rho} 
    + \Theta(\ycut-\rho) \ln \frac1\ycut 
    - \frac34
  \right].
\end{equation}
Modulo the replacement $\zcut \to \ycut$, this is identical to the
result for pruning, Eq.~(\ref{eq:pruning-LO}), and in particular has
two regimes: it is linear in $\ln \frac{1}{\rho}$ when $\rho > \ycut$,
and saturates at a constant value $(\ln \frac1{\ycut} - \frac34)$ for
$\rho < \ycut$.
In contrast to the case of pruning, it is intriguing that this
structure appears rather similar to what is observed in the Monte
Carlo results for quark jets in Fig.~\ref{fig:tagged-mass-MC}.
This would suggest that there are cases where effects beyond LO might
be modest.

\subsection{3-particle configurations}

\begin{figure}
  \centering
  \includegraphics[height=0.25\textwidth]{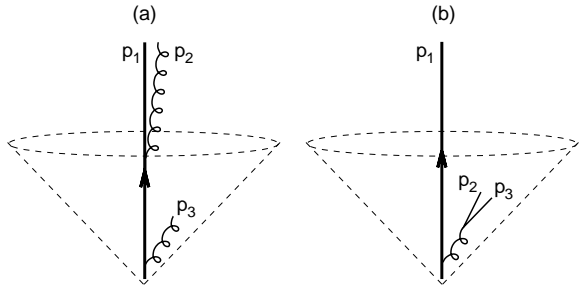}
  \caption{Two characteristic partonic configurations that arise at
    in the tree-level $\order{\as^2}$ contribution. The dashed cone
    provides a schematic representation of the boundary of the jet.}
  \label{fig:partonic-contributions}
\end{figure}

The next step in understanding the mass-drop tagger is to consider
3-particle configurations, where for the first time one encounters the
recursive nature of the tagger and potentially also the dependence on
$\mu$. 

Since we will be mainly interested in logarithmically enhanced
contributions, we can exploit the fact that these come from
configurations in which momenta are ordered in angle and/or energy. 
Some interesting such configurations are illustrated in
Fig.~\ref{fig:partonic-contributions}.

Configuration $(a)$ has the ordering $\theta_{13} \gg \theta_{12}$,
with the ordering sufficiently strong that we can assume $m_\text{jet}
= m_{123} \gg m_{12}$.
Because the jet was clustered with the angular-ordered C/A algorithm,
the MDT first splits the jet into \subjet{12} and \subjet{3}.
If $E_3/E_{12} > \ycut$ then the declustering passes the asymmetry
cut; the strong angular ordering ensures that it also passes the
mass-drop condition and so the jet as a whole is tagged.
If $E_3/E_{12} < \ycut$, then the MDT recurses, into the heavier of
the two subjets, i.e.\ \subjet{12}, which can be
analysed as in the previous, LO section.
The key point here is that in the limit in which $E_3 \ll E_\jet$,
the presence of gluon 3 has no effect on whether the \subjet{12}
system gets tagged.
This is true even though we chose a configuration where $m_{\jet}$ is
dominated by emission $3$.
This was part of the intended design of the MDT: if the jet contains
hard substructure, the tagger should find it, even if there is other
soft structure (including underlying event and pileup) that strongly
affects the original jet mass.
It is possible to show that if one combines the NLO contribution that
comes from configurations like (a) with the corresponding virtual
graphs, one obtains a contribution to $\Sigma^\text{(MDT)}(\rho)$ that
goes as $\as^2 L^2$ for arbitrarily large $L$.
This involves fewer logarithms than any of the plain jet mass,
trimming or pruning.
However it turns out not to be the leading contribution in terms of a
counting of logarithms and therefore we postpone its detailed
discussion.

Configuration (b) in Fig.~\ref{fig:partonic-contributions}
reveals an unintended behaviour of the tagger.
Here we have $\theta_{23} \ll \theta_{12}\simeq \theta_{13}$, so the
first unclustering leads to \subjet{1} and \subjet{23} subjets.
It may happen that the parent gluon of the \subjet{23} subjet was
soft, so that $E_{23} < \ycut E_\jet$.
The jet therefore fails the symmetry requirement at this stage, and so recurses
one step down.
The formulation of the MDT is such that it recurses into the more
massive of the two prongs, i.e.\ only follows the \subjet{23} prong,
even though this is soft.
This was not what was intended in the original design, and is to be
considered a flaw --- in essence one follows the wrong branch.

It is interesting to determine the logarithmic structure that results
from the wrong-branch issue.
Exceptionally, we are going to work in an approximation in which we
treat logarithms of $\ycut$ on the same footing as logarithms of
$\rho$. 
We will, however, neglect terms that do not have the maximal number of
logarithms of either argument.
The wrong-branch distribution can then be written as
\begin{multline}
  \label{eq:MDTflaw}
  \frac{\rho}{\sigma}\frac{d\sigma}{d\rho}^{(\text{MDT,NLO}_\text{flaw})}
  = \rho \; C_F C_A  \left(\frac{\alpha_s}{\pi} \right)^2
  \int 
  \frac{dx}{x}
  \frac{d \theta^2}{\theta^2}  \Theta \left (R^2-
    \theta^2 \right) \Theta\left(\ycut-x \right) 
  \times   \\ \times
  \int^1 \frac{d z}{z} \Theta(z - \ycut)
  \int
  \frac{d\theta_{23}^2}{\theta_{23}^2} \, \delta \left(\rho-z x^2 
    \frac{\theta_{23}^2}{R^2} \right)  \Theta(\theta - \theta_{23})
\end{multline}
where $\theta$ is the angle between \subjet{1} and the \subjet{23}
system, while $x = E_{23}/E_\jet$ and $z = E_{2}/E_{23}$.
In writing the constraints on the angles, we have assumed
strong-ordering of the angles.
We are also working in a soft approximation, $x \ll 1$ and $z \ll 1$.
The answer is non-zero only for $\rho \lesssim \ycut^2$, because $x$
must be less than $\ycut$, while the maximum $\theta_{23}$ angle is of
order $R^2$.\footnote{In the phase-space region where $\theta \sim
  \theta_{23} \sim R$, the approximation of strongly ordered angles is
  inappropriate. The determination of the exact onset of the
  wrong-branch issue would require a full treatment of that region.
  One would also need to go beyond the small-$z$ approximation:
  insofar as the squared jet mass involves a factor $z(1-z)$ rather
  than simply $z$, one would then expect an onset in the neighbourhood
  of $\rho \sim \ycut^2/4$ rather than $\ycut^2$.
  However, in terms of a logarithmic counting, these considerations
  should only affect subleading logarithms.
}
If $\rho \gtrsim \ycut^3$ then the $\ycut$ condition in the second
line of Eq.~(\ref{eq:MDTflaw}) does not play a role, and one obtains
\begin{equation}
  \label{eq:MDTflaw-region1}
  \frac{\rho}{\sigma}\frac{d\sigma}{d\rho}^{(\text{MDT,NLO}_\text{flaw})}
  =
  \frac{C_F C_A}{12} \left(\frac{\alpha_s}{\pi} \right)^2  
  \ln^3 \frac{\ycut^2}{\rho}\,, \qquad
     \ycut^3 \lesssim \rho \lesssim \ycut^2\,,
\end{equation}
otherwise the result is
\begin{multline}
  \label{eq:MDTflaw-region2}
  \frac{\rho}{\sigma}\frac{d\sigma}{d\rho}^{(\text{MDT,NLO}_\text{flaw})}
  =
  \frac{C_F C_A}{4} \left(\frac{\alpha_s}{\pi} \right)^2  
  \left(
   \ln^2 \frac{\ycut^3}{\rho} \ln \frac1{\ycut} \,+ 
      \right.\\ \left.
      +\, \ln \frac{\ycut^2}{\rho}\ln^2 \frac1{\ycut}
      \,-\, \frac23 \ln^3 \frac1{\ycut}
  \right)\,, \quad\;
  \rho \lesssim \ycut^3\,.
\end{multline}
Considering just the asymptotically small-$\rho$ region, which starts
for $\rho \lesssim \ycut^3$, the integrated distribution,
$\Sigma^\text{(MDT)}(\rho)$ has a logarithmic structure $\as^2 L^3 \ln
\frac{1}{\ycut}$, i.e.\ enhanced by $\as L^2$ relative to the LO
result and by a power of $L/\ln\frac{1}{\ycut}$ relative to
configurations of type (a).

\begin{figure}
  \centering
    \includegraphics[width=0.48\textwidth]{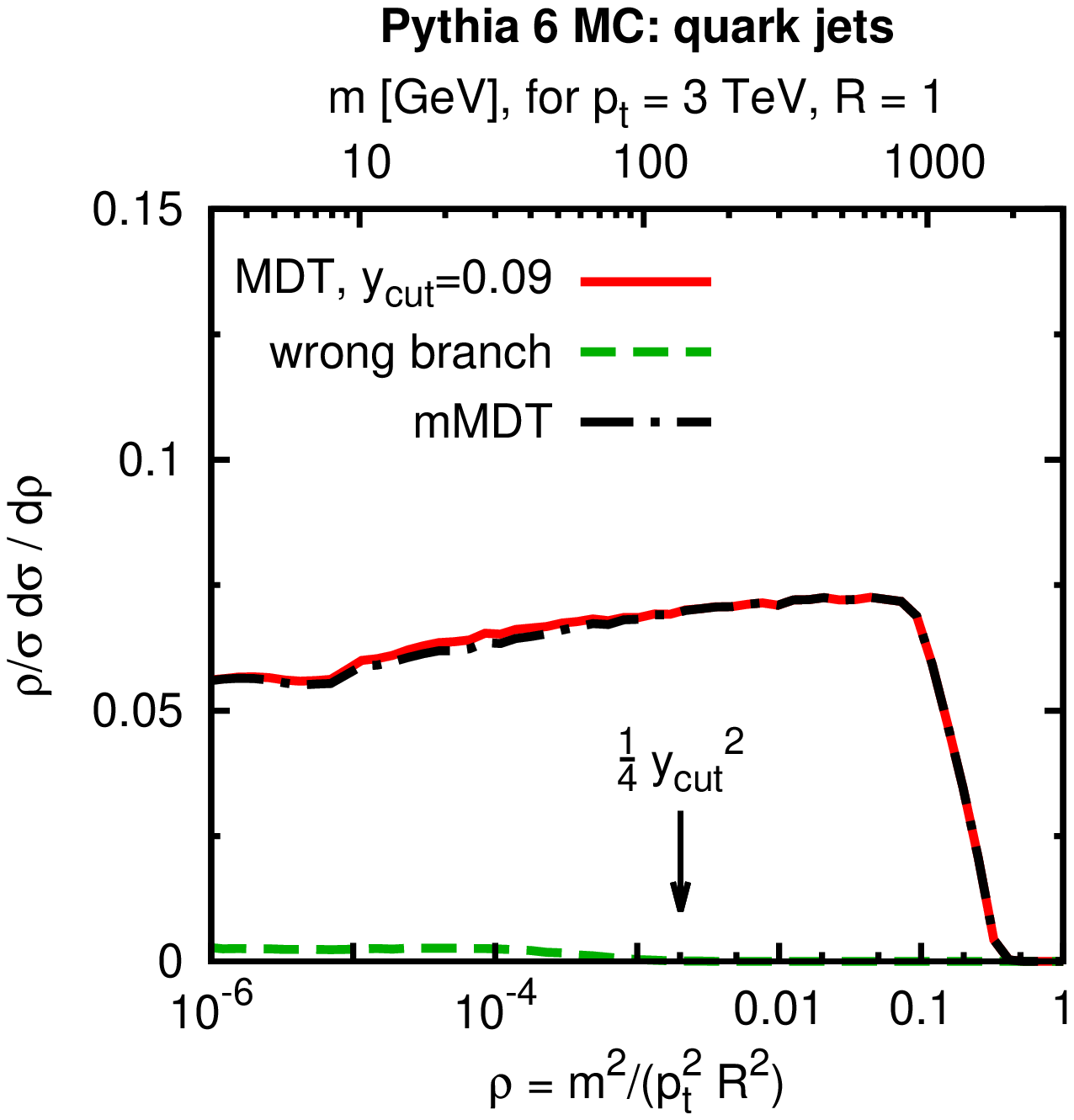}
    \includegraphics[width=0.48\textwidth]{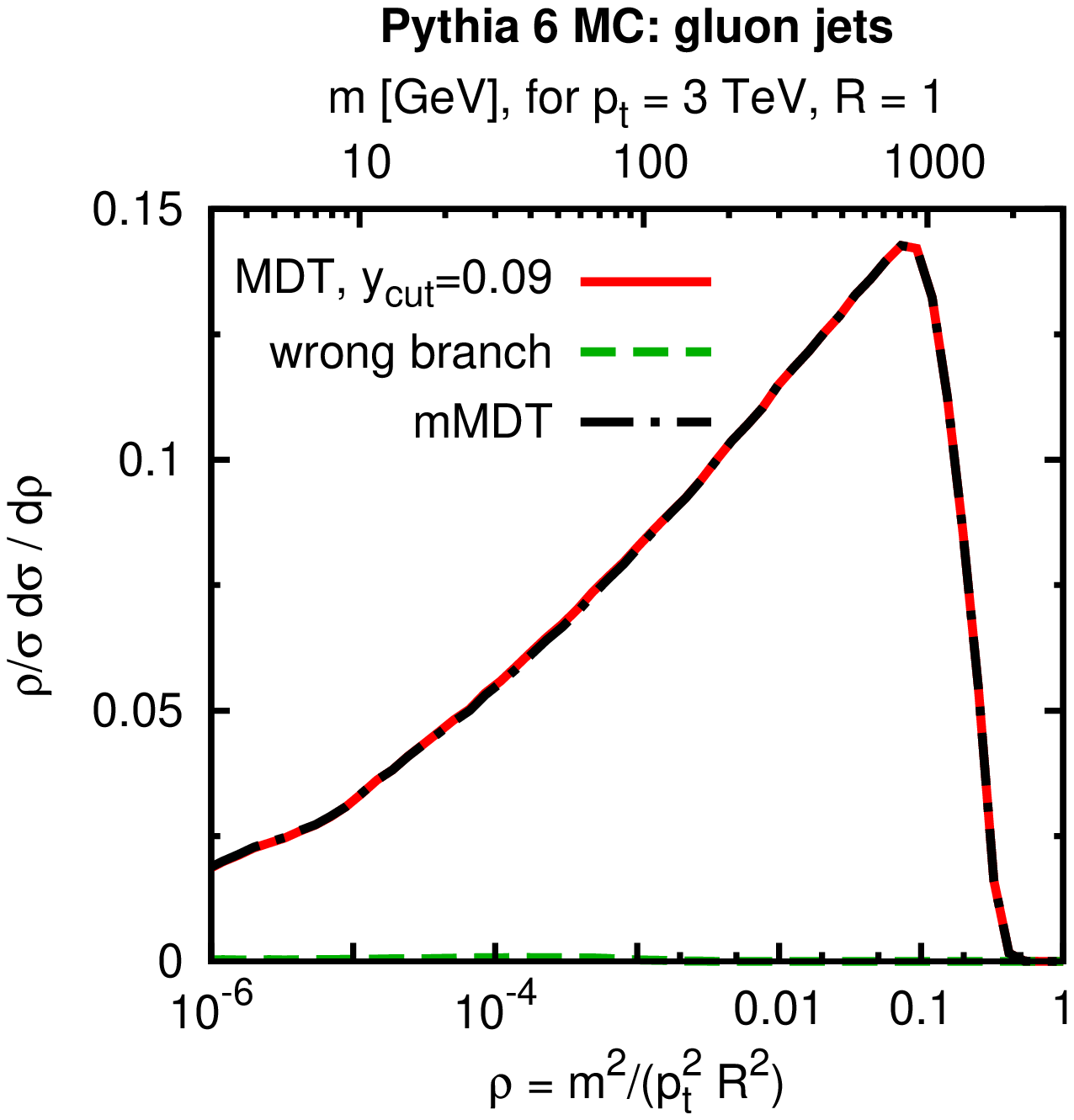}
  \caption{The MDT mass distribution, from Monte Carlo
    simulation (same as Fig.~\ref{fig:tagged-mass-MC}), with the contribution originating from wrong
    branches shown as a dashed line. Wrong branches are those for
    which, at some stage during the declustering, the tagger followed
    a prong whose $m^2 + p_t^2$ was smaller than that of its partner
    prong. }
  \label{fig:wrong-branch}
\end{figure}

Based on the above calculation, one might expect the ``wrong-branch''
contributions to dominate over the LO type behaviour.
In practice they don't.
Part of the reason for this is visible in the fixed-order result:
these terms set in only for relatively small values of jet mass, $\rho
\lesssim \ycut^2$, with a small coefficient, and the logarithm itself
is reduced in size because it involves either $\ycut^2/\rho$ or $
\ycut^3/\rho$, depending on the region.
Another part of the reason is that at higher orders the wrong-branch
contribution involves a Sudakov-type suppression, coming from the
probability that the harder prong of the jet was less massive than the
softer one, even though it has an energy that is at least a factor of
$1/\ycut$ larger than the softer prong.
The small contribution from the wrong-branch configurations is
illustrated in Fig.~\ref{fig:wrong-branch}, obtained in Monte Carlo
simulation, where events with a wrong-branch tag are defined as those
for which at some stage during the declustering, the tagger followed a
prong whose $m^2 + p_t^2$ was smaller than that of its partner prong.

While the wrong branch issue is numerically small, it is an
undesirable characteristic of the MDT and calls for being eliminated. 
Rather than pursuing a full (and non-trivial) calculation of the
resummed mass distribution for the MDT, we therefore propose in the
next section that the MDT be modified.

\section{Modified Mass-Drop Tagger}
\label{sec:mMDT}

The modification of the mass-drop tagger that we propose is to replace
step \ref{item:MDT-bad-step} of the definition on
p.~\pageref{item:MDT-bad-step}, with
\begin{enumerate}
\item[\ref{item:MDT-bad-step}.] 
  Otherwise redefine $j$ to be that of $j_1$ and $j_2$ with the larger
  transverse mass ($m^2 + p_t^2$) and go back to step 1 (unless $j$
  consists of just a single particle, in which case the original jet
  is deemed untagged).
\end{enumerate}
At leading order, since there is no recursion, this modified MDT (mMDT)
behaves identically to the original MDT.
However, in the case of configurations like those of
Fig.~\ref{fig:partonic-contributions}b, the tagger 
will follow the \subjet{1} branch rather than the \subjet{23} branch
thus eliminating the wrong-branch issues and the associated
terms in 
Eq.~(\ref{eq:MDTflaw}).

Fig.~\ref{fig:wrong-branch} includes the tagged-mass spectrum from the
modified mass-drop tagger in Monte Carlo simulation. One sees
that, phenomenologically, the modification is a minor one, as can be
checked also on events where the jet stems from a resonance decay
(i.e.\ signal rather than background).

\subsection{All-order tagged-mass distribution}
\label{sec:mmdt-all-orders}

\begin{figure}
  \centering
  \includegraphics[width=0.32\textwidth]{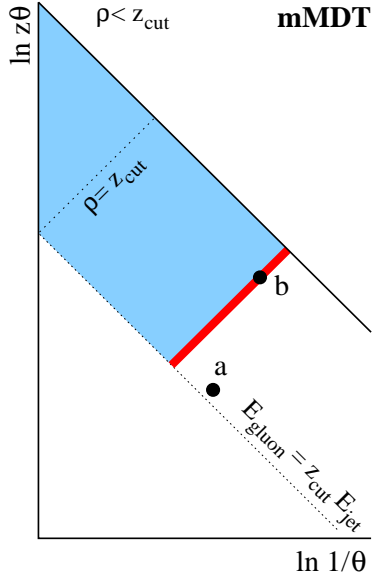}\hfill
  \caption{Lund kinematic diagram for the modified Mass-Drop Tagger
    (mMDT), concentrating on the region $\rho < \zcut$.
    The emissions labelled $a$ and $b$ are relevant to the discussion
    of angular versus mass ordering and the treatment of terms
    involving powers of $\as \ln^2 \ycut$. }
  \label{fig:mMDT-lund}
\end{figure}

Not only does the mMDT eliminate the wrong-branch issue, but it also
turns out to greatly facilitate the resummation of the tagged mass
distribution.

As usual, we will work in the limit in which $\ycut$ is small,
but $\as \ln \ycut$ is also small.
To avoid complicating our formulae with excessive $\Theta$-functions,
we will only quote explicit results in the plateau region of the LO
calculation, i.e.\ $\rho < \ycut$. For $\rho > \ycut$, one simply
obtains the plain jet-mass distribution.

It is useful to carry out the calculation in an angular ordered
formulation, reflecting the inherent angular ordering that is present
in the unclustering sequence followed by the tagger, a consequence of
the fact that it is based on the C/A algorithm.
We consider any number $n$ of emissions, strongly ordered in angle, $\theta_i
\ll \theta_{i-1}$, in configurations such that the $n^\text{th}$
emission has a momentum fraction greater than $\ycut$, while all the
others, at larger angles, have momentum fractions smaller than $\ycut$.
The latter are simply unclustered and discarded by the mMDT and it is
only when it reaches gluon $n$, the first with a momentum fraction
greater than $\ycut$, that it tags the
structure.
This leads to the following all-order result for the mass distribution:
\begin{multline}
  \label{eq:MDT-resum-step-1}
  \frac{1}{\sigma}\frac{d \sigma}{d \rho}^\text{(mMDT)}
  = 
  \sum_{n=1}^\infty 
  \int
  \frac{\alpha_s C_F}{\pi} 
  dz_n\, p_{gq}(z_n)\frac{d\theta_n^2}{\theta_n^2} 
  \Theta\left(z_n  - \ycut  \right)
  \delta \left(\rho-z_n  \frac{\theta_n^2}{R^2}\right)
  \Theta(\theta_{n-1} - \theta_{n})\\
  \times
  \prod_{i=1}^{n-1}
    \int
    \frac{\alpha_s C_F}{\pi} 
    dz_i \, p_{gq}(z_i)\frac{d\theta_i^2}{\theta_i^2} 
    \left[
      \Theta\left(\ycut - z_i   \right)
      -1
    \right]
    \Theta(\theta_{i-1} - \theta_{i})\,,
\end{multline}
In this formula, $z_i$ is the fraction of energy carried by gluon $i$
relative to that of the original jet. 
Because $\ycut \ll 1$, all emissions $i < n$ carry away only a
negligible fraction of the jet's energy, so that one can 
consider the jet as always having the same energy even after multiple
declusterings.
As well as including real emissions, we have
accounted for virtual corrections, the $-1$ contribution in the square
brackets; from unitarity considerations, these can be treated as
having the same phase-space integration as the real corrections, but
obviously without the constraint $z_i < \ycut$ imposed by the mass drop tagger.

The terms in square brackets in Eq.~(\ref{eq:MDT-resum-step-1}) can be
rewritten $-\Theta(z_i - \ycut)$. 
This makes it clear that all the $z_i$ in the integrals are restricted
to be larger than $\ycut$.
Insofar as we neglect logarithms of $\ycut$, we can then replace the
ordering of $\theta_i$ with an ordering in the variable $\rho_i \equiv
z_i \theta^2_i / R^2$, allowing us to rewrite
Eq.~(\ref{eq:MDT-resum-step-1}) in terms of integrals over (strongly)
ordered $\rho_i$ values, i.e. $\rho_i < \rho_{i-1}$.
The result for the integral of the $\rho$ distribution is then
straightforward to express as an exponential,
\begin{subequations}
  \label{eq:modMDT-rho-ordering}
  \begin{align}
    \Sigma^\text{(mMDT)} (\rho)&= \exp \left[ - \int_\rho^1
      \frac{d\rho'}{\rho'} \int^1_{\max(\ycut,\rho')} dz \, p_{gq}(z)
      \frac{C_F }{\pi}\, \alpha_s( z \rho' p_t^2 R^2)
    \right]\,,\\
    &= \exp \left[ -D(\max(\ycut,\rho)) 
                  - S(\ycut,\rho) \Theta(\ycut - \rho) \right]
  \end{align}
\end{subequations}
where we have now explicitly written in the scale for the coupling and
taken care of the modified $z$ integration limit for $\rho' > \ycut$.

As usual, it can be convenient to examine
Eq.~(\ref{eq:modMDT-rho-ordering}) in the fixed coupling
approximation.
It is given by
\begin{equation} 
  \label{eq:modMDT-fixed-coupling}
  \Sigma^\text{(mMDT)}(\rho) = 
  \exp \left [  
    - \frac{\alpha_s C_F}{\pi} 
    \left(
        \ln  \frac{\ycut}{\rho} \ln \frac{1}{\ycut}
      \,-\, \frac34 \ln \frac{1}{\rho}
      \,+\, \frac12 \ln^2 \frac{1}{\ycut}
    \right)
  \right]\,, 
  \quad (\text{for}\;  \rho < \ycut)\,,
\end{equation}
which is simply the exponential of the integral of
the LO result, Eq.~(\ref{eq:MDT-LO}).

Eq.~(\ref{eq:modMDT-rho-ordering}) corresponds to evaluating the
probability for excluding the shaded region shown in
Fig.~\ref{fig:mMDT-lund}.
From this, and the explicit fixed-coupling form,
Eq.~(\ref{eq:modMDT-fixed-coupling}), it is straightforward to see
that the most logarithmically divergent term in $\Sigma^\text{(mMDT)}$ at any order
in $\as$ is $\as^n L^n$, i.e.\ there are no terms beyond single
logarithms.
Considering that all other taggers had terms $\as^n L^p$ with $p$ up
to $2n$ or $2n-1$, this is a striking result.

Note that the strong ordering approximation for $\rho_i$ values that
is implicit in obtaining Eq.~(\ref{eq:modMDT-rho-ordering}) is the
main reason why we are able to neglect the effect of the mass-drop
condition in the tagger: for $\mu$ not too small, each time that one
unclusters a subjet $j$ into a $j_1$ and $j_2$, if $z > \ycut$, then one
knows that $m_{j_1} \ll m_j$ and so the mass-drop condition  $m_{j_1}
< \mu m_j$ is automatically satisfied.
Of course, for finite $\mu$ values, there is a relative order $\as$
probability that $m_{j_1} > \mu m_{j}$, so causing the mass-drop
condition to fail.
Insofar as we control terms $\as^n \ln^{n} \rho$ in $\Sigma^\text{(mMDT)}$, this
corresponds to corrections $\as^{n+1} \ln^{n} \rho$, which are beyond
our accuracy.

It is interesting that Eq.~(\ref{eq:modMDT-rho-ordering}), evaluated
with a coupling that freezes in the infrared, tells us that every jet
should be successfully mass-drop tagged, albeit possibly with a very
small tagged mass. 
In practice, confinement modifies this picture and in Monte Carlo
studies at hadron-level about 90\% of jets pass the mMDT procedure.

So far we have concentrated on a limit where $\ycut \ll 1$, while at
the same time neglecting logarithms of $\ycut$. It is interesting to
explore what happens when we go beyond this limit.
For sufficiently small $\ycut$, one might also aim to control terms
$(\as \ln^2 \ycut)^m (\as \ln \rho)^n$ for any $m,n$.
In this case a potential subtlety is that one should account for the
difference between angular and mass ordering, because given some
emission $a$ with $z > \ycut$, there is a probability $\sim \as \ln^2 \ycut$
of having a second emission $b$ with $z > \ycut$, at a smaller angle
than $a$ but contributing more than $a$ to the jet mass.
Such a configuration is illustrated in Fig.~\ref{fig:mMDT-lund}.
Here, emission $a$ will be unclustered before emission $b$. 
Its contribution to the squared mass $m_{a1}^2$ will in general be
much smaller than that from $b$, $m_{b1}^2$.
Consequently $m_{ab1} - m_{b1} \ll m_{ab1}$, i.e.\ there is no
substantial mass drop when unclustering $a$. 
Emission $a$ is therefore discarded and it is only when $b$
is unclustered that the jet is tagged.
This type of configuration might appear to complicate the treatment of
the tagger, but actually it simply implies that it is irrelevant
whether emission $a$ is present or not. 
For this reason, we believe that Eq.~(\ref{eq:modMDT-rho-ordering}),
written in terms of mass ordering, is correct for all terms $(\as
\ln^2 \ycut)^m (\as \ln \rho)^n$.
Accordingly, we have chosen to explicitly include terms that are subleading
in a counting of powers of $\ln \rho$, but $\ln^2 \ycut$-enhanced, in
our expressions Eqs.~(\ref{eq:modMDT-rho-ordering}),
(\ref{eq:modMDT-fixed-coupling}).\footnote{ For pruning and trimming,
  where for small $\zcut$ we explicitly control terms $\as^n
  \ln^{2n-q} \rho$ ($q=0,1$ for trimming and $\anomalouspruning$,
  $q=1,2$ for $\sanepruning$), it is possible that our formulae also
  control all terms $(\as \ln^2 \zcut)^m \as^n \ln^{2n-q}
  \rho$. However we leave the detailed verification of this conjecture
  to future work.  }
We believe the result is identical also for $\mu=1$: 
there will be an infinitesimal mass drop when
emission $a$ is unclustered, which is now sufficient to trigger the
mass-drop condition; however,
the masses $m_{ab1}$ and $m_{b1}$ differ little in most of the
relevant phase space, so that once again it is irrelevant whether
emission $a$ is present or not. 

It is also possible to examine the mass-drop tagger for moderate $\ycut$
values.
One of the key new features that arises at single-logarithmic accuracy
in this limit is that one now discards emissions with moderate $z$,
and these have a finite probability for modifying the flavour of the
remaining hard prong.
Therefore Eq.~(\ref{eq:modMDT-rho-ordering}) needs to be extended to
account for a matrix structure in flavour space.
This, and other aspects of the moderate-$\ycut$ case, are discussed in
detail in appendix~\ref{sec:finite-ycut-modMD}.

\subsection{Absence of non-global logarithms}
\label{sec:mMDT-non-global}

As we have already observed, there are no terms in the integrated
tagged mass distribution of the form $\as^n \ln^m \rho$ with $m > n$.
In other words, there is at most one  logarithm of $\rho$ for each
power of $\as$.
It is to our knowledge the first time that a jet-mass type observable
is found with this property.
The reason that there are only single logarithms is that the mMDT
completely removes contributions from soft emissions,
i.e.\ one is left only with collinear divergences, but not
soft-collinear ones, or pure soft ones.

The absence of pure soft divergences has a particularly interesting
consequence, namely the absence of non-global logarithms.
As we explained in section~\ref{sec:plain-jet-mass}, non-global
logarithms are potentially problematic.
They typically arise from situations where a soft emission outside a
(sub)jet emits a yet softer emission into the (sub)jet.
Soft emissions inside the jet are systematically discarded by mMDT
(or, in the situations where they're kept, don't affect the final
tagged jet mass) and so the non-global logarithms are eliminated.
The same mechanism ensures the absence of related ``clustering''
logarithms~\cite{Appleby:2002ke,Delenda:2006nf}.
This makes the mMDT particularly interesting, as the only infrared and
collinear safe single-jet observable that can be straightforwardly
calculated to single logarithmic accuracy with the full $N_C$
dependence.
It also suggests that the mMDT should be given priority in
calculations aiming for accuracy beyond single logarithms.


\subsection{Comparison with Monte Carlo results}\label{sec:mMDT-MC}
\begin{figure}
  \centering
  \includegraphics[width=0.49\textwidth]{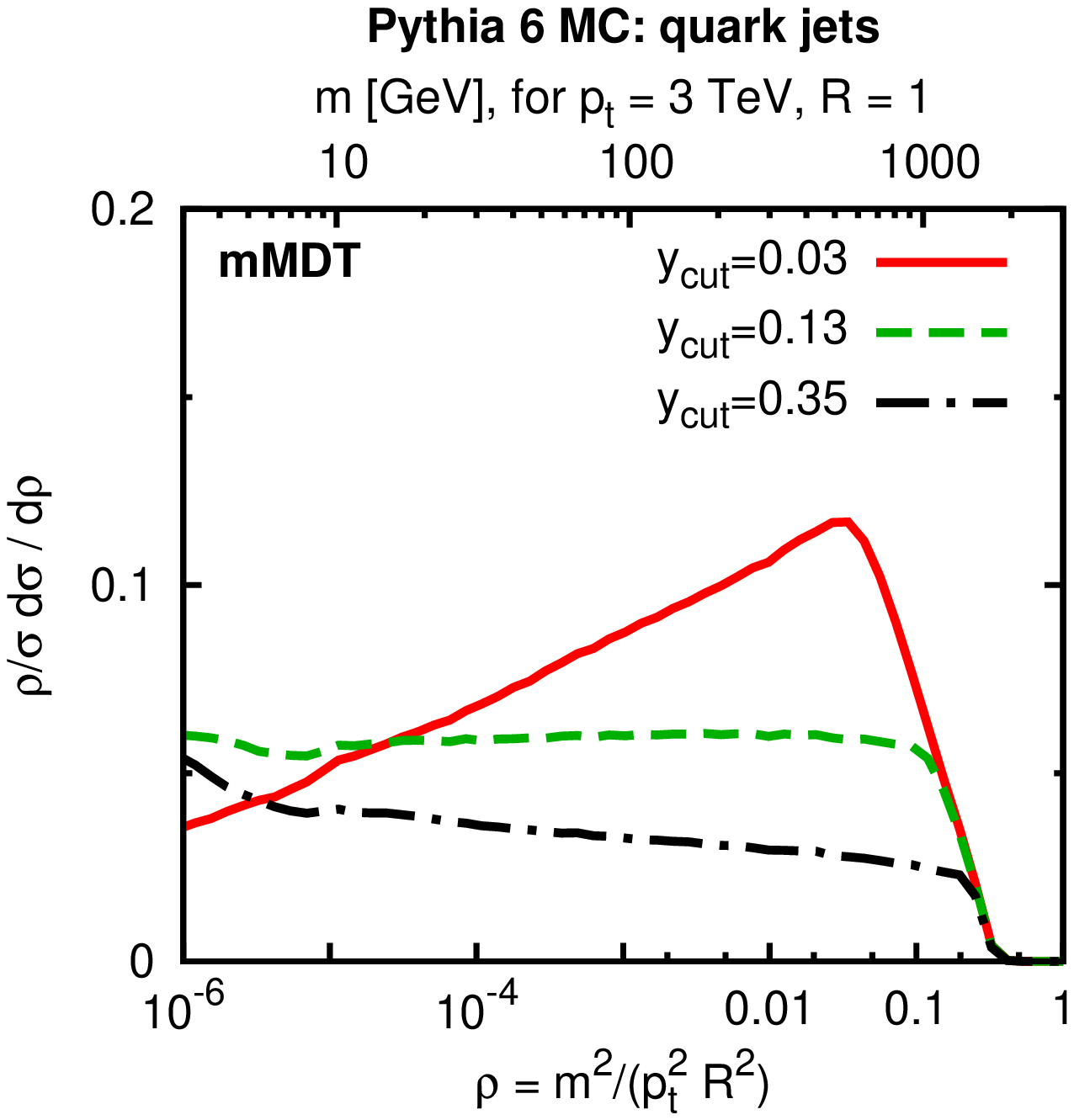}\hfill
  \includegraphics[width=0.49\textwidth]{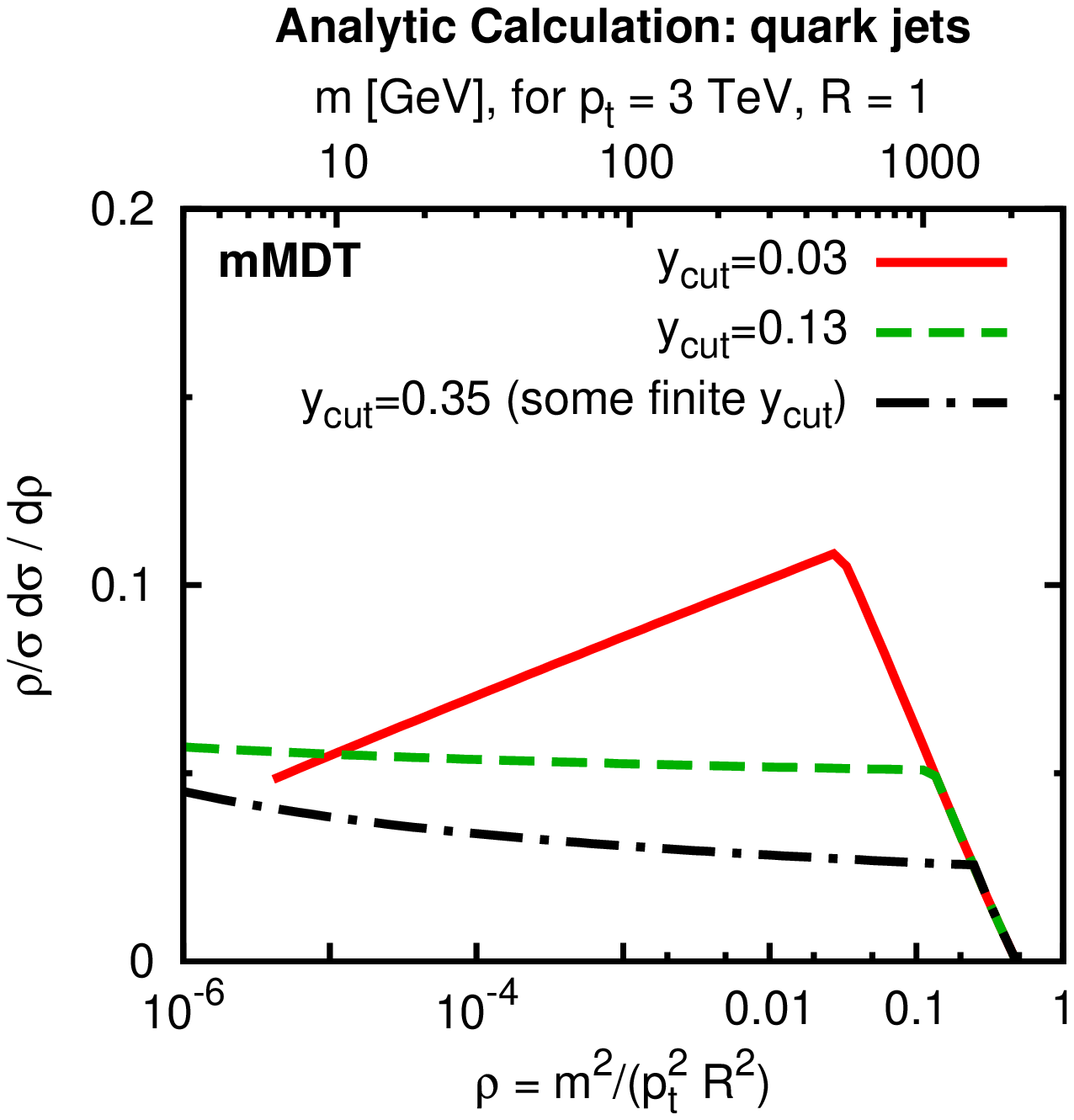}\\[5pt]
  \includegraphics[width=0.49\textwidth]{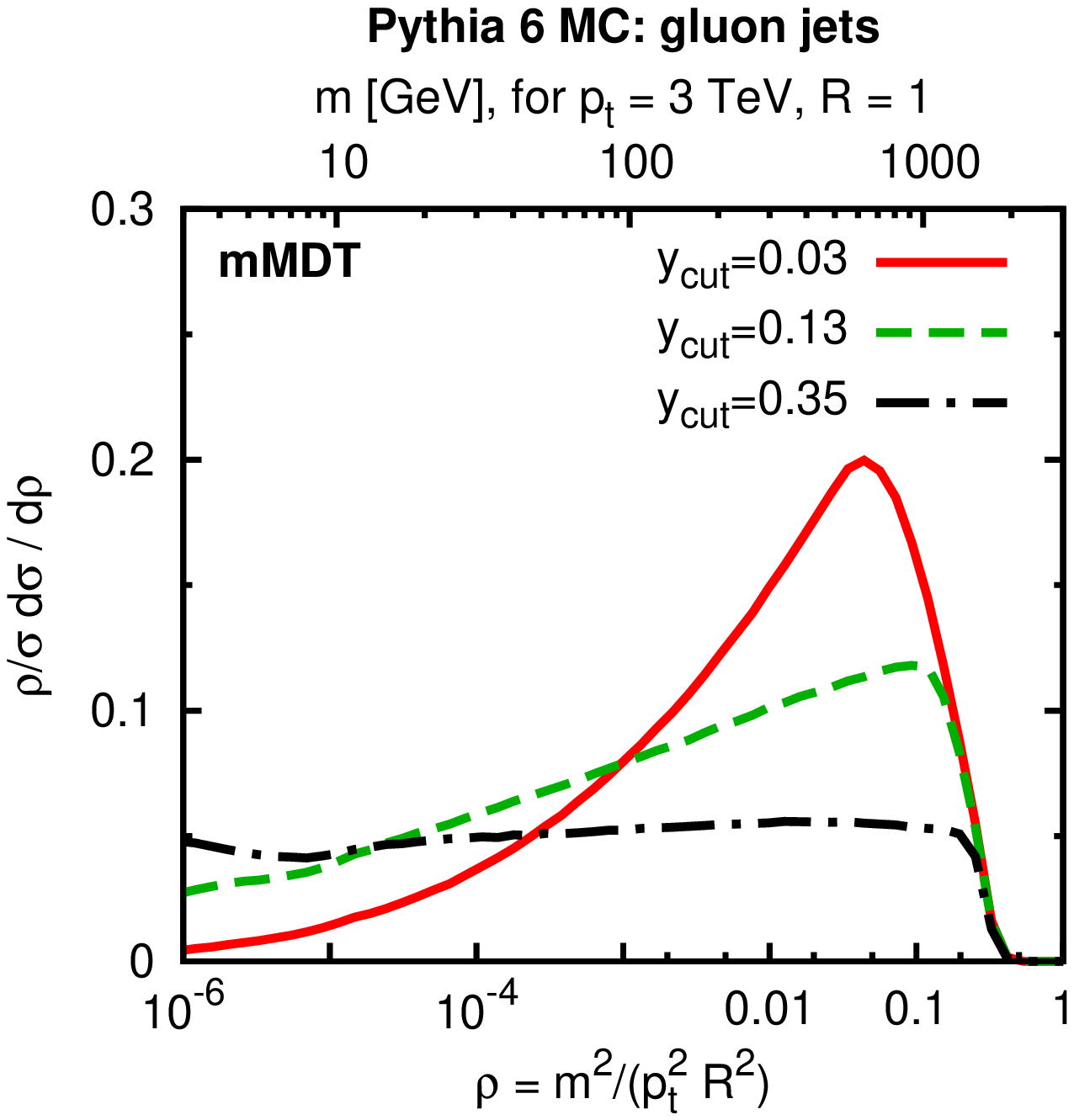}\hfill
  \includegraphics[width=0.49\textwidth]{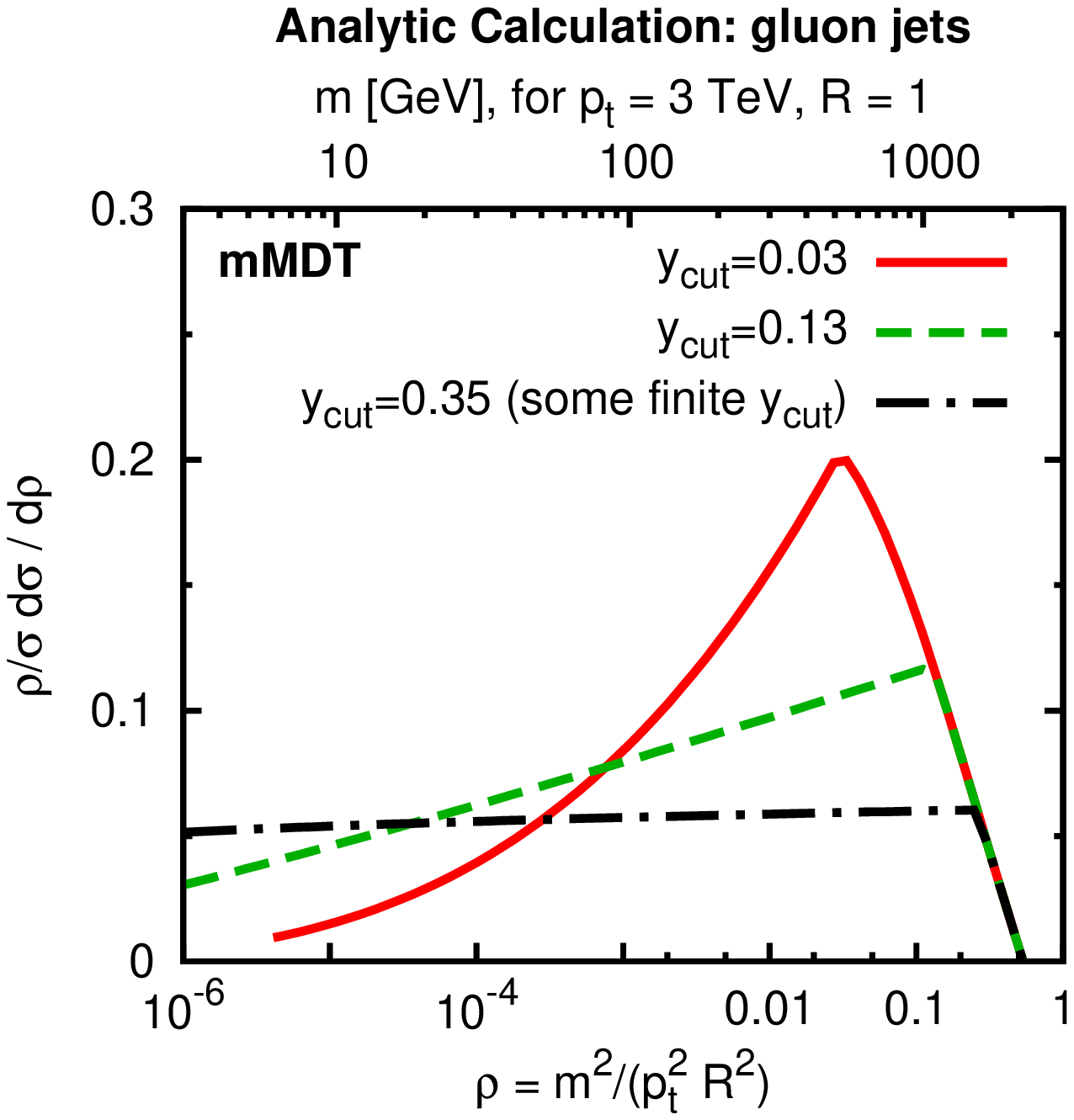}
  \caption{%
    Comparison of Monte Carlo (left panels) and analytic results
    (right panels) for the modified mass-drop tagger (mMDT).  The
    upper panels are for quark jets, the lower panels for gluon
    jets.  %
    Three values of $\ycut$ are illustrated, while $\mu$ is always
    taken to be $0.67$ (its precise value has no impact on the results,
    as long as it is not substantially smaller than this).
    The details of the MC event generation are as for
    Fig.~\ref{fig:tagged-mass-MC}.  }
  \label{fig:mMDT-multiple-ycut}
\end{figure}
Our analytical results are shown in Fig.~\ref{fig:mMDT-multiple-ycut}
(right-hand plots) compared to parton-level Monte Carlo predictions
with Pythia 6 (left, virtuality ordered shower).
The upper panels show the results for quark jets, the lower panels for
gluon jets.
Three choices of $\ycut$ are shown.
The agreement between Monte Carlo and the analytical results is striking. 
In particular, we note that there are two particular values of asymmetry parameter, namely $\ycut=0.13$ for quark-initiated jets, and $\ycut=0.35$ in the case of gluon-initiated jets, for which the mMDT mass distribution is essentially flat. We will come back to this observation in section~\ref{sec:background}, where we discuss background shapes in more detail.

Note that for the $\ycut=0.35$ choice, the analytical results have been
supplemented with a subset of the finite $\ycut$ effects,
specifically, those that are flavour-diagonal. Further details are
given in appendix~\ref{sec:finite-ycut-modMD}.
Residual small differences between the Monte Carlo and analytical
results for $\ycut=0.13$ are in part due to the fact that we have left
out finite $\ycut$ effects there.

\subsection{Dependence on $\mu$ parameter}
\begin{figure}
  \centering
  \includegraphics[width=0.49\textwidth]{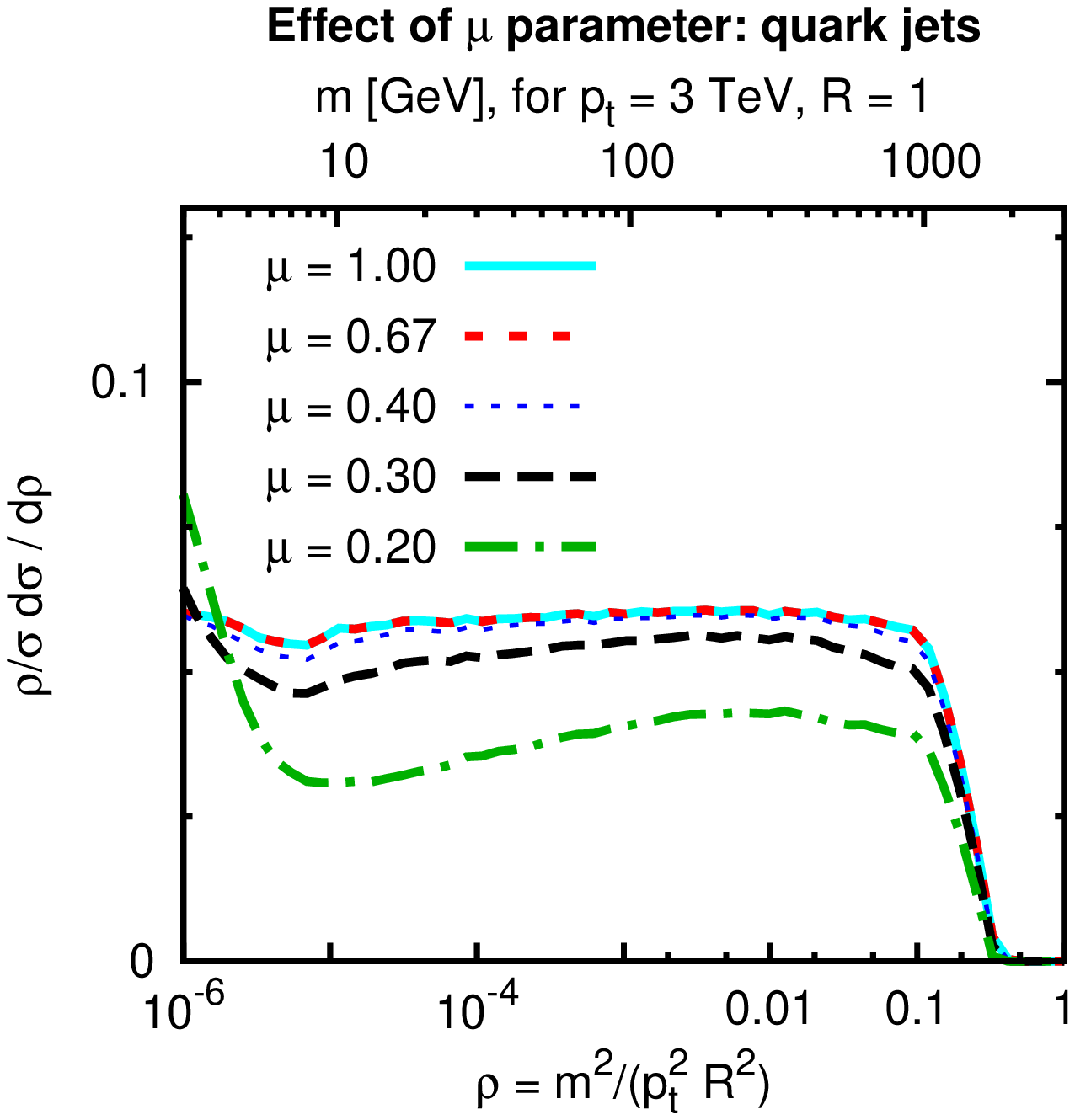}\hfill
  \includegraphics[width=0.49\textwidth]{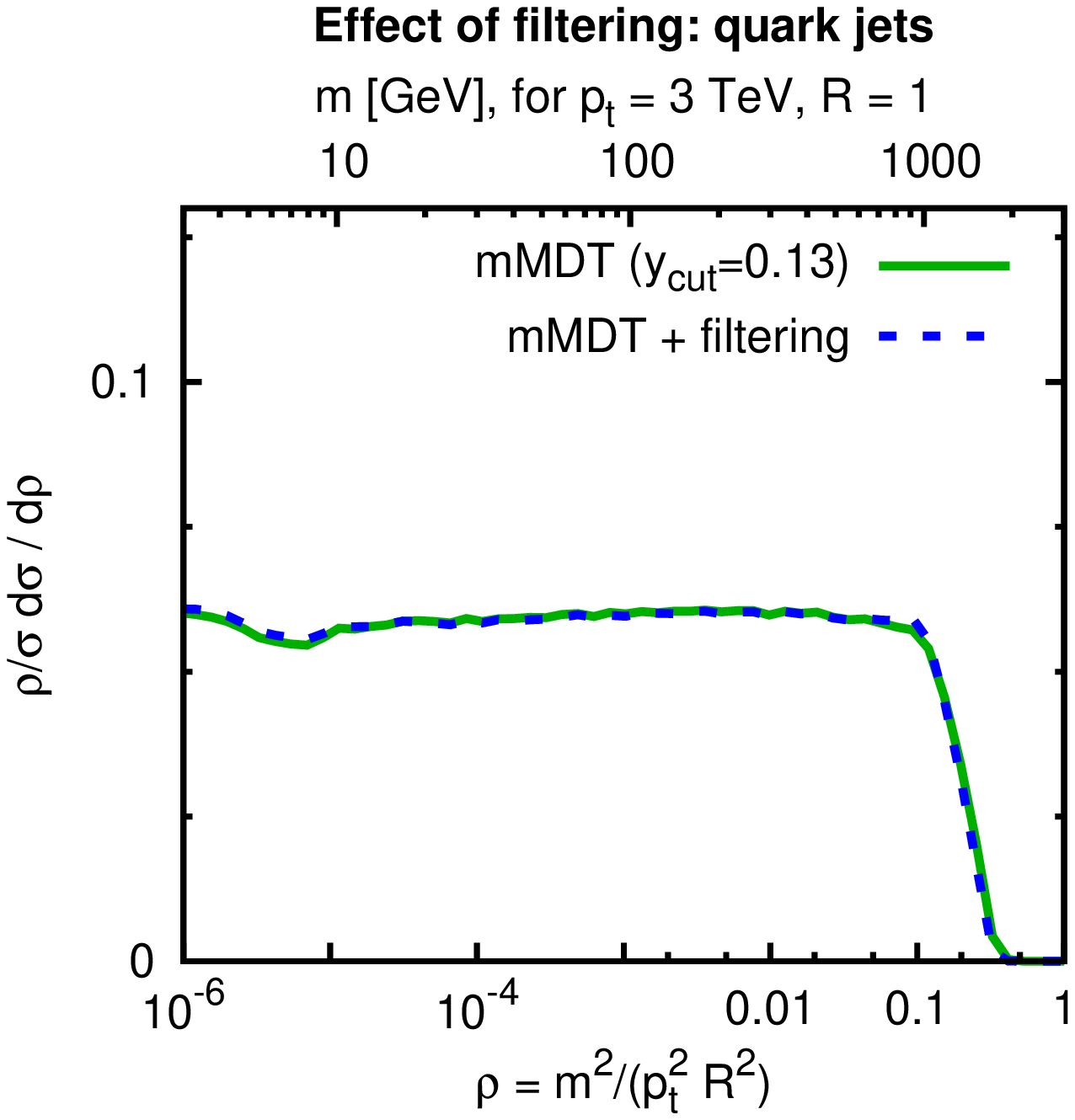}  
  \caption{ MC study of the impact of the mass-drop parameter ($\mu$)
    on the mMDT mass distribution (left panel) and of filtering (right
    panel) for quark jets.  
    Filtering is carried out with $\Rfilt = \min(\frac12
    \Delta_{12},0.3)$ and $\nfilt=3$.
    The details of the MC event generation are
    as for Fig.~\ref{fig:tagged-mass-MC}. 
    Gluon-initiated jets behave in a similar way.}
  \label{fig:mMDT-massdrop}
\end{figure}

As we have already discussed in section~\ref{sec:mmdt-all-orders}, the
dependence of the mass-drop parameter $\mu$ enters beyond the
single-logarithmic accuracy we achieve for mMDT. 
Fig.~\ref{fig:mMDT-massdrop} (left panel) shows the results of a
simple Monte Carlo study to numerically investigate the impact of the
mass-drop parameter on the tagged mass distribution.
One sees that for $0.4 \lesssim \mu \le 1$ there is essentially no
dependence on $\mu$.
For smaller values of $\mu$ the background tagging rate drops.
This is caused by contributions that are subleading in terms of the
number of logarithms of $\rho$, but enhanced by powers of $\ln^2 \mu$,
and associated with the Sudakov suppression for requiring that each of
the two prongs of the tagged jet have a very small mass.

In light of these theoretical and Monte Carlo observations it seems
that one could use mMDT entirely without any mass-drop condition.
We believe that this simplification of the tagger deserves further
investigation in view of possibly becoming the main recommended
variant of mMDT.\footnote{This would of course leave ``modified Mass
  Drop Tagger'' as a somewhat inappropriate name!}

\subsection{Interplay with filtering}
\label{sec:filtering}

The mass-drop tagger is often used together with a filtering
procedure, which reduces sensitivity to underlying event and pileup.
In its original incarnation a filtering radius $R_\text{filt}$ was
chosen equal to $\min(\Delta_{12}/2,0.3)$~\cite{Butterworth:2008iy},
where $\Delta_{12}$ is the 
angular separation between the two prongs of the jet after tagging
(for brevity, we call this the tagged jet).
The tagged jet was then reclustered with radius
$R_\text{filt}$, and only its $\nfilt$ hardest prongs are kept.

From the point of a general analytical discussion of the effect of
filtering, it is immaterial whether one use $R_\text{filt} =
\min(\Delta_{12}/2,0.3)$ or simply some moderate fixed fraction of
$\Delta_{12}$.\footnote{An extensive analytical study of the optimal
  choice for signal reconstruction was given by Rubin in
  Ref.~\cite{Rubin:2010fc}.}
What matters more is the choice of $\nfilt$: for a tagged jet
with $n$ particles, filtering will always leave the jet unmodified if
$n \le \nfilt$.
It is only if the jet has more than $\nfilt$ subprongs on an
angular scale $R_\text{filt}$ that filtering will change its mass.
This occurs with relative probability $\as^{\nfilt-1}$
(e.g. for $\nfilt=3$ there must be at least two additional
gluons in order for filtering to discard anything).

Naively one would therefore think that filtering introduces a
modification at order N$^{\nfilt-1}$LL. 
However one should keep in mind that filtering doesn't cause the jet
to be discarded, but instead simply changes its mass.
Suppose, for instance, that it reduces the mass by some factor $f$
with a probability $\as^{\nfilt-1}$.
Given a pre-filtering integrated mass distribution of $\Sigma(\rho) =
\sum_n c_n \as^n L^n$, the post-filtering distribution will be
\begin{subequations}
  \label{eq:post-filt}
  \begin{align}
    \Sigma^\text{(filt)}(\rho) &= \Sigma(\rho) +
    \as^{\nfilt-1}\left(\Sigma(\rho/f^2) - \Sigma(\rho)\right)
    \\
    &= \sum_n c_n \as^n L^n + \sum_n c_n \as^{\nfilt + n - 1}
    \left[ (L + 2\ln f)^n - L^n \right]
    \label{eq:post-filt-b}
  \end{align}
\end{subequations}
The right-hand term of Eq.~(\ref{eq:post-filt-b}) goes as
$\as^{\nfilt + n - 1} L^{n-1}$, i.e.\ it is N$^{\nfilt}$LL. 
Accordingly, with the common choice $\nfilt = 3$, it is unlikely
that there will be a need to perturbatively calculate filtering's
impact on the background in the near future!

We can verify this conclusion numerically with the help of a Monte
Carlo study.
This is shown in Fig.~\ref{fig:mMDT-massdrop} (right), where mMDT mass
distributions are compared with and without filtering, using $\nfilt =
3$. 
The difference between them is hardly perceptible.


\subsection{Calculability at fixed order}\label{sec:FO}
An interesting consequence of the presence of only single logarithms
relates to the extent to which fixed-order calculations are reliable. 
For observables with terms $\as^n L^{2n}$, fixed-order perturbation
theory breaks down when $ L \sim 1/\sqrt{\as}$ and becomes unreliable
somewhat earlier.
Instead, for observables whose most divergent terms are $\as^n L^{n}$,
the breakdown occurs when $L \sim 1/\as$, i.e.\ fixed-order
perturbation theory has a parametrically larger domain of
applicability.
We have not investigated the behaviour of the fixed-order predictions
in detail, however such a study would be worthwhile and is
straightforward to perform to NLO in the jet mass distribution with
tools such as MCFM~\cite{Campbell:2002tg} and
NLOJet++~\cite{Nagy:2003tz}.

\section{Phenomenological considerations} \label{sec:pheno}

\subsection{Comparisons between taggers}
\label{sec:perturbative-comparisons}

We have commented in previous sections on similarities between the
taggers for regions of intermediate tagged mass.
In particular if one chooses $\ycut = \frac{\zcut}{1-\zcut}$, then one
expects trimming and pruning to be nearly identical to mMDT in the
regions $\rho > \zcut (R_\text{sub}/R)^2$  and $\rho > \zcut^2$
respectively. 

\begin{figure}[h]
  \centering
  \includegraphics[width=0.49\textwidth]{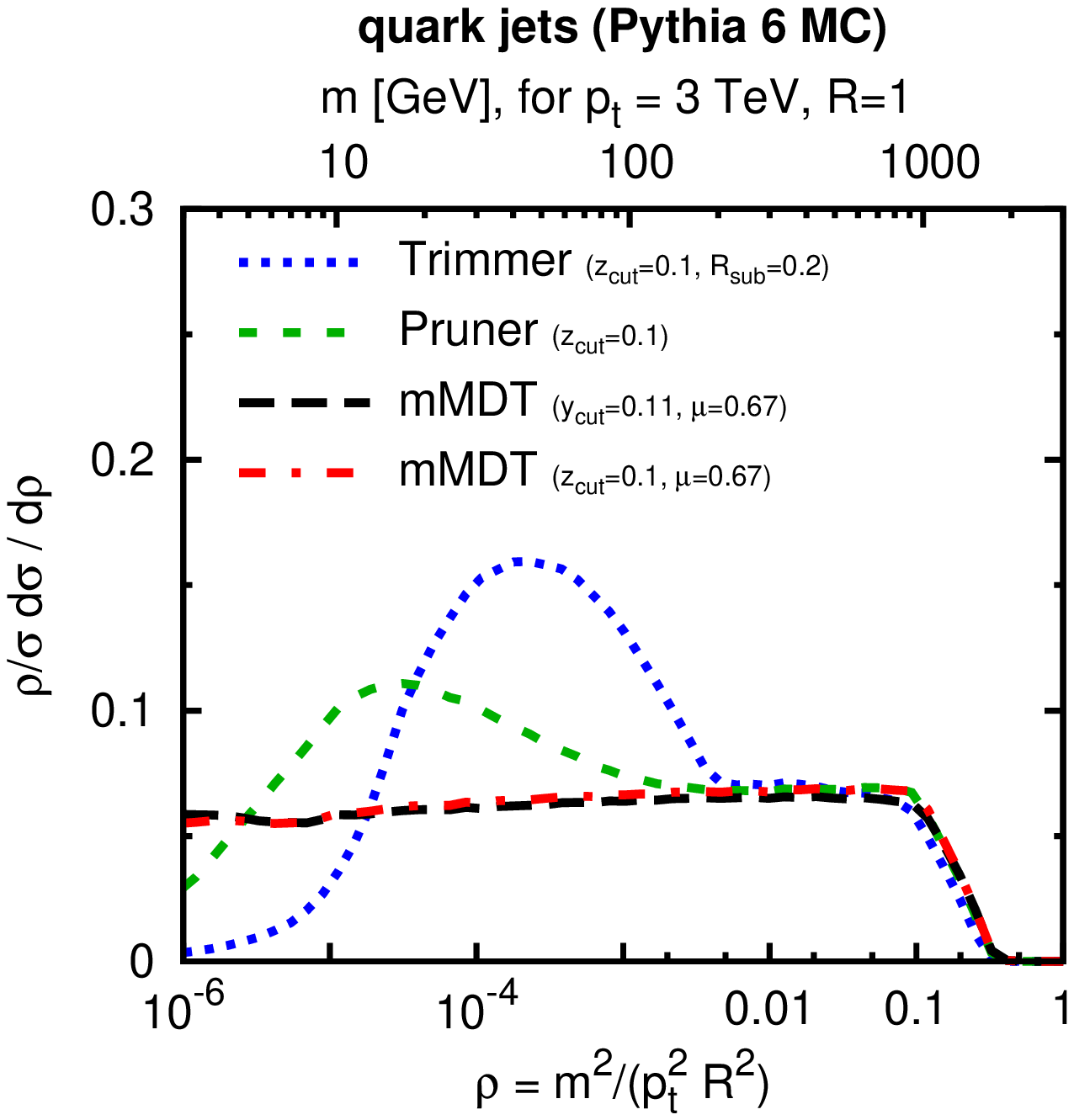}\hfill
  \includegraphics[width=0.49\textwidth]{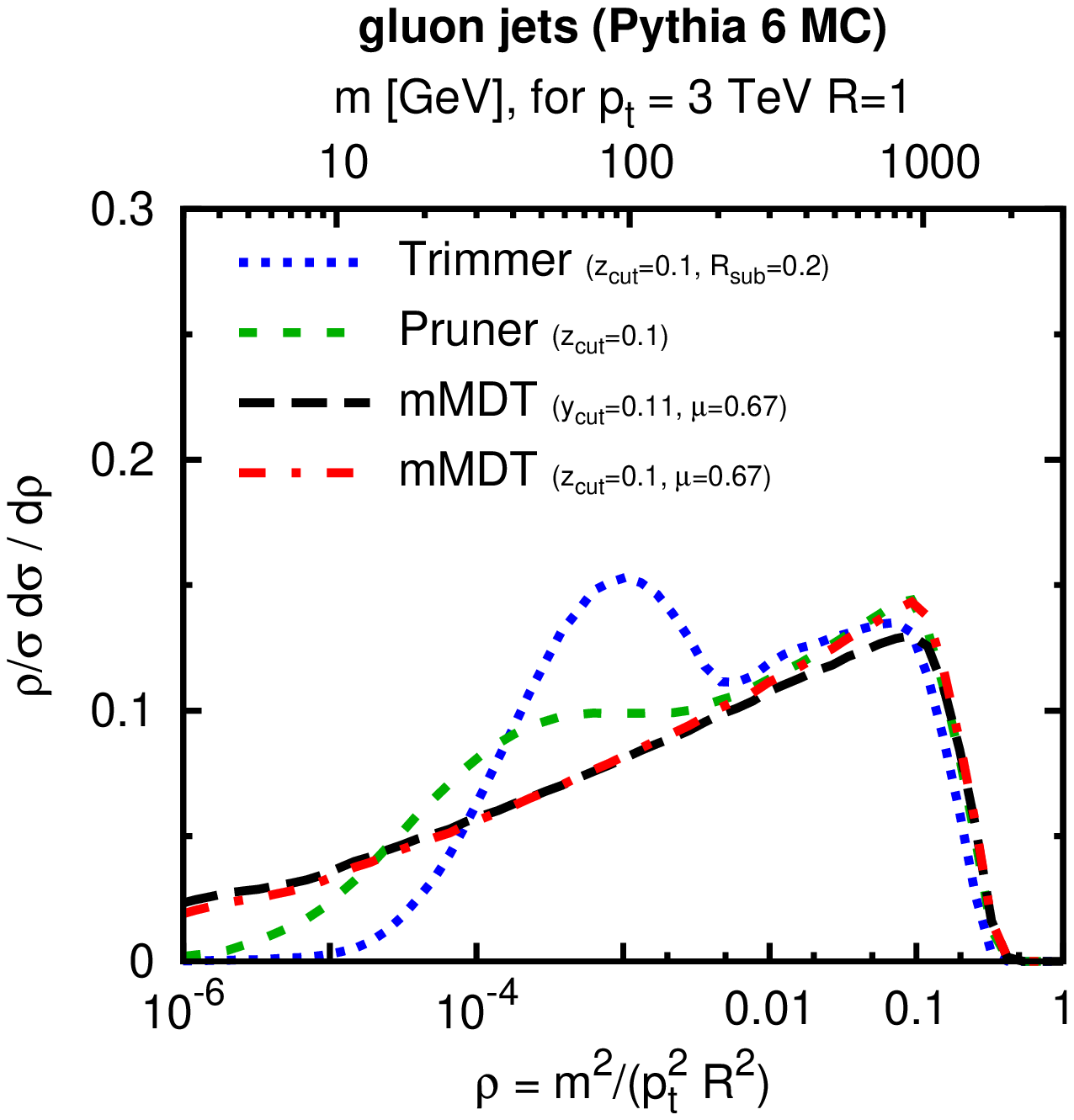}\\[5pt]
  \caption{%
    Comparisons of Monte Carlo distributions for trimming, pruning and
    mMDT for equivalent parameters, i.e.\ identical $\zcut=0.1$ for
    trimming and pruning, while for mMDT we use $\ycut =
    \zcut/(1-\zcut)$. We also show the $\zcut$-variant of mMDT defined
    in the text, with $\zcut=0.1$. 
    The details of the MC event generation are as for
    Fig.~\ref{fig:tagged-mass-MC}. 
  }
  \label{fig:perturbative-comparisons}
\end{figure}

Choosing $\ycut = 0.11$ and $\zcut=0.1$, this feature is evident in
Fig.~\ref{fig:perturbative-comparisons}. 
There are remaining small differences between the tools, and in
particular in the gluon case, for $\rho < \zcut$ one sees that
trimming and pruning are closer to each other than either is to mMDT.
With the help of further Monte Carlo studies, we have traced the
difference to fact that both trimming and pruning directly cut on
transverse momentum fractions (albeit normalised slightly
differently), while mMDT cuts on a ratio of a $k_t$-distance to a
mass, which only indirectly translates to a cut on momentum fractions.
If, for instance, in step 2 of the definition of (m)MDT one replaces
the cut
$y = \min(p_{tj_1}^2, p_{tj_2}^2) \Delta R_{j_1 j_2}^2/ m_j^2 > \ycut$
with $\min(p_{tj_1}, p_{tj_2})/(p_{tj_1} + p_{tj_2}) > \zcut$, then
the small differences between mMDT and pruning in the region $\rho > \zcut^2$
disappear almost entirely, as can be seen.
It is straightforward to show that this change does not affect the
resummation at the order we have considered.

These observations are important, because previous discussions that
have commented on differences between groomers (e.g.~\cite{Boost2010})
were considering them with non-equivalent parameters.
As we see here, a suitable choice of parameters is essential for the
comparisons to be as informative as possible.

Among the groomers examined in Ref.~\cite{Boost2010}, there was also
filtering (without the mass-drop procedure and with a fixed
$\Rfilt$). While we have not investigated plain filtering in a similar
level of detail to trimming, pruning and mMDT, preliminary
investigations suggest that it leads to a background jet mass
distribution that is very similar to that for the plain jet mass, in
particular as concerns the leading-log structure $\as^n L^{2n}$.

\subsection{Background shapes}\label{sec:background}

From the point of view of searches with a small signal-to-background
ratio, the reliability of the prediction for the background and
especially its shape is crucial.

The background may be predicted with the aid of perturbation theory,
for which our resummation, merged with fixed-order calculations, would
be the state-of-the-art.
Alternatively, backgrounds may be predicted with data-driven methods. 
One example of such a method is to measure the background mass
distribution to the left and right of an expected W/Z or H mass peak
and use that to predict the background mass distribution in the peak
location. 
One may also take the shape of the background for moderate
$p_t$ jets, and attempt to use it to predict the shape for higher
$p_t$ jets.
From this point of view the structures present in the mass
distribution are of importance: for example Sudakov peaks, as they
appear in the normal jet mass, in trimming and in pruning, can
considerably complicate data-driven methods: they prevent one from
reliably interpolating the background between two sidebands, because
the peak may lie over one of the sidebands, or even worse, in between
them; they also make it more complicated to use a mass distribution at
one $p_t$ to predict the distribution at another $p_t$, because
Sudakov peak positions depend on the jet $p_t$.\footnote{One might of
  course instead use $\rho$ distributions, which are more stable with
  respect to changes in the jet $p_t$. }

The (modified) mass-drop tagger is particularly interesting in this respect for
two reasons. 
Firstly it is free of Sudakov peaks. 
Secondly it has an interesting feature that can be seen by expanding
Eq.~(\ref{eq:modMDT-rho-ordering}) to second order in the coupling,
restricting our attention to the region $\rho < \ycut$:
\begin{equation}
  \label{eq:mMDT-2nd-order}
  \frac{\rho}{\sigma}\frac{d \sigma}{d \rho}^\text{(mMDT)} = 
  \frac{\as C_F}{\pi} \ln \frac{e^{-\frac34}}{\ycut} \left[1 + 
  \frac{\as}{\pi} \ln \frac1{\rho} \left( 
    \beta_0 - C_F \ln \frac{e^{-\frac34}}{\ycut}
  \right)
  + \cdots
  \right]
\end{equation}
where $\beta_0 = (11 C_A - 2n_f)/12$.
Relative to the LO formula, Eq.~(\ref{eq:MDT-LO}), running coupling
effects (the $\beta_0$ term) cause the the distribution to increase
for low $\rho$, while the exponentiation in
Eq.~(\ref{eq:modMDT-rho-ordering}) brings a (single-logarithmic)
Sudakov type suppression.
For a specific value of $\ycut$, $\exp(-\frac34 -
\frac{\beta_0}{C_F})$
in the case of quark jets, those two effects cancel, leaving a mass
spectrum that is to a good approximation independent of $\rho$, a
property that is potentially valuable in data-driven background
estimates.
For $n_f = 5$ the relevant $\ycut$ value is $\ycut =
e^{-\frac{35}{16}} \simeq 0.11$.
Note that this is determined in the small-$\ycut$ approximation, which
is subject to corrections of relative $\order{\ycut}$.
Those corrections lead to a slight increase of the critical $\ycut$
value that is needed for flatness, which is consistent with the
practical observation of flatness for quark jets in
Fig.~\ref{fig:mMDT-multiple-ycut} at $\ycut\simeq0.13$.

Fig.~\ref{fig:mMDT-multiple-ycut} is also consistent with
the expectation from Eq.~(\ref{eq:mMDT-2nd-order}) that for small
$\ycut$ the mass distribution will tend to fall off towards small
$\rho$, with the slope being dominated by the Sudakov term;
conversely, for large $\ycut$ the distribution is more likely to
increase towards small $\rho$, with the slope being dominated by the
running-coupling term.
For gluon jets the $C_F$ coefficients are replaced by $C_A$ (and $\frac34$
by $\beta_0/C_A = \frac{23}{36}$). 
This causes the Sudakov-induced term to be relatively more important,
hence the tendency to decrease more steeply towards small $\rho$ and
the need for a larger $\ycut$ value in order to obtain a flat
distribution.

\subsection{Non-perturbative effects}
\label{sec:non-perturbative}

While the main aim of this work has been to understand perturbative
effects in the taggers, it is important to also be aware of the extent
to which they may be affected by non-perturbative contributions.

\subsubsection{Limit of perturbative calculation}
\label{sec:PT-limit}
One simple study is to determine, for each tagger, the
non-perturbative transition point, below which our calculations
start to probe the non-perturbative region.
One can define the transition point as the highest mass for which the
coupling, in any of the integrals, must be evaluated below some
non-perturbative transition scale $\mu_\text{NP}$. One can imagine
$\mu_\text{NP}$ to be of order $1\GeV$.

For the normal jet mass, the transition point can be evaluated by
considering an emission $i$ with $E_i \theta_i = \muNP$.
The squared jet mass is $m^2 = E_i  E_{\jet} \theta_i^2$ and so the
transition point is found taking the largest possible value for
$\theta$, which gives $m^2 \simeq \muNP  E_\text{jet} R$. In
longitudinally-invariant variables, this reads 
\begin{equation}
  \label{eq:m2-NP-plain}
  m^2 \simeq \muNP \, p_{t,\text{jet}}\, R\,,
  \qquad\quad\text{(plain jet mass)}.
\end{equation}
Note that this scale grows with the jet $p_t$, so that even apparently
large masses, $m \gg \Lambda_\text{QCD}$, may in fact be driven by
non-perturbative physics.
For a $3\TeV$ jet with $R=1$, taking $\muNP = 1\GeV$, the
non-perturbative region corresponds to $m \lesssim 55\GeV$,
disturbingly close to the electroweak scale!

To obtain the transition point for trimming, one simply replaces $R$
with $\Rsub$, giving 
\begin{equation}
  \label{eq:m2-NP-trimmed}
  m^2 \simeq \muNP \, p_{t,\text{jet}}\, \Rsub\,,
  \qquad\quad\text{(trimming)},
\end{equation}
assuming that this lies in the region $\rho < r^2 \zcut$, which
usually will be the case for sufficiently high $p_t$ jets.
For our canonical $3\TeV$, $R=1$ jet, taking $\Rsub = 0.2$ tells us
that the non-perturbative region is $m \lesssim 25\GeV$.

For both $\sane$- and $\anomalouspruning$, the non-perturbative transition
region is formally in the same location as for the plain jet mass.
This is because of the integrals over $\rhofat$,
Eqs.~(\ref{eq:sane-prune-result-v2}), (\ref{eq:anom-prune-result}),
whose lower limits can be as low as $\rho$.
Note, however, that the onset of the non-perturbative effects may be
substantially different, because the fraction of the answer that is
associated with the non-perturbative region, as well as the interplay
between real and virtual components, are different compared to the
plain jet mass.

Finally, for the modified mass-drop tagger, we first observe that
the smallest scale in the coupling will occur when the momentum
fraction of the tagged splitting is $z \simeq \ycut$.
The squared mass of the jet is then $m^2 \simeq \ycut E_\text{jet}^2
\theta^2$. 
Substituting the condition for the emission to be non perturbative,
$\ycut^2 E_\text{jet}^2 \theta^2 = \muNP^2$, leads to a transition
point of
\begin{equation}
  \label{eq:m2-NP-mMDT}
  m^2 \simeq \frac{\muNP^2}{\ycut}\,,\qquad\qquad\text{(mMDT)}.
\end{equation}
Note that in contrast with the cases seen above, this transition point
is independent of the jet $p_t$, and genuinely close to the
non-perturbative region.
Taking $\ycut = 0.1$, it corresponds to a scale of about
$3\GeV$.%
\footnote{
  The unmodified mass-drop tagger is more subtle, because
  non-perturbative effects can influence the likelihood of following
  the right v.\ wrong branches. 
  As a result, non-perturbative effects can set in, at least formally,
  at the same scale as for the plain jet mass, i.e.\ $\mu_\text{NP}\,
  p_t \,R$.
  In practice, given that the wrong branch issue is phenomenologically
  minor, this is unlikely to lead to substantially enhanced
  non-perturbative effects relative to the mMDT, however it is a
  relevant consideration from a calculational point of view.  }

\subsubsection{Monte Carlo study of hadronisation}
\label{sec:hadr-MC}

\begin{figure}[h]
  \includegraphics[width=0.49\textwidth]{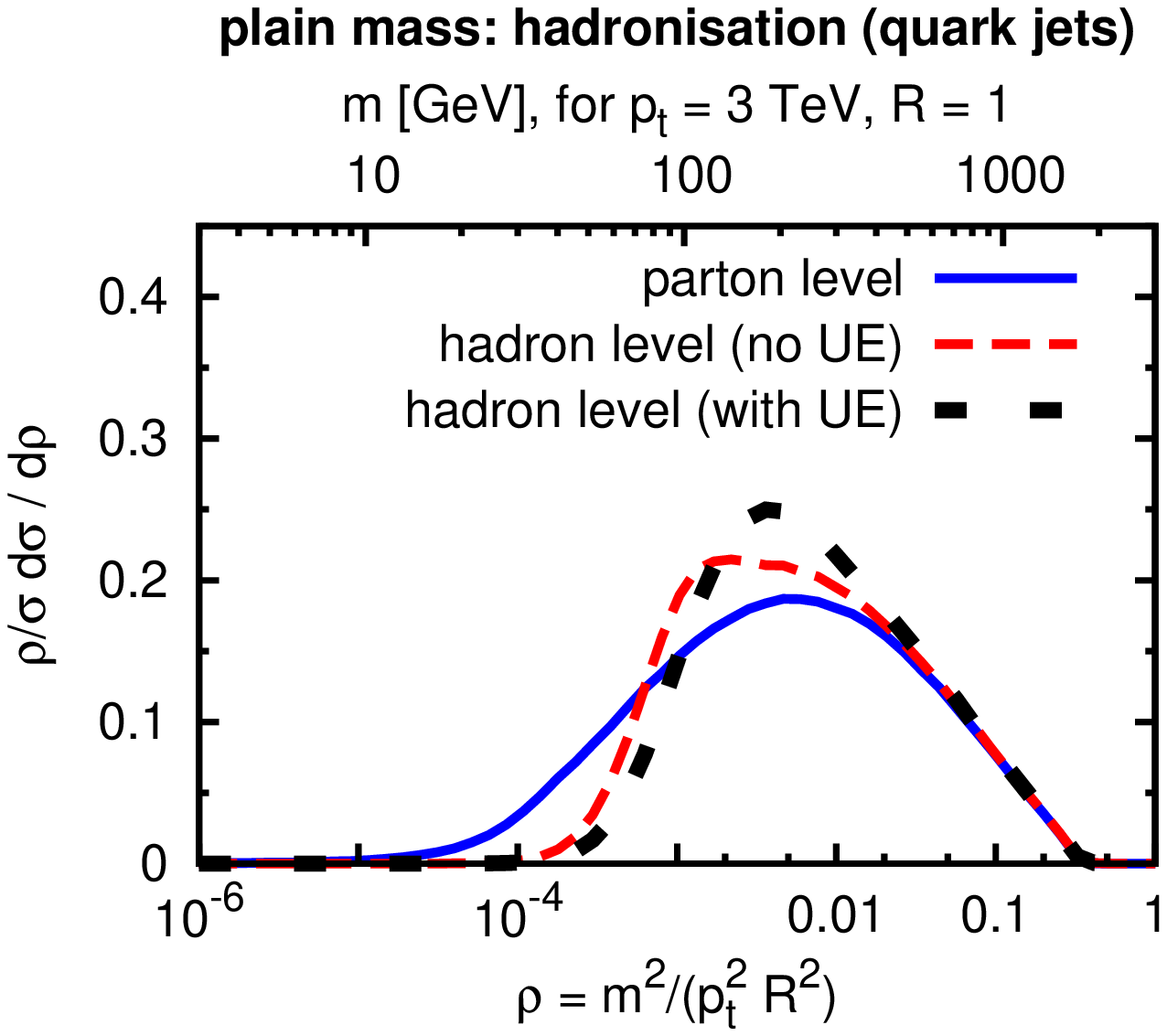}%
  \hfill
  \includegraphics[width=0.49\textwidth]{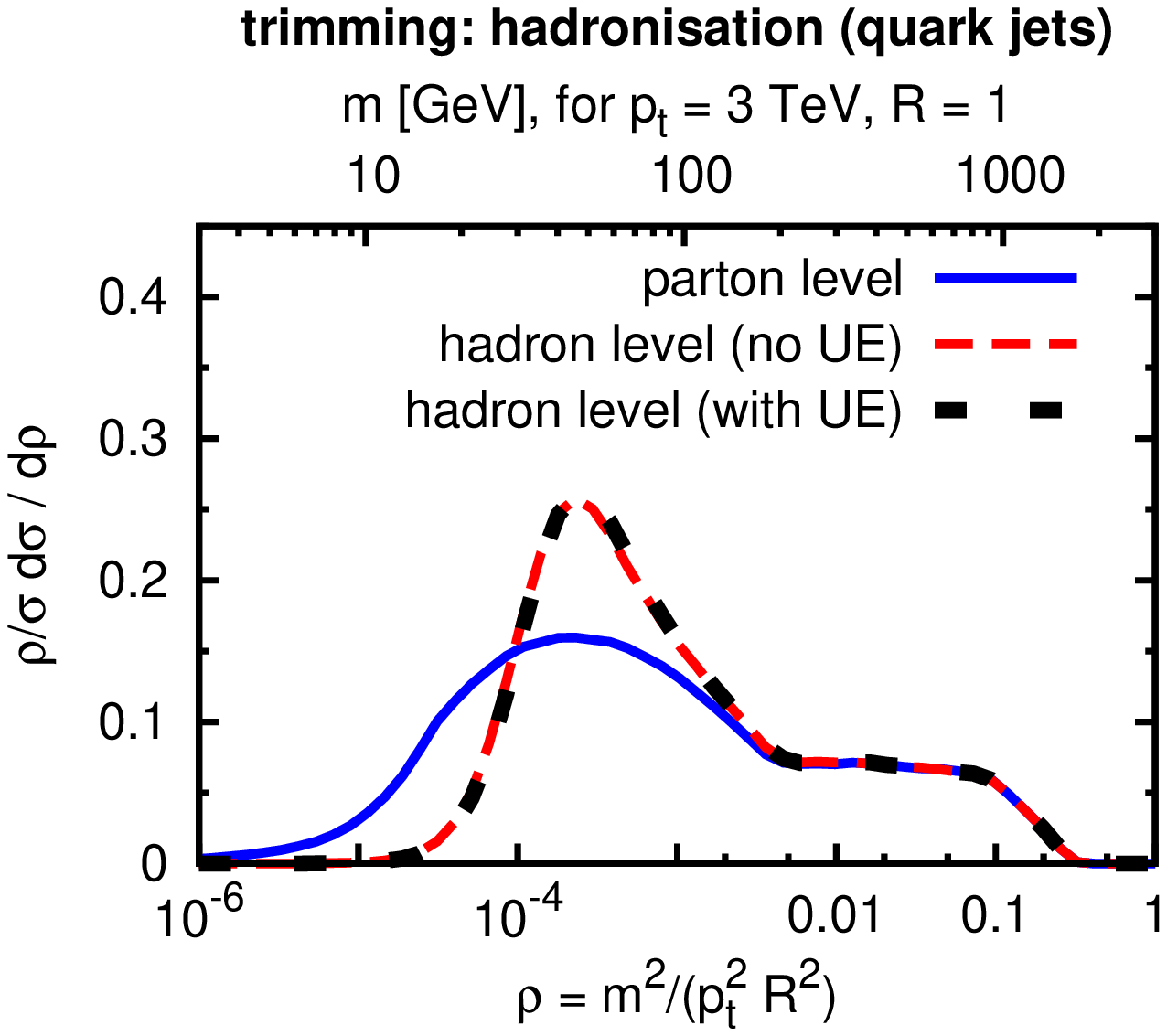}%
  \\[5pt]
  \includegraphics[width=0.49\textwidth]{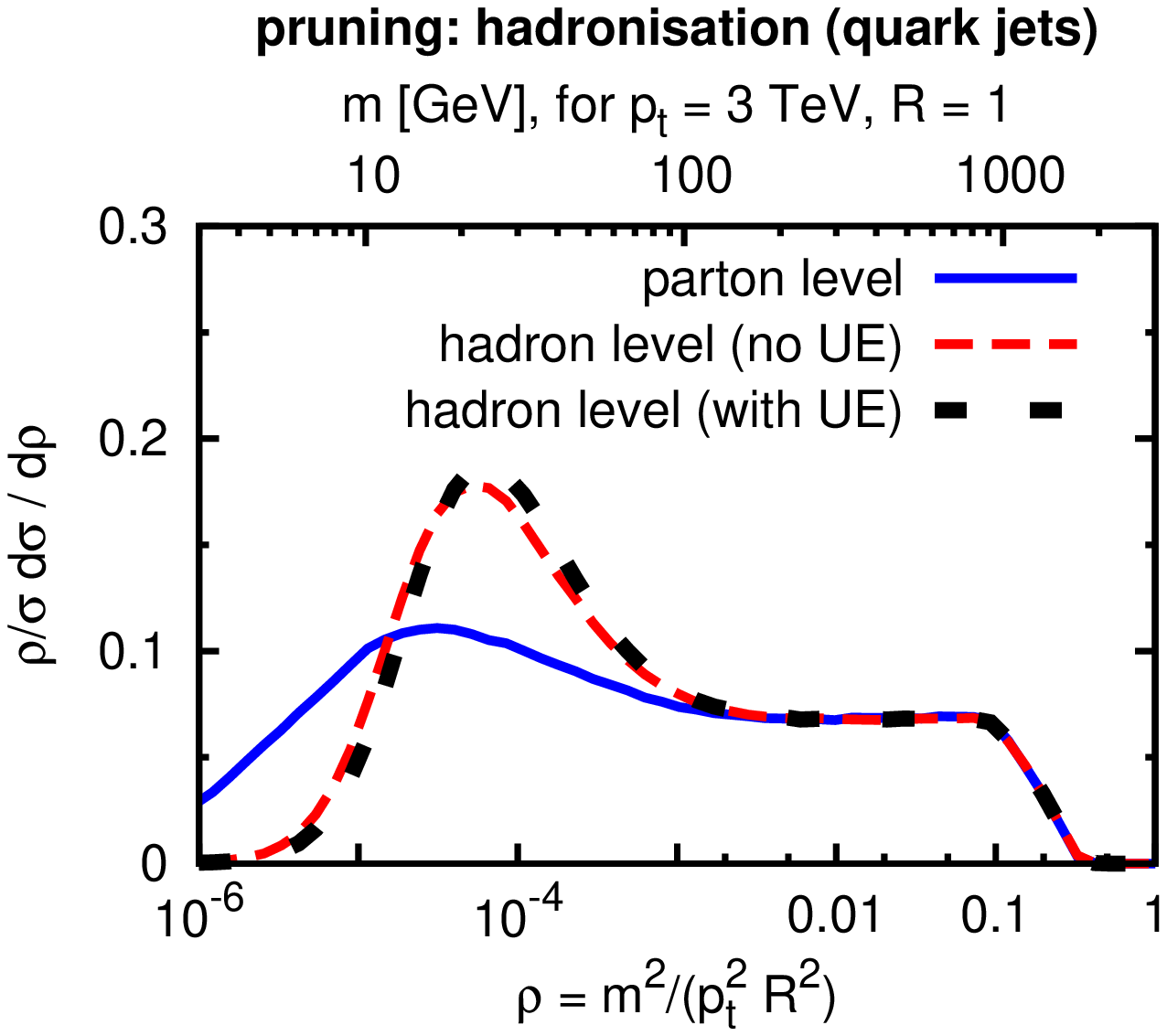}%
  \hfill
  \includegraphics[width=0.49\textwidth]{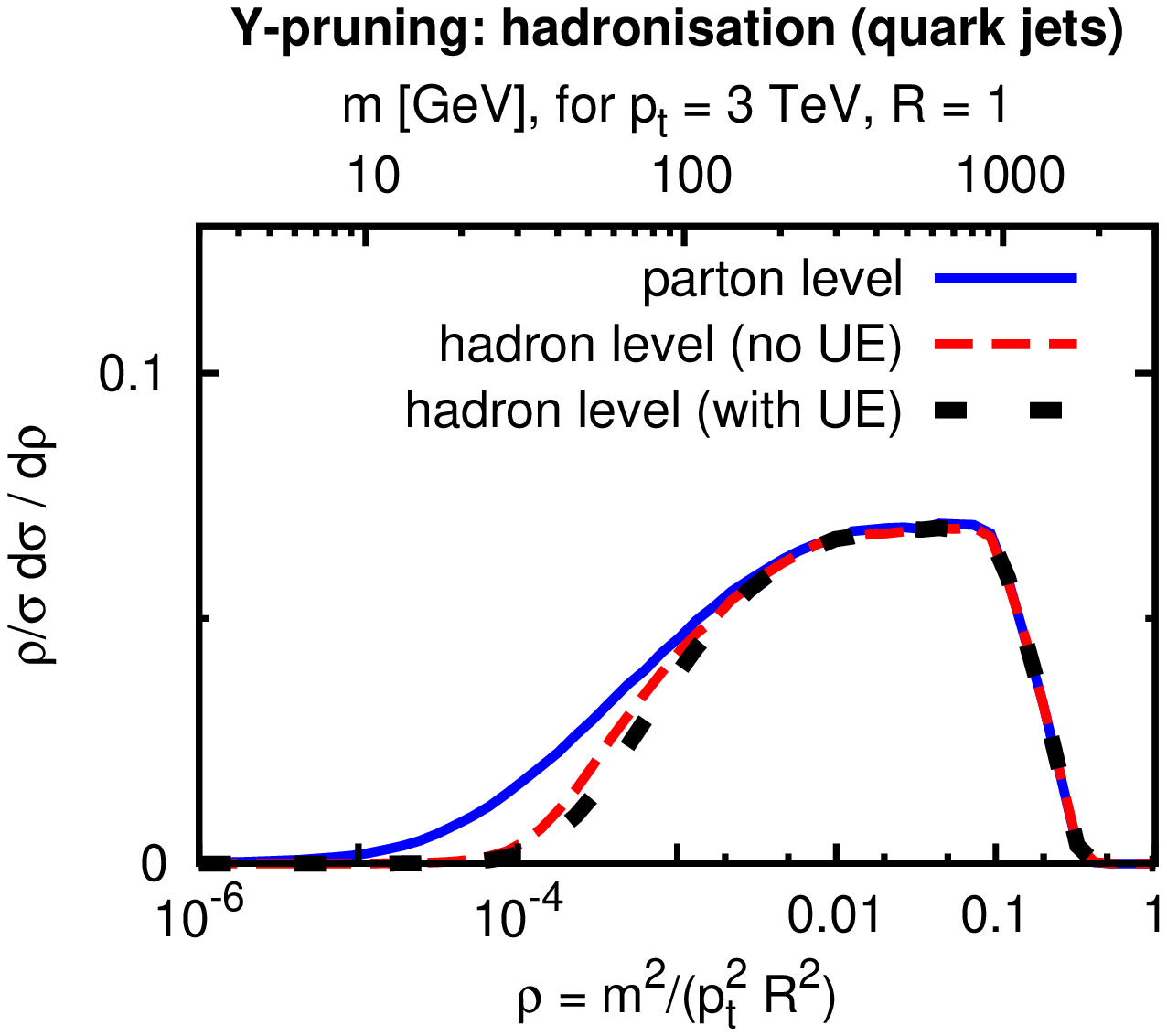}%
  \\[5pt]
  \includegraphics[width=0.49\textwidth]{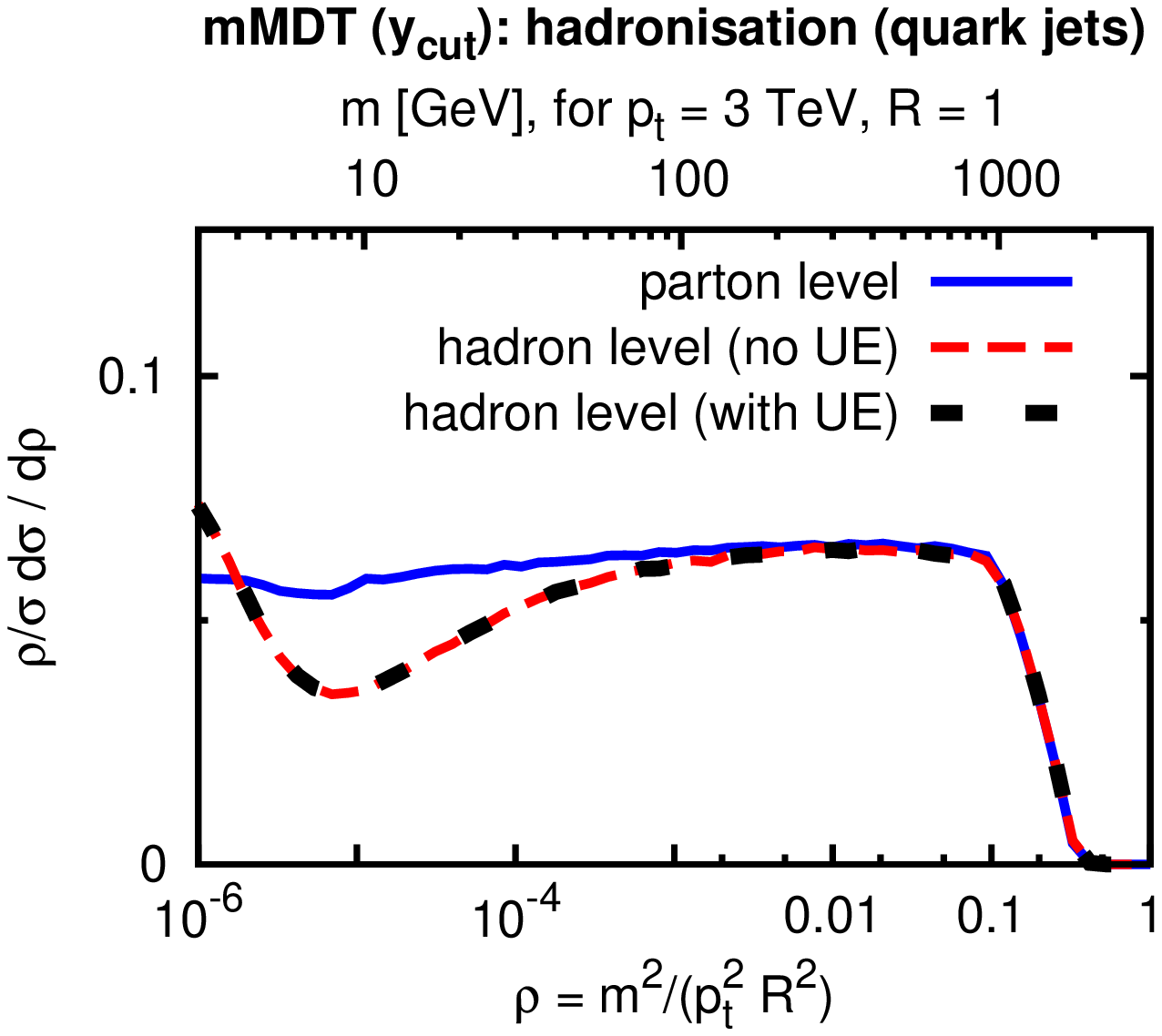}%
  \hfill
  \includegraphics[width=0.49\textwidth]{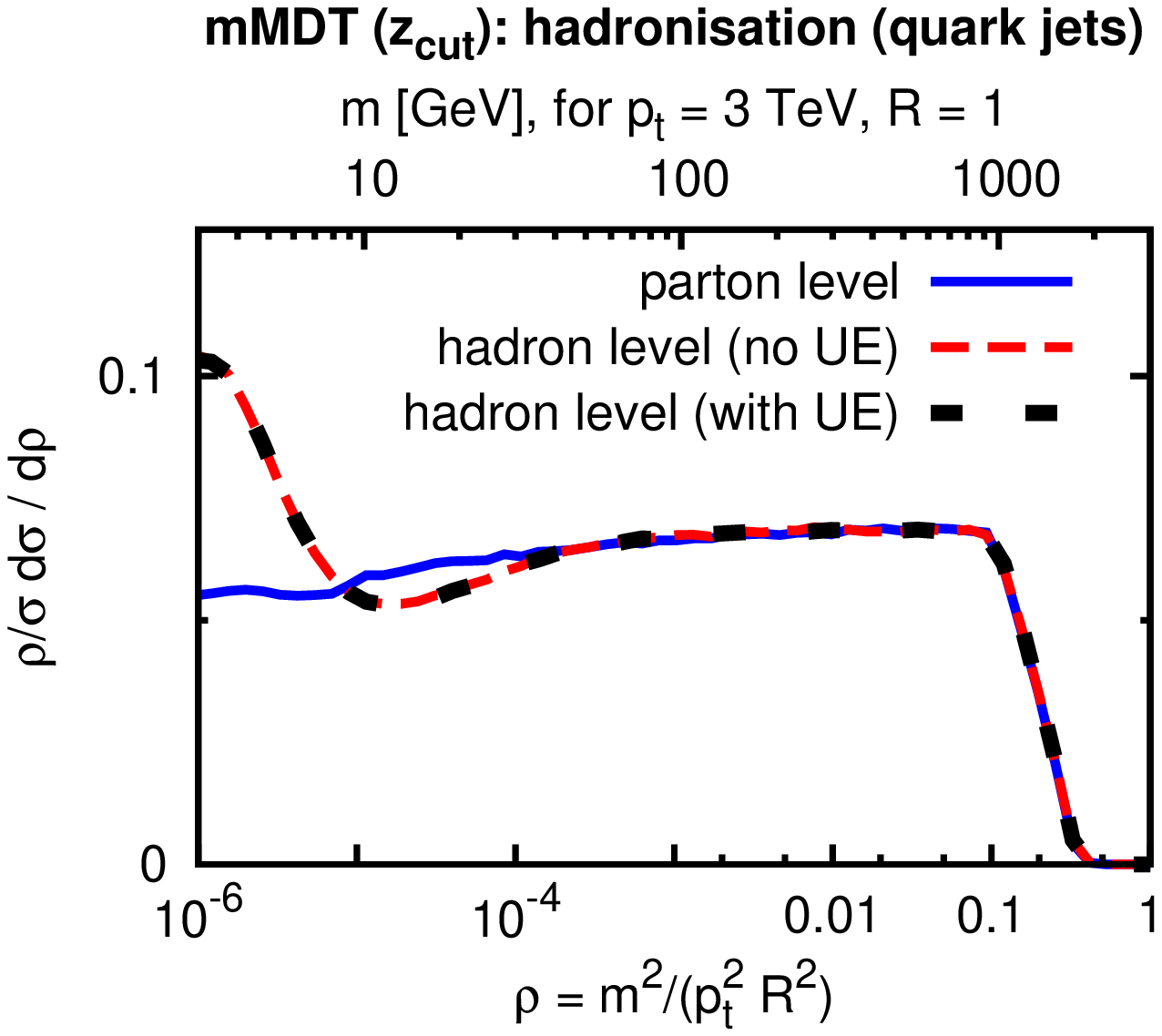}%
  \caption{The impact of hadronisation and the underlying event (UE)
    on the mass spectra for different taggers.
    The details of the MC event generation are as for
    Fig.~\ref{fig:tagged-mass-MC}. }   \label{fig:non-perturbative-comparisons}
\end{figure}

\begin{figure}[ht]
  \centering
  \includegraphics[width=0.49\textwidth]{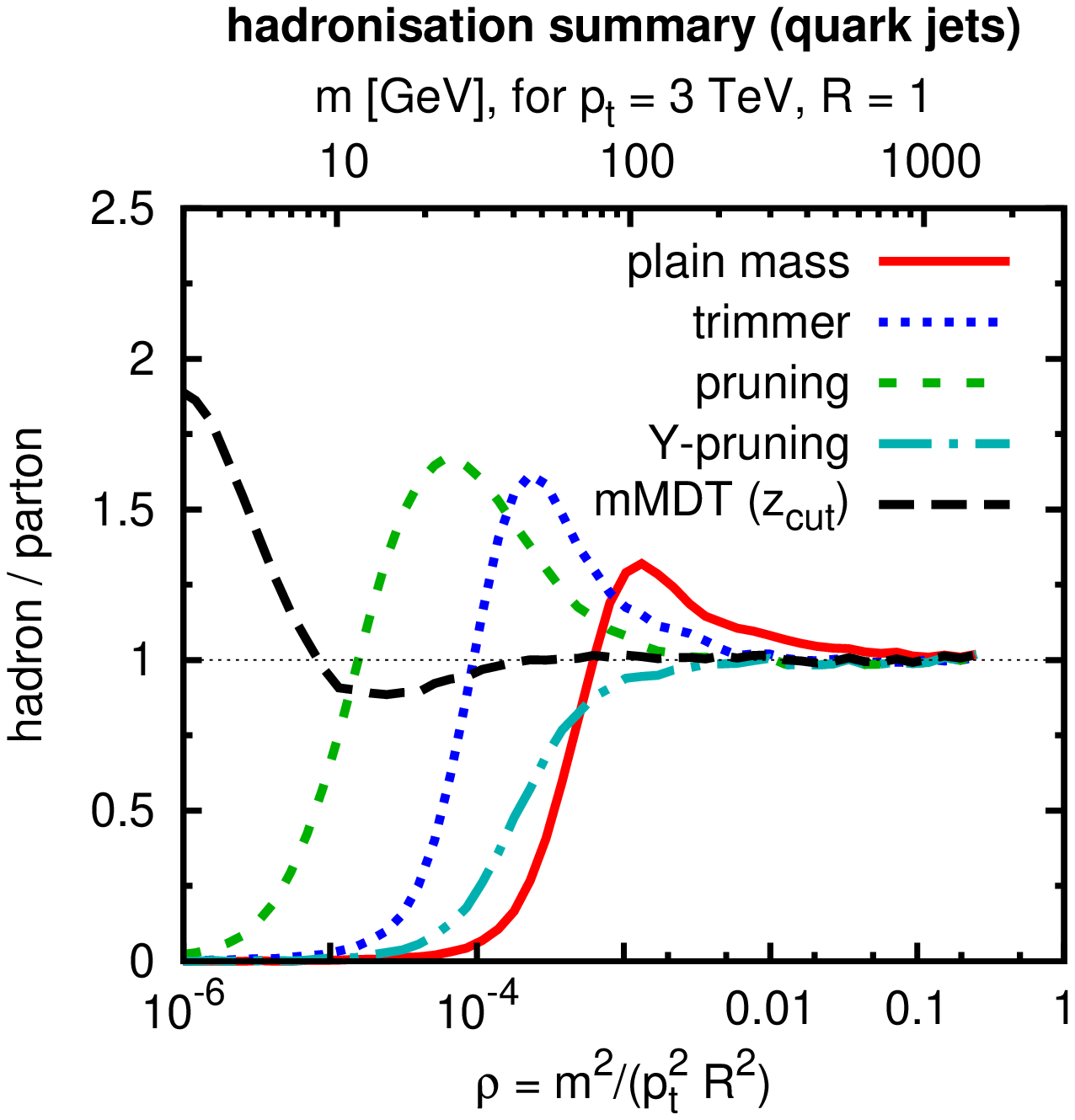}
  \hfill
    \includegraphics[width=0.49\textwidth]{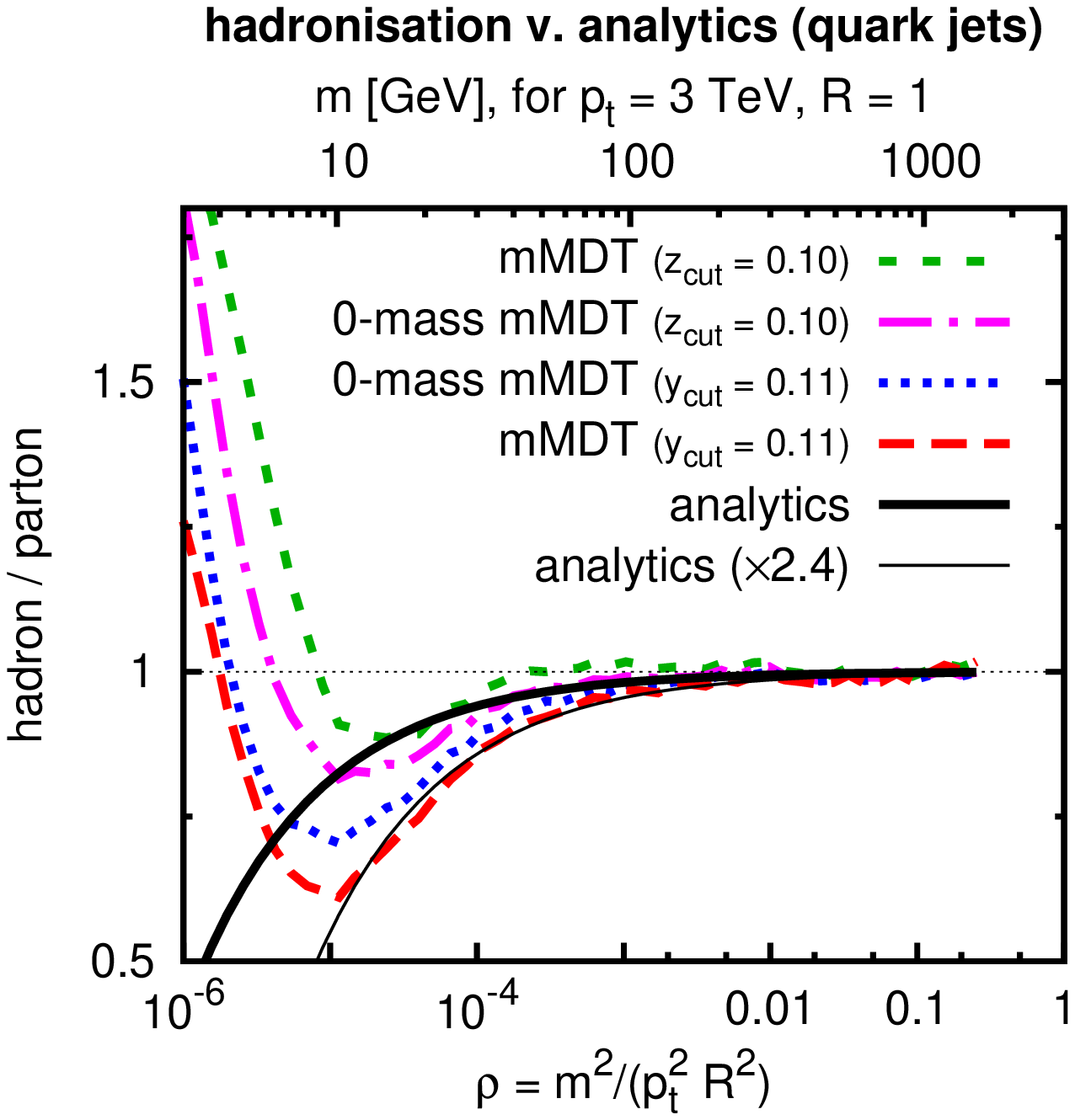}\\[5pt]
  \begin{minipage}{0.49\linewidth}
  \includegraphics[width=\textwidth]{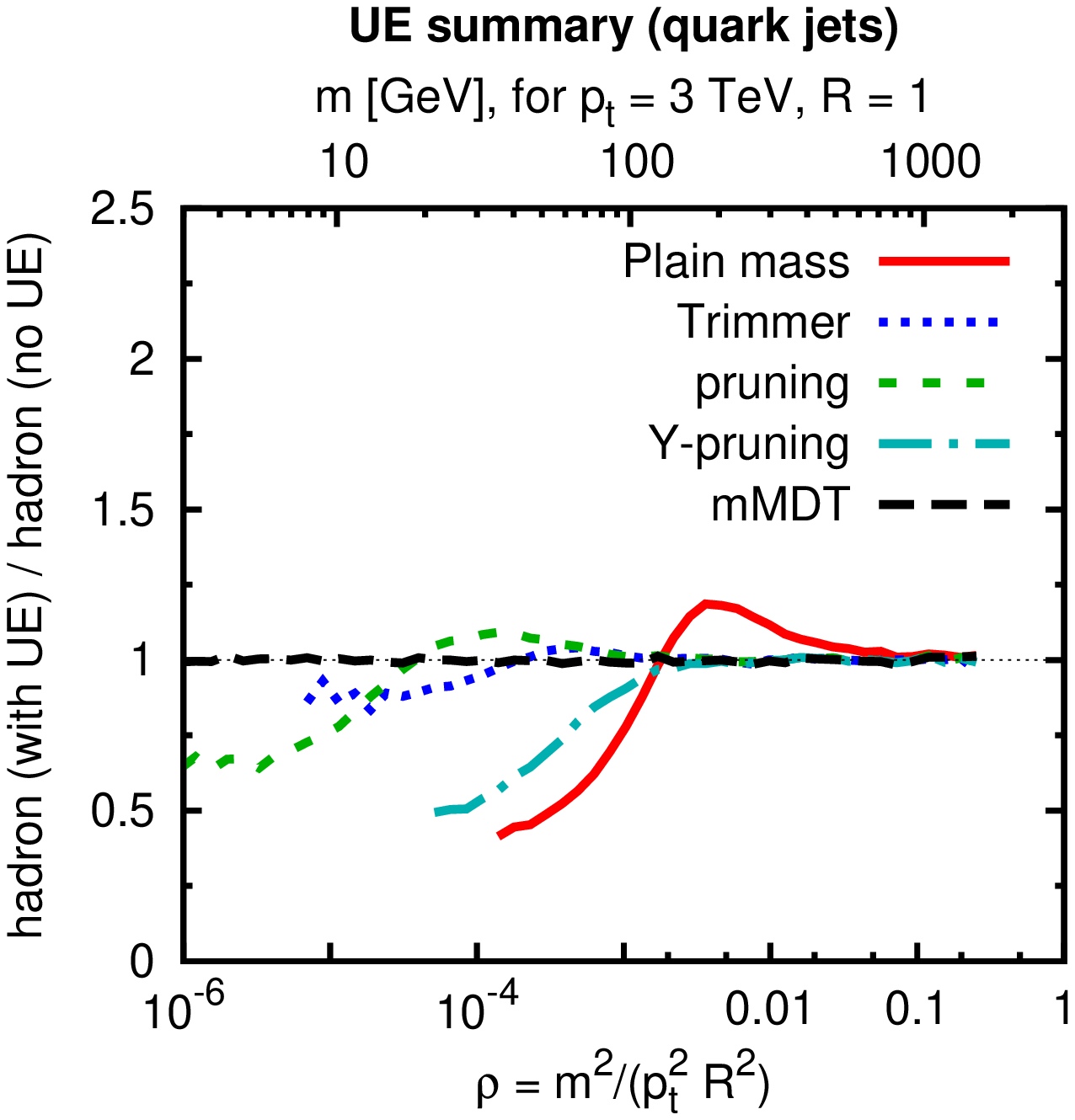}
  \end{minipage}  \hfill
  \begin{minipage}{0.49\linewidth} 
    \caption{Top left: ratio of hadron-level (without UE) to parton level
    results for various groomers and taggers. 
    Top right: ratio of hadron-level (without UE, both with finite and zero hadron masses) to parton level
    results for mMDT (with $\ycut$ and $\zcut$) and comparison to the
    analytic calculation of hadronisation corrections from
    section~\ref{sec:hadr-analytics}.
    Bottom: ratio of hadron
    levels with and without UE.
    The details of the MC event generation are as for
    Fig.~\ref{fig:tagged-mass-MC}.     
  } \label{fig:NP-ratios}
    \end{minipage}
\end{figure}

It is instructive to supplement the above discussion with Monte Carlo
studies of the effect of hadronisation.
Figure~\ref{fig:non-perturbative-comparisons} shows the mass
distributions at parton-level, hadron-level without underlying event
(UE) and hadron-level with UE, for plain jet mass, trimming, full and
$\sanepruning$ and mMDT using either a $\ycut$ or a $\zcut$.
Figure~\ref{fig:NP-ratios} shows the corresponding ratios of hadron
and parton-level distributions.

Let us first concentrate on the effect of hadronisation.
For any given mass, the plain jet mass is
the most strongly affected by hadronisation, with $25\%$ corrections
even for jet masses of $100\GeV$, in the neighbourhood of the peak
region. This scale is about twice that estimated as the limit of the
perturbative calculation in section~\ref{sec:PT-limit},\footnote{The belief that jet mass
  peaks are beyond perturbative control is widespread, though this
  statement usually holds for the peak of $d\sigma/dm$ or
  $d\sigma/dm^2$. Here we are instead considering $m^2d\sigma/dm^2$,
  whose peak is at much larger mass values. It is therefore somewhat
  surprising that there are still substantial effects.}
which itself was large because it scales as $\sqrt{p_t}$, as given in
Eq.~(\ref{eq:m2-NP-plain}).

We anticipated that trimming should only be affected by
non-perturbative physics at a somewhat smaller mass than for ungroomed
jets.
This is indeed what we see (most clearly in the top left panel of
Fig.~\ref{fig:NP-ratios}). 
Still, trimming's peak region is strongly affected, even more so than
for the plain jet mass, which is a consequence of the non-trivial
interplay between the change in perturbative peak position and the
change in non-perturbative effects as one goes from plain to trimmed
jet mass.

While pruning nominally has non-perturbative effects setting in at the
same mass as the plain jet mass, we argued that their onset might in
practice be somewhat different, as is indeed observed: it appears not
too dissimilar to trimming.
$\sanepruning$ looks somewhat different because it doesn't have a
Sudakov peak, however from Fig.~\ref{fig:NP-ratios} it is clear that
the order of magnitude of hadronisation effects is similar in full pruning and
$\sanepruning$.

As expected, it is the mMDT that has the smallest hadronisation
corrections, with non-trivial structure appearing at about $10\GeV$,
about three times the scale estimated in section~\ref{sec:PT-limit}
for the limit of the perturbative calculation.
The impact of hadronisation for mMDT depends somewhat on whether it is
used with a $\ycut$ or $\zcut$, and for the latter in particular
hadronisation remains very modest all the way down to $10\GeV$.

\subsubsection{Analytic hadronisation estimate for mMDT}
\label{sec:hadr-analytics}
It is worthwhile examining whether the form of the onset of
hadronisation for mMDT, above $10\GeV$, can be explained at least
qualitatively.
Multiple effects can play a role: for example, hadronisation was
argued in~\cite{Dasgupta:2007wa} to shift a given jet's squared mass by
an amount $\delta m^2 \simeq C\, \Lambda_\text{NP}\, p_t R$, where $C$ is
either $C_F$ or $C_A$ and $\Lambda_\text{NP} \sim 0.4\GeV$.
Hadronisation is also believed to change a jet's (or a prong's)
momentum, shifting it by an amount $\delta p_t \simeq - C
\Lambda_\text{NP}/R$~\cite{Korchemsky:1994is,Dasgupta:2007wa}.
(The numbers are given here for the anti-$k_t$ algorithm with $R \ll
1$ and in the
case of the jet mass assume a scheme in which hadron masses are
neglected; the $p_t$ shift result for the $k_t$ algorithm is given in
Ref.~\cite{Dasgupta:2009tm}; the other cases, including for the C/A
algorithm, have yet to be calculated).

For a tagger one needs to work out the interplay between hadronisation
and the tagging procedure. 
For example, let us consider the shift in jet mass, in the case of a
quark jet.\footnote{We are grateful to Jesse Thaler for useful
  discussions on this point.}
The action of the tagger is such that the average effective radius of
a tagged jet is a function of the tagged jet mass itself, $R_\text{eff} \sim
f(\ycut) m/p_t$, where 
\begin{equation}
  \label{eq:1}
  f(\ycut) = \frac{\int_{\ycut}^{1-\ycut} dz \, p_{gq}(z) \left[z(1-z)\right]^{-\frac12}}{ \left.
      \int_{\ycut}^{1-\ycut} dz\, p_{gq}(z) \right.}\, ,
\end{equation}
for quark-initiated jets. For $\ycut
\simeq 0.1$, $f(\ycut) \simeq 2.5$. 
Thus we obtain
\begin{equation}
  \label{eq:deltam2=NP}
  \delta m^2 \simeq C_F \, f(\ycut)\, \Lambda_\text{NP} \, m\, 
  \qquad 
  \longrightarrow
  \qquad
  \delta m \simeq \frac12\, C_F \, f(\ycut)\, \Lambda_\text{NP}\,.
\end{equation}
For cases where $d\sigma/dm$ scales as $1/m$, this leads to a correction
\begin{equation}
  \label{eq:NP-correction-factor-mass-shift}
  \frac{d\sigma}{dm}^{\NP} = \frac{d\sigma}{dm}^{\PT} \left(1 +
    \frac12\, C_F \, f(\ycut)\, \frac{\Lambda_\text{NP}}{m} \right)\,.
\end{equation}

Next, let us consider the effect of the $p_t$ shift. 
This is most relevant in cases where one of the prongs, at parton
level, has a momentum such that it just passes the $\ycut$ asymmetry
requirement.
After hadronisation its $p_t$ is reduced, and so it may no longer pass
that requirement.
That leads to a drop in efficiency, which can be evaluated as follows.
The effect will be relevant for asymmetric splittings, where the
softer prong's momentum fraction is $z \sim \ycut$.
The effective jet radius will be of order $\frac{m}{p_t}
\ycut^{-\frac12}$, and so the absolute change in the prong's $p_t$
will be $-C_A \Lambda_\NP  \ycut^{1/2}\frac{p_t}{m}$.
This leads to a change in the momentum fraction (relative to original
jet) for the softer prong
of $-C_A \frac{\Lambda_\NP}{m} \ycut^{1/2}$.
Note the $C_A$ colour factor here, since the soft prong will almost
always be a gluon.
Given that the perturbative tagging efficiency is equal to the
integral over the splitting function down to momentum fractions
$\simeq \ycut$, the non-perturbative correction can be evaluated by
estimating how the integral changes when requiring a momentum fraction
greater than $\ycut + C_A \frac{\Lambda_\NP}{m} \ycut^{1/2}$.
This gives us
\begin{subequations}
  \label{eq:NP-correction-factor-pt-shift}
  \begin{align}
    \frac{d\sigma^{\NP}}{dm} &\simeq \frac{d\sigma^{\PT}}{dm} \;
    \frac{\ln \left(\ycut + C_A \frac{\Lambda_\NP}{m}
        \ycut^{1/2}\right)+ \frac34}{\ln \ycut + \frac34}\,,
    \\
    &\simeq \frac{d\sigma^{\PT}}{dm} \; \left(1 - C_A \frac{
        \ycut^{-1/2}}{\ln 1/\ycut - \frac34} \, \frac{\Lambda_\NP}{m}
    \right).
  \end{align}
\end{subequations}
One element that we have neglected here is that if hadronisation
causes a (sub)jet with mass $m_1$ to fail the $\ycut$ (or $\zcut$) requirement,
then mMDT continues to recurse into the harder prong. This
will populate the lower mass region and the jet might then tagged as
having mass $m_2 \ll m_1$.
The contribution from this effect to masses of order $m_2$ will be
proportional to $\as \Lambda_\NP/m_1$, whereas the direct correction
to masses of order $m_2$ will be proportional to $\Lambda_\NP/m_2$,
which is parametrically larger.

The dependence of the hadronisation correction on $m$ is identical in
Eqs.~(\ref{eq:NP-correction-factor-mass-shift}) and
(\ref{eq:NP-correction-factor-pt-shift}), with only the coefficient
changing.
Interestingly the corrections depend just on the jet mass, and not on
the jet $p_t$; this is characteristically different from the situation
for plain jet mass.

Numerically it is the negative contribution from the $p_t$ shift that
dominates over the mass shift.
Considerable caution is needed, however, as concerns the actual
numerical prediction from these formulae: we have ignored hadron-mass
effects, which are known to be
substantial~\cite{Salam:2001bd,Mateu:2012nk}; we have ignored the
(complicated) issue that the two-pronged structure of the jet will
undoubtedly modify the pattern of hadronisation corrections relative
to the calculations of \cite{Dasgupta:2007wa}, both for
the overall jet mass and the prong transverse momentum; we have also
ignored the differences between mMDT with a $\ycut$ and a $\zcut$,
even though we have seen that they have different non-perturbative
effects, possibly because $y$'s definition involves the jet mass,
which is itself subject to further corrections.
Accordingly, it is probably only the overall $\Lambda_\NP/m$ scaling
in Eqs.~(\ref{eq:NP-correction-factor-mass-shift}) and
(\ref{eq:NP-correction-factor-pt-shift}) that can be considered robust.

Despite these caveats, it is still interesting to compare the result
of Eqs.~(\ref{eq:NP-correction-factor-mass-shift}) and
(\ref{eq:NP-correction-factor-pt-shift}) to the Monte Carlo results. 
This is done in the top-right plot of Fig.~\ref{fig:NP-ratios}.
The plot shows the Monte Carlo results for both the $\ycut$- and
$\zcut$-based mMDT.
For the results labelled ``0-mass,'' all particles' 4-momenta have
been transformed (before clustering) so as to have zero mass, while
maintaining their $p_t$, rapidity and azimuth.
The figure also shows our analytical result, as well as a variant
where the hadronisation corrections have been rescaled by an
(arbitrarily chosen) factor of $2.4$.
All the Monte Carlo results seem to be roughly consistent with our
predicted $\Lambda_\NP/m$ scaling down to $\order{10\GeV}$.
However the normalisation of the hadronisation correction appears to
be very sensitive to the details of the tagger and the input
particles. 
The version of mMDT formulated in terms of a $\zcut$ and with massless
input particles appears to agree reasonably well with our prediction.
This may just be a coincidence, though it is also true that this is
the variant for which our estimates above were most likely to be
reasonable. 

A final comment concerns the absolute size of the hadronisation
corrections for the $\zcut$-based mMDT variants: in the region of
phenomenological interest, it seems that hadronisation is just a
couple of percent.
This suggests that these mMDT variants may be optimally suited to
high-precision studies, both in new physics searches, and possibly
also even applications such measurements of the strong coupling.

\subsubsection{Underlying event}
\label{sec:UE}

A discussion of non-perturbative effects would not be complete without
considering the underlying event (UE), whose impact for each tagger
can be seen in Fig.~\ref{fig:non-perturbative-comparisons}, with a
summary in the bottom plot of Fig.~\ref{fig:NP-ratios}. 
The jet mass is the most strongly affected, while all the
groomed/tagged results show a significantly reduced UE sensitivity,
which was part of the intention in their design.
For trimming and pruning this sensitivity remains genuinely small
throughout the phenomenologically relevant region, and in particular
significantly smaller than the hadronisation corrections.
For mMDT the dependence on UE is almost imperceptible, at or below the
$1\%$ level for all jet masses.

For $\sanepruning$ the UE sensitivity is not negligible: this is because
the UE can significantly increase the original jet's mass and the
resulting pruning radius.  
Consequently, a jet that was classified as $\sane$-pruned without UE, may
be reclassified as $\anomalous$-pruned.
The overall pruning rate increases slightly (because for $\anomalous$-pruned jets the $z$-cut is turned off), while the $\sane$-pruned rate is
noticeably decreased.
This sensitivity to UE is perhaps the one main disadvantage of
$\sanepruning$, and is, we believe, inherent to any approach that
effectively relies on the original jet mass to help discriminate
between colour singlet signals and colour triplet/octet backgrounds.

One should be aware that the above pattern of UE dependence does
depend on the jet transverse momentum.
For example, mMDT was originally designed in conjunction with
filtering in order to reduce the effect of UE. 
This appears not to be necessary here, but had we considered jets with
transverse momentum of a couple of hundred GeV, as was the context for
the original MDT+filtering study, then the much larger effective
radius for the tagged jet would have led to noticeable UE effects in
the absence of filtering.

\subsection{Choice of Monte Carlo Generator}
\label{sec:choice-monte-carlo}

\begin{figure}[ht]
  \centering
  \includegraphics[width=0.49\textwidth]{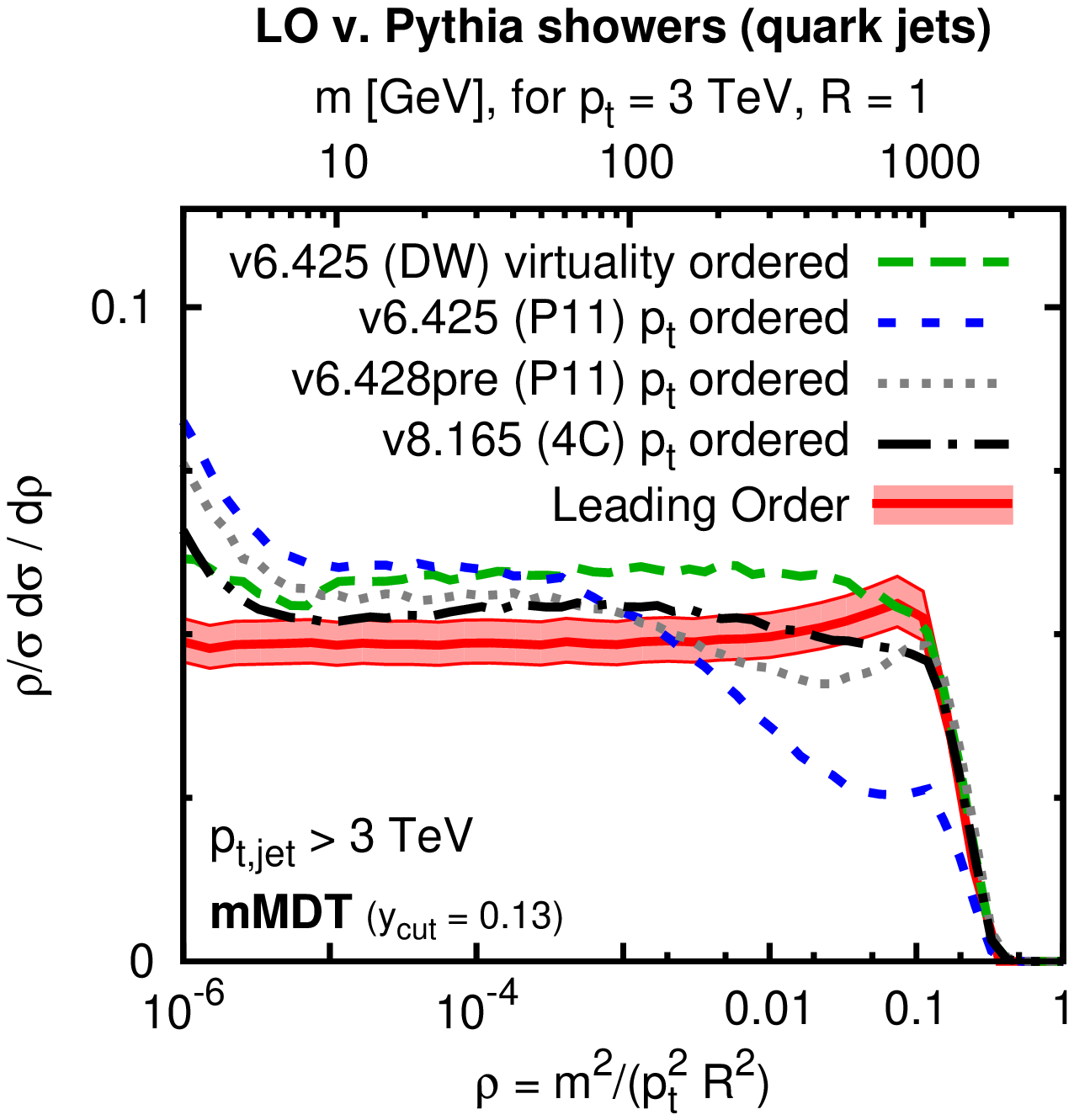}
  \hfill
  \includegraphics[width=0.49\textwidth]{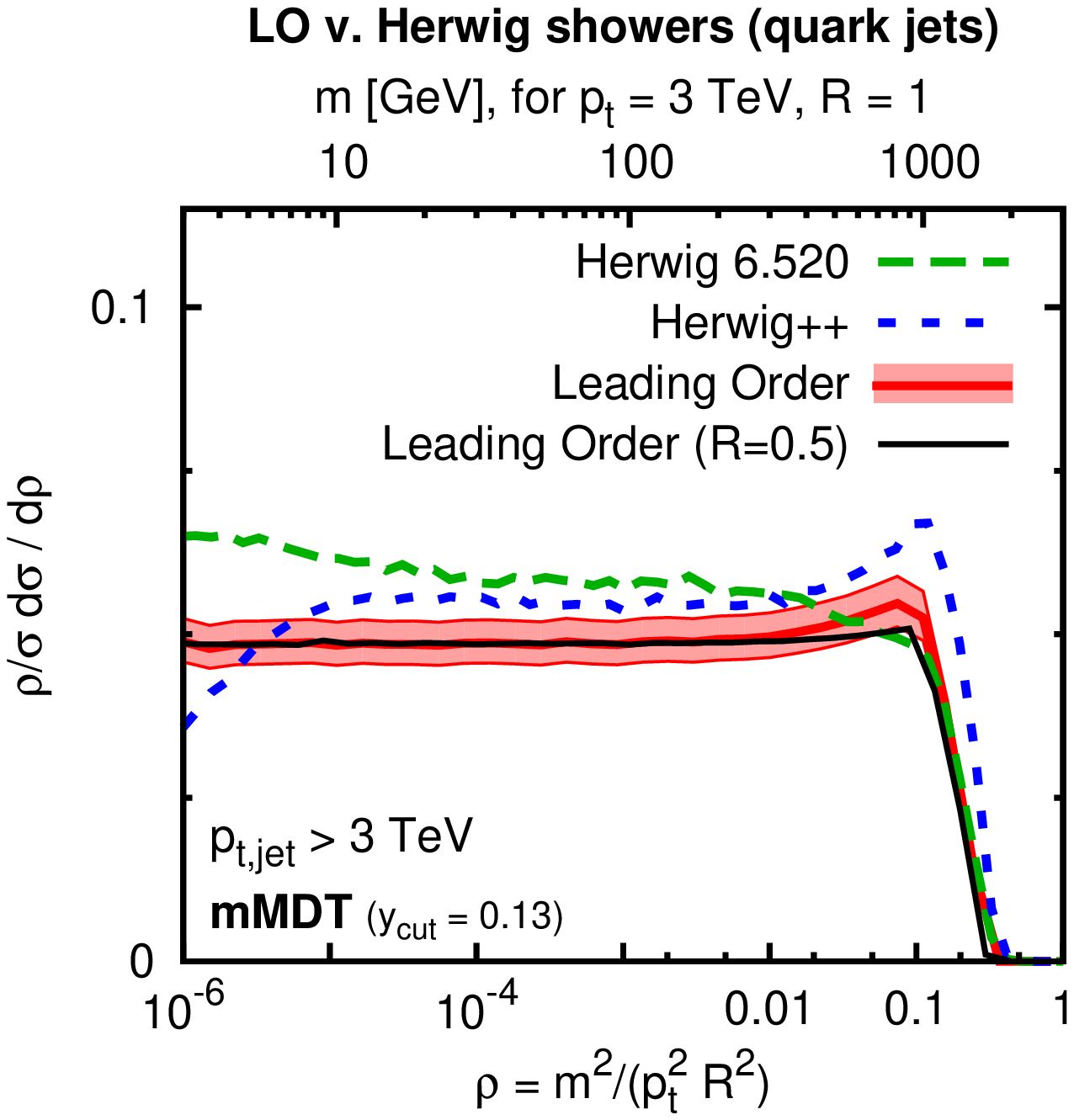}\\[5pt]
  \begin{minipage}{0.49\linewidth}
    \includegraphics[width=\textwidth]{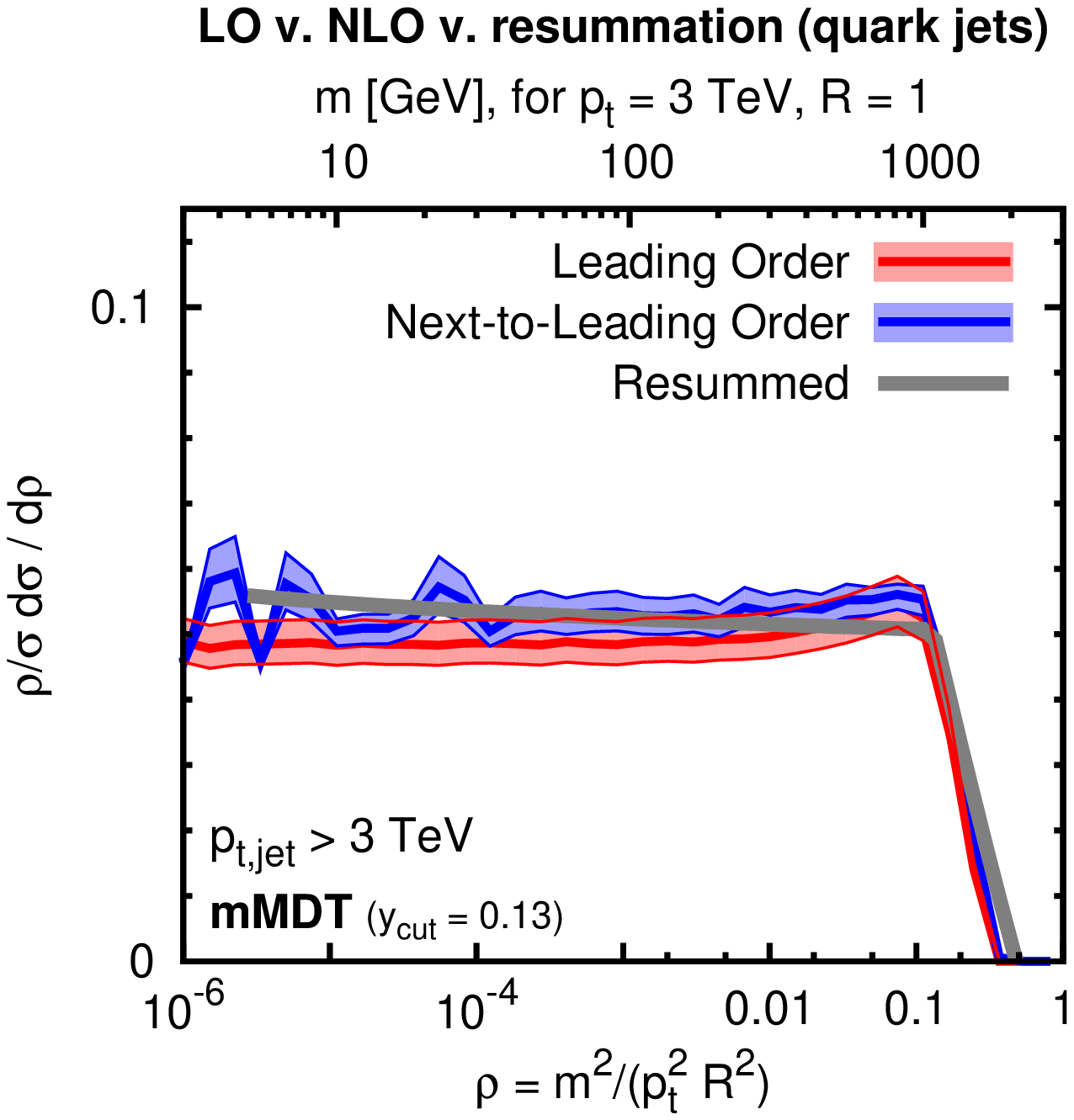} 
  \end{minipage}  \hfill
  \begin{minipage}{0.49\linewidth} 
    \caption{Mass distributions for mMDT tagged jets, comparing
      different parton-shower generators, the resummation and the exact
      leading-order and next-to-leading order
      results (obtained with NLOJet++~\cite{Nagy:2003tz} with the
      MSTW2008 NLO PDF set~\cite{Martin:2009iq}), with a central
      scale choice $\mu_R = \mu_F = p_t$ and simultaneous scale
      variations by a factor of two.
      Unless otherwise specified the curves correspond to a jet radius
      of $R=1$.
      The Monte Carlo results have been obtained at parton level, with
      the underlying event turned off. See text for further details.
    }
    \label{fig:MC-comparisons}
  \end{minipage}
\end{figure}

Throughout this work, we have regularly compared our analytical
results for the tagged mass distributions with the output of Monte
Carlo parton shower simulations from Pythia 6.425~\cite{Pythia6},
with the DW tune~\cite{DW} of its virtuality-ordered shower.
We have generally found good agreement between our analytics and
the Pythia parton-shower simulations.
It is also of interest to check whether the agreement is equally good
when using different parton showers.
To do so, we concentrate on the mMDT
mass distribution, in the case of quark-initiated jets, for
$\ycut=0.13$, at the parton level.

The top-left plot of figure~\ref{fig:MC-comparisons} shows the
comparison between the different showers in Pythia~6 and Pythia~8: the
virtuality-ordered one in Pythia~6, our default, and the $p_t$-ordered
one in Pythia~6~\cite{PythiaPtOrdered} (in the Perugia~2011~\cite{Skands:2010ak}
tune) and the $p_t$-ordered shower from Pythia~8~\cite{Pythia8} (in the 4C
tune~\cite{Corke:2010yf}). 
The top-right plot
shows the mMDT mass distribution obtained with
the angular-ordered showers from Herwig~6.520~\cite{Corcella:2002jc}
and Herwig++~2.6.3~\cite{Herwig++,Arnold:2012fq,Lonnblad:1998cq} in their default
tunes. 
The Monte Carlo curves are obtained with a generation cut of
$p_t>2.2\TeV$ applied to the $qq\to qq$ hard process, and the tagging
analysis is then carried out on all jets with
$p_t>3\TeV$.\footnote{While it is clear that having one generator cut
  and a higher subsequent jet selection cut is the correct thing to
  do, it is also computationally more expensive.
  In all the other plots of this paper, we have simply used a
  generator cut of $3\TeV$, and always examined the two leading jets.
  We have verified that these two procedures give essentially
  identical results, both for Pythia~6.4's virtuality ordered shower
  and for Herwig~6.520.
  In contrast, for the $p_t$ ordered showers in Pythia~6.4 and
  Pythia~8, the two procedures give visibly different results, and it
  is mandatory to use the procedure with staggered generation and
  selection cuts.}
All plots include the full leading order
(LO) result obtained with the program NLOJet++~\cite{Nagy:2003tz}.
The fixed-order calculation is important in that it enables us to
check the distributions for large masses, where resummation may not be
appropriate.
We ensure a high purity of quark-initiated jets in the fixed-order
calculations by setting the incoming gluon parton distribution
functions to zero.

The plots in the top row figure~\ref{fig:MC-comparisons} show that
nearly all the Monte Carlo generators are in reasonable agreement with
each other, with our resummation and with the LO calculation.
The one exception is the $p_t$-ordered shower in Pythia~6.245, which
predicts a noticeably different shape for the distribution, both at
small and large masses.
We have checked that this characteristic holds also in another
widespread tune of the $p_t$-ordered shower, Z2~\cite{Field:2010bc}.
This significant difference relative to our calculations and the other
generators appears to be limited to situations where the jet
transverse momenta are close to the kinematic limit. 
We have checked that similar differences appear also for the other
substructure tools considered in this paper.
Following discussions with the authors of Pythia, they provided us
with code for a modified version of the $p_t$-ordered shower, which
resolves an issue in which the hardness of the final-state shower could be
affected by the presence (or not) of soft initial-state emissions.
Results with this modified shower are shown in
Fig.~\ref{fig:MC-comparisons} (top-left) as a dotted curve, labelled
v6.428pre, and one observes a clear improvement in the agreement with
other tools. 
This example illustrates the value of analytical understanding in
situations such as this where Monte Carlo results from various
generators differ noticeably.

We note that the LO curve exhibits non-trivial structure
(a small bump) in the vicinity of $\rho=0.1$. 
This structure is absent in most of the Monte Carlo results, as well as
in the results obtained from our analytical calculation (it is however
present for Herwig++, and somewhat stronger than in the LO result).
We believe that it is driven by the precise structure of hard
large-angle radiation: this can be thought of as having a significant
hard initial-state radiation contribution, neglected in our calculations
and only approximately present in the parton showers.
To confirm this hypothesis we also show the LO calculation for a jet
of radius $R=0.5$ (left-hand plot), which should reduce the
initial-state radiation contribution.
Indeed, the structure at $\rho\simeq 0.1$ is much less pronounced.
We expect that if we had carried out simulations with tools such
as MC@NLO~\cite{Frixione:2002ik} or POWHEG~\cite{Nason:2004rx} (or
alternatively CKKW~\cite{Catani:2001cc} or MLM~\cite{Alwall:2007fs}
matching), these would have correctly accounted for this type of
large-mass structure, without significantly modifying the results at
lower $\rho$.
It would be interesting to verify this expectation, however such a
study is beyond the scope of this work.

Finally, the bottom-left plot shows our resummed prediction and the
NLO result.  
As discussed in section~\ref{sec:background}, the choice $\ycut=0.13$
minimises higher-order corrections and hence the all-order result is
dominated by the LO contribution, even at relatively small masses.
This property is confirmed in the NLO calculation, whose central value
is just within the scale uncertainty band of the LO
calculation.\footnote{Scale uncertainties have been obtained through
  simultaneous variation of renormalisation and factorisation scales
  by a factor of two around a central value taken equal to the $p_t$
  of the leading jet.
  The scales are kept identical in the (3-jet@NLO) differential mMDT
  cross section and in the (2-jet@NLO) normalisation cross section. 
  Note the following caveat when varying factorisation scales: the
  variation of the quark densities is a function also of the gluon
  densities, however the matrix elements involving incoming gluons are
  all discarded, in order to obtain mainly quark jets; therefore
  factorisation scale dependence is not expected to cancel exactly at
  NLO, in contrast with the situation for a normal NLO calculation. }
%

\subsection{Effect of the taggers on signal--background discrimination}\label{sec:signal}

We have so far considered only the question of how the various
taggers/groomers behave for backgrounds, i.e.\ quark or gluon-induced
jets.
A key question for evaluating the performance of taggers is also that
of how they fare on signal jets, for example $W$, $Z$ or Higgs-bosons.
The basic, known tree-level result, is that for the decay of a scalar
particle, the tagging efficiency of a tagger like mMDT is essentially
\begin{equation}
  \label{eq:signal-efficiency}
  \epsilon_S = \int_{\zcut}^{1-\zcut} dz \; P_{H \to q\bar q}(z) = 1 - 2\zcut\,,
\end{equation}
where the results makes use of the fact that $P_{H \to q\bar q}(z) =
1$. As usual, in the small $\zcut$ limit, $\ycut$ and $\zcut$ are
interchangeable.
The same result holds for pruning (original and $\sanepruning$), modulo
corrections associated with initial-state radiation (ISR).
For trimming, the result depends on $m/p_t$, and is $1-2\zcut$ for
$\rho > \zcut r^2$ and tends to $1$ for asymptotically smaller $m/p_t$
(again, modulo corrections from ISR). 
Of course, the tagging always needs to be performed in a given mass
window, and these estimates assume that the mass window is
sufficiently wide relative to any loss of mass resolution caused by
ISR, UE and pileup (the width was studied in detail for MDT with
filtering by Rubin in Ref.~\cite{Rubin:2010fc}).

\begin{figure}[t]
  \centering
  \begin{minipage}{0.49\linewidth}
    \includegraphics[width=\textwidth]{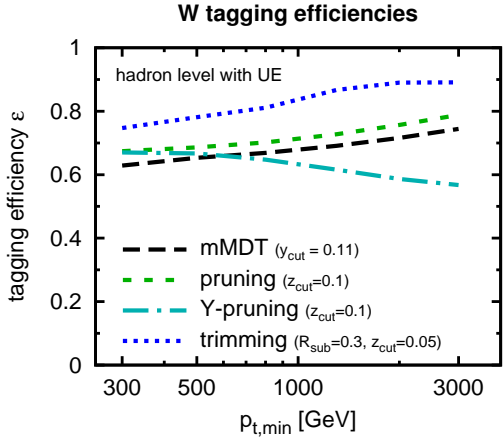}
  \end{minipage}\hfill
  \begin{minipage}{0.49\linewidth}
    \caption{%
      Efficiencies for tagging
      hadronically-decaying $W$'s, for a range of taggers/groomers,
      shown as a function of the $W$ transverse
      momentum generation cut in the Monte Carlo samples (Pythia~6, DW
      tune).
      Further details are given in the text.  
      \label{fig:SB-signal-eff}
    }
  \end{minipage}
\end{figure}

Fig.~\ref{fig:SB-signal-eff} shows tagging efficiencies obtained with
Pythia~6 (DW tune) at hadron level (with UE).
They have been obtained in $WZ$ events, with the $Z$ decaying
leptonically and the $W$ hadronically.
The tagger is applied to the hardest jet in the event, which is deemed
tagged if its final mass is in the window $64$--$96\GeV$.
The fraction of jets that were tagged is shown as a function of a
minimum $p_t$ cut applied on the $q\bar q \to W Z$ hard event in the
simulation.
As expected, the tagging efficiencies are fairly independent of the
$p_{t,\min}$ choice, and reasonably consistent with the $1 - 2\zcut$
expectation. 
The differences that one sees relative to that expectation have two
main origins. 
Firstly Eq.~(\ref{eq:signal-efficiency}) holds at tree-level. 
It receives $\order{\as}$ corrections from gluon radiation off the $W
\to q\bar q'$ system.  
Monte Carlo simulation suggests these effects are responsible,
roughly, for a $10\%$ reduction in the tagging efficiencies.
Secondly, Eq.~(\ref{eq:signal-efficiency}) was for unpolarized
decays. 
By studying leptonic decays of the $W$ in the $pp \to WZ$ process, one
finds that the degree of polarization is $p_t$ dependent, and the
expected tree-level tagging-efficiency ranges from about $76\%$ at low
$p_t$ to $84\%$ at high $p_t$.
These two effects explain the bulk of the modest differences between
Fig.~\ref{fig:SB-signal-eff} and the result of
Eq.~(\ref{eq:signal-efficiency}). 
However, the main conclusion that one draws
from Fig.~\ref{fig:SB-signal-eff} is that the ultimate performance of
the different taggers will be driven by their effect on the background
rather than by the fine details of their interplay with signal events.
This provides an a posteriori justification of our choice to
concentrate our study on background jets.

\begin{figure}[h]
  \centering
  \includegraphics[width=0.49\textwidth]{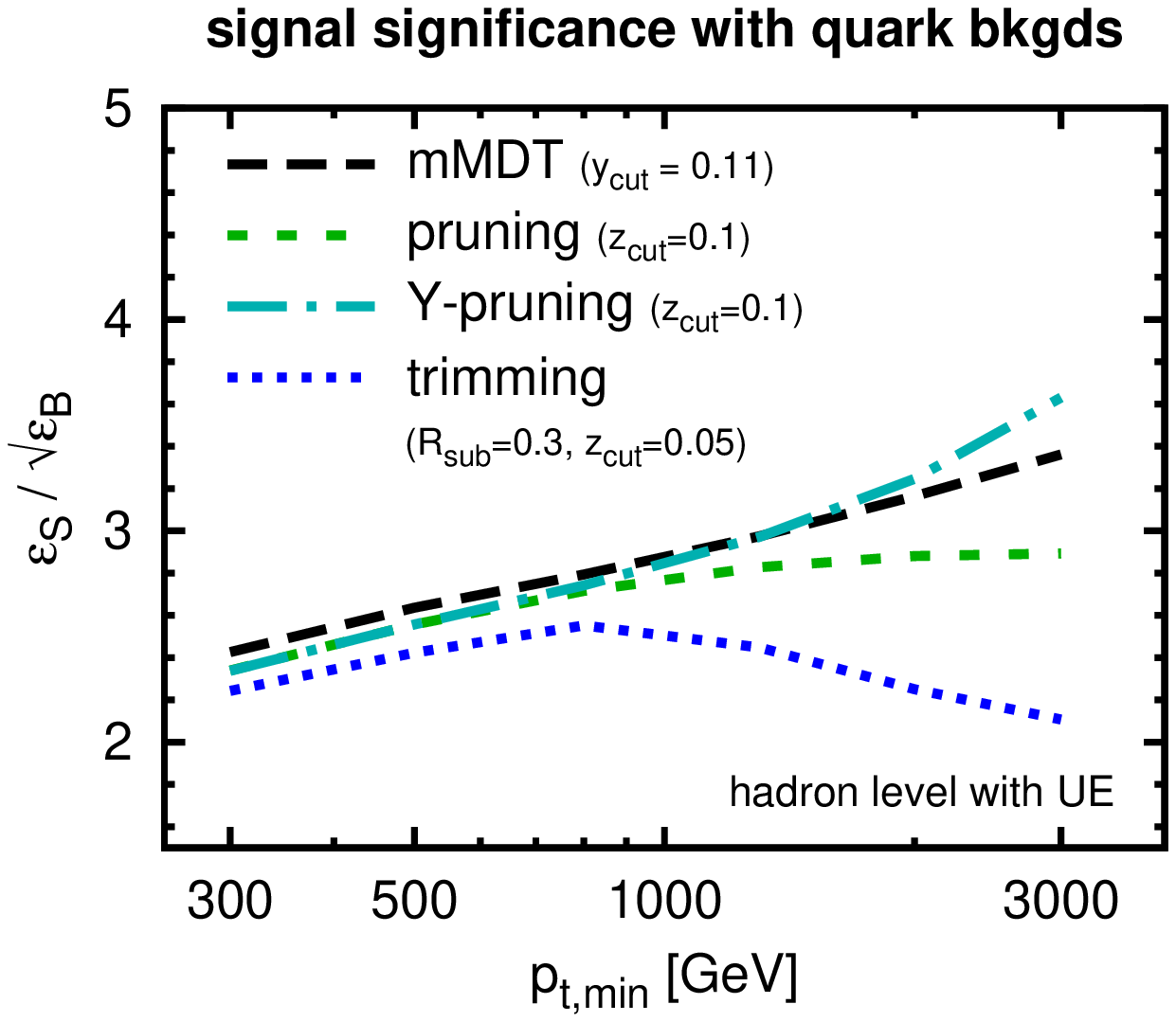}\hfill
  \includegraphics[width=0.49\textwidth]{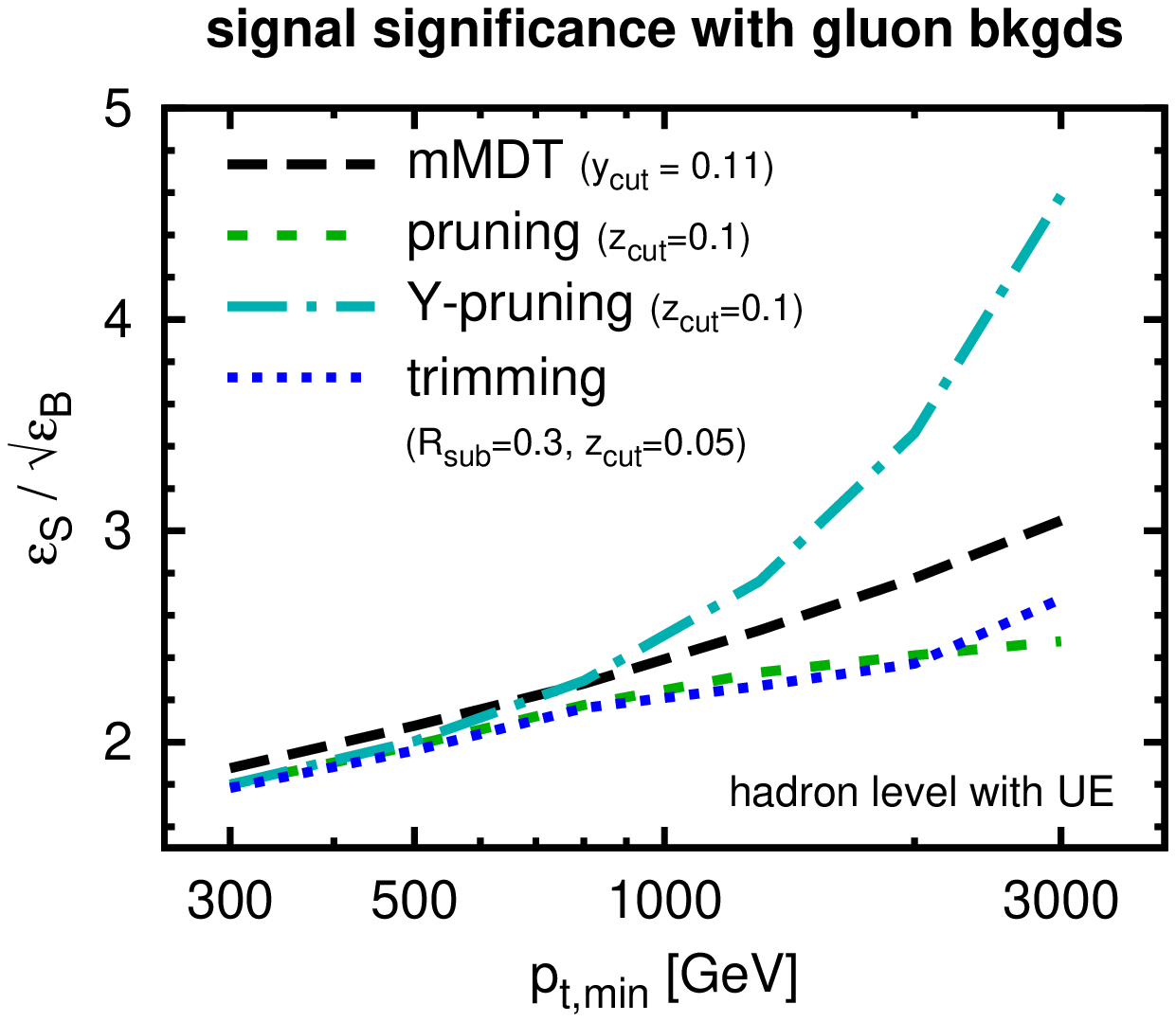}\\[5pt]
  \caption{%
    The significance obtained for tagging signal ($W$'s) versus
    background, defined as $\epsilon_S/\sqrt{\epsilon_B}$, for a range
    of taggers/groomers, shown
    as a function of the transverse
    momentum generation cut in the Monte Carlo samples (Pythia~6, DW
    tune)
    Further details are given in the text.  }
  \label{fig:SB-summary}
\end{figure}

Figure~\ref{fig:SB-summary} shows the overall performance of the
different taggers quantified as $S = \epsilon_S / \sqrt{\epsilon_B}$, which
is proportional to the signal significance that can be obtained with a
given tagger.
Here $\epsilon_B$ is the fraction of quark (left plot) or gluon (right
plot) jets that are tagged and pass the mass cut.

Let us start by discussing mMDT. 
Its signal significance $S$ grows with $p_t$. 
This is driven by three modest effects combining together: the signal
efficiency increases at high $p_t$; the background tagging rate is, in
a first approximation, proportional to $\as(p_t)$, which decreases at
high $p_t$; 
and for our choice $\ycut = 0.11$, the tagging rate decreases slightly
for decreasing $m/p_t$ (cf.\ Fig.~\ref{fig:mMDT-multiple-ycut}).
The signal significance is lower for gluon backgrounds than for
quark backgrounds, which is simply a consequence of the $C_A$ v.\
$C_F$ colour factor in the leading-order background tagging rate. 
This is partially compensated for at high $p_t$ by the steeper $m/p_t$
dependence in the gluon case.

Next, consider trimming. 
At low $p_t$ it has a slightly lower significance than mMDT, mainly
because the particular $\zcut$ we've used is slightly non-optimal for
tagging purposes.
However, its main relevant feature is the drop in significance
relative to the mMDT curve for $p_t \gtrsim 800\GeV$. 
This corresponds to a $\rho$ value of $0.01$, which is to be compared
to the point $\rho = r^2 \zcut = 0.0045$ in Eqs.~(\ref{eq:trimming-LO}),
(\ref{eq:trimming-LL-small-mass-FC}) at which the background starts to
grow and develop a low-mass Sudakov peak.
cf.~Eq.~(\ref{eq:trimming-LL-small-mass-FC}).
The departure from mMDT is less pronounced in the gluon case than in
the quark case because the stronger Sudakov suppression from the $C_A$
colour factor reduces the height of the low-mass background Sudakov peak.

Finally, we examine pruning. 
Like trimming, pruning has a low-mass Sudakov peak, but it develops
only for lower masses than for trimming, and accordingly the drop in
performance of pruning relative to mMDT is mitigated.
Most interesting, perhaps, is $\sanepruning$. 
Its background enjoys a double-logarithmic Sudakov suppression for
small $m/p_t$, due to the factor $e^{-D(\rho)}$ in
Eq.~(\ref{eq:sane-prune-fixed-coupling}).
The analogous effect for the signal is, we believe,
single-logarithmic, hence the modest reduction in signal yields in
Fig.~\ref{fig:SB-signal-eff}.
Overall the background suppression dominates, leading to improved
tagging significance at high $p_t$.
This is most striking in the gluon case, because of the $C_A$ colour
factor in the $e^{-D(\rho)}$ Sudakov suppression.
Despite this apparent advantage, one should be aware of a
defect of $\sanepruning$, namely that at high $p_t$ the $\sane$/$\anomalous$
classification can be significantly affected by underlying event and
pileup, because of the way in which they modify the original jet
mass and the resulting pruning radius.
It remains of interest to develop a tagger that exploits the same
double-logarithmic background suppression while not suffering from
this drawback.\footnote{In this context it may be beneficial to study
  a range of variables, such as $N$-subjettiness \cite{Thaler:2010tr}
  and energy correlations~\cite{Larkoski:2013eya}, or even
  combinations of observables as done in
  Refs~\cite{Soper:2010xk,Cui:2010km}. It is also of interest to
  examine observables specifically designed to show sensitivity to
  colour flows, such as pull~\cite{Gallicchio:2010sw} and
  dipolarity~\cite{Hook:2011cq}, though it is not immediately apparent
  that these exploit differences in the double logarithmic structure.
  It would also, of course, be interesting to extend our analysis to
  other types of method such as template tagging~\cite{Almeida:2010pa}.
}

\section{Conclusions}
\label{sec:conclusions}

\begin{table}
  \centering
  \begin{tabular}{lccccc}\toprule
             & highest logs  & transition(s) & Sudakov peak & NGLs & NP:
             $m^2 \lesssim$\\
    \midrule
    plain mass & $\as^n L^{2n\phantom{-1}}$ & ---  & $L\simeq
    1/\sqrt{\asbar}$ & yes & $\mu_\text{NP}\, p_t \,R$\\
    \midrule 
    trimming & $\as^n L^{2n\phantom{-1}}$ & $\zcut$, $ r^2 \zcut$ 
    &    $L\simeq 1/\sqrt{\asbar} - 2\ln r$ & yes & $\mu_\text{NP}\, p_t \,\Rsub$\\
    pruning & $\as^n L^{2n\phantom{-1}}$ & $\zcut$, $\zcut^2$    &
    $L\simeq 2.3/\sqrt{\asbar}$ & yes &  $\mu_\text{NP}\, p_t \,R$\\
    MDT     & $\as^n L^{2n-1}$ & $\ycut$, $\frac14\ycut^2$, $\ycut^3$
    & --- & yes& 
    $\mu_\text{NP}\, p_t \,R$
    \\
    \midrule 
    $\sanepruning$ & $\as^n L^{2n-1}$ & $\zcut$    & (Sudakov tail) &
    yes &  $\mu_\text{NP}\, p_t \,R$\\
    mMDT    & $\as^n L^{n\phantom{2-1}}$  & $\ycut$               &
    --- & no & $\mu_\text{NP}^2/\ycut$\\
    \bottomrule
  \end{tabular}
  \caption{Table summarising the main features for the plain jet mass,
    the three original taggers of our study and the two variants
    introduced here.
    In all cases, $L = \ln \frac{1}{\rho} = \ln \frac{R^2
      p_t^2}{m^2}$, $r = \Rsub/R$
    and the log counting applies to the region below the smallest
    transition point. The transition points themselves are given
    as $\rho$ values.
    Sudakov peak positions are quoted for $d\sigma/d L$; they are
    expressed in terms of $\asbar \equiv
    \as C_F/\pi$ for quark jets and $\asbar \equiv \as C_A/\pi$ for gluon
    jets and neglect corrections of $\order{1}$.
    ``NGLs'' stands for non-global logarithms.
    The last column
    indicates the mass-squared below which the non-perturbative (NP) region
    starts, with $\mu_\text{NP}$ parametrising the scale where perturbation
    theory is deemed to break down.
  }
  \label{tab:summary}
\end{table}

In this paper we have developed an extensive analytical understanding
of the action of widely used boosted-object taggers and groomers on
quark and gluon jets.

We initially intended to study three methods:
trimming, pruning and the mass-drop tagger (MDT).
The lessons that we learnt there led us to introduce new variants,
$\sanepruning$ and the modified mass-drop tagger (mMDT).
The key features of the different taggers are summarised in
table~\ref{tab:summary}. 
We found, analytically, that the taggers are similar in certain
phase-space regions and different in others, identified the transition
points between these regions and carried out resummations of the
dominant logarithms of $p_t/m$ to all orders.

One tagger has emerged as special, mMDT, in that 
it eliminates all sensitivity to the soft divergences of QCD.
As a result its dominant logarithms are $\as^n L^n$, entirely of
collinear origin.
It is the first time, to our knowledge, that such a feature is
observed, and indeed all the other taggers involve terms with more
logarithms than powers of $\as$.
One consequence of having just single, collinear logarithms is that
the complex non-global (and
super-leading~\cite{Forshaw:2006fk}) logarithms are absent.
Another is that fixed-order calculations have an enhanced range of
validity, up to $L \ll 1/\as$ rather than $L\ll 1/\sqrt{\as}$.
The modified mass-drop tagger is also the least affected by non-perturbative
corrections.
Finally the $\ycut$ parameter of the tagger can be chosen so as to ensure a
mass distribution that is nearly flat, which can facilitate the
reliable identification of small signals.
Intriguingly, the mass-drop parameter appears to be largely redundant,
which suggests that one might further simplify the tagger by
eliminating it, while retaining all of the tagger's attractive features.
Also of interest is the $\sane$ variant of pruning.
This is the only one of these simple taggers to derive a significant
advantage from the difference in net colour between electroweak
signals and QCD backgrounds. 
That advantage comes at the cost of enhanced UE and pileup
sensitivity, and it remains to be seen if this drawback can be
alleviated. 

This article forms part of a wider project to gain an understanding of
the behaviour of taggers on both signals and backgrounds.
Such an understanding is important to help ensure that these tools are
used as robustly as possible and to gain insight into the similarities
and differences between tools.
We saw explicitly, in section~\ref{sec:choice-monte-carlo}, how our
results helped identify issues in Monte Carlo generators, and in
section~\ref{sec:signal}, how they gave us a powerful tool to
understand signal-background discrimination performance as a function
of jet $p_t$.

We look forward to continued future work on this subject. This may
include the extension of our analysis to signal processes, higher
accuracy calculations for the taggers, measurements and
phenomenological comparisons especially for mMDT, and the study of a
wider range of observables.
We believe that such work will provide solid foundations for the field
of jet substructure and help guide its future development.

\acknowledgments
MD and SM would like to thank the CERN Theory Group for hospitality
during parts of this work.
SM would like to thank Michael Spannowsky for many useful discussions.
We would all like to acknowledge useful discussions with Matteo
Cacciari, Steve Ellis, Steve Mrenna, Sebastian Sapeta, Mike Seymour,
Torbj\"orn Sj\"ostrand, Peter Skands, Gregory Soyez, Jesse Thaler and
Jon Walsh and are grateful to an anonymous referee for insightful
comments.
We also thank David Grellscheid and Leif L\"onnblad
for assistance with ThePEG.
Finally, we are grateful for support from the French Agence Nationale de
la Recherche, under grant ANR-09-BLAN-0060, from the European
Commission under ITN grant LHCPhenoNet, PITN-GA-2010-264564, from
ERC advanced grant Higgs@LHC and from the UK's STFC.


\appendix

\section{Formulae for gluon jets}
\label{sec:gluon-jets}
In the main text we explicitly derived resummed expressions for
quark-initiated jets. Analogous expressions for gluon jets can be
easily obtained by replacing the colour factor $C_F$ with $C_A$ and
considering gluon splittings rather than quark ones, which amounts to
the substitution $p_{gq}\to p_{xg}\equiv\left( \frac12 p_{gg}+\frac{T_R n_f}{C_A}
  p_{qg}\right)$, where the reduced splitting functions are defined by
\begin{subequations}
  \label{eq:reduced-splitting}
  \begin{align}
    p_{qg}(z)& =\frac12(z^2 + (1-z)^2)\,,\\
    p_{gg}(z)& = 2\frac{1-z}{z} + z(1-z)\,,\\
    p_{gq}(z)& =  \frac{1+(1-z)^2}{2z}\,,
  \end{align}
\end{subequations}
Note that, exploiting the symmetry $z\leftrightarrow (1-z)$ of the  $g\to gg$ splitting, $p_{gg}$ has been conveniently written in such a way that it only exhibits a singularity for $z\to0$.

We can define now the equivalents of Eq.~(\ref{eq:jet-mass-Delta}) and Eq.~(\ref{eq:Sab}) for gluon-induced jets:
\begin{equation}
  \label{eq:jet-mass-Delta-gluon}
  D_g(\rho) = 
  \int_\rho^1 \frac{d\rho'}{\rho'} 
  \int_{\rho'}^1 dz\,\,  p_{xg}(z)\,
  \frac{\as(z \rho' R^2 p_t^2 ) C_A}{\pi}\,,
\end{equation}
\begin{equation}
  \label{eq:Sab-gluon}
    S_g(a,b) = \int^a_b \frac{d\rho'}{\rho'} \int_{\zcut}^1
    dz \, p_{xg}(z) \, \frac{\as(z \rho' R^2 p_t^2 ) C_A}{\pi}\,,
\end{equation}
It is then easy to write down the resummed expressions for the mass distribution of gluon-induced jets, for each of the cases considered in this paper, i.e.\ plain jet mass, trimming, pruning and mMDT. As in the main part of the paper, we report results in the small-$\zcut$ ($\ycut$) limit.
\subsection{Plain jet mass}
The resummed expression for the integrated distribution of the plain
jet mass, in the case of gluon jets is given by
\label{sec:gluon-jets-plain}
\begin{equation}
  \label{eq:Sigma-plain-jet-mass-gluon}
  \Sigma_g(\rho) = e^{-D_g(\rho)} \cdot \frac{e^{-\gamma_E
     D_g'(\rho)}}{\Gamma(1 +D_g'(\rho))} \cdot \cN_g(\rho)\,,
\end{equation}
where $\cN_g(\rho)$ contains non-global logarithms and clustering logarithms. The above expression is to be compared to the case of quark-initiated jets, Eq.~(\ref{eq:Sigma-plain-jet-mass}).
\subsection{Trimming}
\label{sec:gluon-jets-trim}
In the case of trimming, the all-order integrated mass distribution for gluon jets reads
\begin{multline}
  \label{eq:trimming-LL-small-mass-gluon}
  \Sigma_g^\text{(trim)} (\rho)
  =
  \exp\Bigg[ 
    - D_g(\max(\zcut,\rho)) 
    - S_g(\zcut,\rho) \Theta(\zcut - \rho)
    \\ \left.
    - \Theta(\zcut r^2 - \rho) \int_\rho^{\zcut r^2} \frac{d\rho'}{\rho'}
    \int^{\zcut}_{\rho'/r^2} \frac{dz}{z} \frac{\as(z \rho' R^2 p_t^2 ) C_A}{\pi}
  \right]\,,
\end{multline}
which is to be compared to the result for quark-initiated jets in Eq.~(\ref{eq:trimming-LL-small-mass}).
\subsection{Pruning}
\label{sec:gluon-jets-prune}
In the case of pruning, the result is most naturally written for the differential jet mass distribution. 
For $\rho<\zcut$, the $\sane$ and $\anomalous$ components of pruning for gluon jets read
\begin{equation}
  \label{eq:sane-prune-result-gluon}
  \frac{\rho}{\sigma_g} \frac{d\sigma_g^\saneprune}{d\rho} = 
  \int_{\zcut}^1 dz \; p_{xg}(z) \;   e^{-D_g\left(\min(\zcut,\frac{\rho}{z})\right) 
      - S_g\left(\min(\zcut,\frac{\rho}{z}),\rho \right)} 
  \frac{\as (z \rho R^2 \, p_t^2) \, C_A}{\pi} \,,
  \end{equation}
\begin{multline}
  \label{eq:anom-prune-result-gluon}
  \frac{\rho}{\sigma_g} \frac{d\sigma_g^\anomprune}{d\rho} = 
  \int_\rho^{\zcut}
  \frac{d\rhofat}{\rhofat}
  \left(e^{-D_g(\rhofat)} \int_{\rhofat}^{\zcut} \frac{dz'}{z'}
    \frac{\as (\rhofat z'\, 
    p_t^2 R^2) \, C_A}{\pi} \right)
  \times
  \\
  \times
  e^{- S_g\left(\rhofat,\rho \right)}
  \int_{\rho/\rhofat}^1 dz \; p_{xg}(z)  \frac{\as (\rho z\,
    p_t^2 R^2) \, C_A}{\pi}
  \left[
    \Theta\left(\frac{\rho}{\rho_{\text{fat}}} - \zcut\right)
    + \right.
    \\
    \left. +
    \Theta\left(\zcut - \frac{\rho}{\rho_{\text{fat}}}\right)
    \exp\left(-\int_\rho^{\zcut\rho_{\text{fat}}} \frac{d\rho'}{\rho'}
    \int_{\rho'/\rhofat}^{\zcut} \frac{dz'}{z'}
    \frac{C_A}{\pi} \as(\rho' z' p_t^2 R^2)
    \right)
  \right]\,.
\end{multline}
  The above expressions are to be compared to the results for quark-initiated jets in Eq.~(\ref{eq:sane-prune-result}) and Eq.~(\ref{eq:anom-prune-result}), respectively.

\subsection{mMDT}
\label{sec:gluon-jets-mMDT}
Finally, the mMDT integrated mass distribution for gluon jets is
\begin{equation}
  \label{eq:modMDT-rho-ordering-gluon}
    \Sigma_g^\text{(mMDT)}     = \exp \left[ 
    - D_g(\max(\ycut,\rho)) 
    - S_g(\ycut,\rho) \Theta(\ycut - \rho) \right]\,.
\end{equation}
which is to be compared to the result for quark-initiated jets in Eq.~(\ref{eq:modMDT-rho-ordering}).

\section{Finite-$\ycut$ effects for the mMDT}
\label{sec:finite-ycut-modMD}

Without the assumption $\ycut \ll 1$, two additional
complications would have entered the derivation of
section~\ref{sec:mmdt-all-orders}. 
Firstly, a $q \to qg$ splitting can result in the gluon being the
harder of the two prongs, so that subsequent declustering follows the
gluon rather than the quark; this occurs with a probability $\sim
\ycut$, and so for finite $\ycut$ one must then include also $g\to gg$
and $g \to q\bar q$ splittings, even for a quark-induced jet. The
resulting effect enters at single-logarithmic accuracy, as we shall
see in detail below.
Secondly, the energy of parton $n$ is scaled by a factor $x_n$
relative to the original jet, where $x_n \equiv
(1-z_1)\ldots(1-z_{n-1})$ and $z_i$ is the fraction of the leading
parton's momentum carried away by emission $i$ (i.e.\ normalised to
the momentum of the parton just before that emission).
Since we had $\ycut \ll 1$ and all $z_i <
\ycut$ (for $i < n$) we automatically had $x_{n} = 1$ and we could
therefore drop it.
This is no longer the case for finite $\ycut$, though we believe the
effect is relevant only for terms $\as^n L^{n-1}$, i.e.\ beyond single
logarithmic accuracy, based on an argument analogous to that given for
filtering in section~\ref{sec:filtering}: suppose there is a probability
$p(x,\as L)$ for there to be a modification by a factor $x$ (of order
$1$) of the tagged-jet $p_t$, and correspondingly of the tagged jet
mass.
Therefore $\Sigma^{(\text{full }x)}(\rho) = \Sigma^{(x=1)}(\rho ) +  \int dx
[\Sigma^{(x=1)}(\rho/x^2) - \Sigma^{(x=1)}(\rho)] p(x, \as L)$, where
$\Sigma^{(x=1)}(\rho)$ is the resummed distribution obtained with the
approximation $x_n=1$.
The factor in square brackets is subleading, and therefore
$\Sigma^{(\text{full }x)}(\rho)$ is identical to $\Sigma^{(x=1)}(\rho
)$ at single logarithmic accuracy.

Let us now examine how to include the flavour changing effects to
single logarithmic accuracy for finite $\ycut$ ($\le 1$).
One simply extends Eqs.~(\ref{eq:modMDT-rho-ordering}),
(\ref{eq:modMDT-fixed-coupling}) to have a matrix structure in flavour
space.
First one defines 
\begin{subequations}
\begin{align}
  S_{q} & =  C_F \int dz\; p_{gq}(z) \; \Theta\left(\frac{z}{1-z} - \ycut\right) 
                              \; \Theta\left(\frac{1-z}{z} - \ycut\right)\,,
  \\                                   
  S_{g} & =  C_A \int dz \;  p_{xg}(z) \;
                                   \Theta\left(\frac{z}{1-z} - \ycut\right) 
                                   \Theta\left(\frac{1-z}{z} - \ycut\right)\,,
  \\                                   
  S_{q\to g} & = C_F\int dz \; p_{gq}(z) 
                              \Theta\left(\ycut - \frac{1-z}{z}\right)\,,
  \\                                   
  S_{g\to q} & = T_R  n_f\int dz \; p_{qg}(z) 
                    \left[ \Theta\left(\ycut - \frac{1-z}{z}\right)
                         + \Theta\left(\ycut - \frac{z}{1-z}\right)
                    \right]\,.
\end{align}
\end{subequations}
Then, the result (in a fixed-coupling approximation) is given by
\begin{equation}
  \label{eq:modMDT-fixed-coupling-full-flavour}
  \frac{\rho}{\sigma}\frac{d \sigma}{d \rho}^\text{(mMDT)} = 
  \frac{\as}{\pi}\; 
  \left(S_q \;\;\; S_g\right)\cdot
  \exp
  \left[
    \frac{\as}{\pi} \ln \frac1\rho
  \left(
  \begin{array}{cc}
    -S_q - S_{q \to g} & S_{g \to q}       \\
    S_{q\to g}         & -S_g - S_{g \to q}
  \end{array} \right)
  \right]
  \left(
  \begin{array}{c}
     I_q \\ I_g
  \end{array} \right)\,,
\end{equation}
where $I_{q,(g)}$ is the initial fraction of quarks and gluons. The
extension to running coupling is trivial.
%

\section{Y-trimming and (m)MDT with an $R_{\min}$ cut}
\label{sec:Ytrimming}

In discussions about this work, a question that has repeatedly
arisen is whether there is a modification of trimming analogous to the
``Y'' pruning requirement.
The most obvious modification, ``Y-trimming'', is to request that
trimming find at least two subjets that pass the trimming cuts.
The behaviour of Y-trimming is, however, qualitatively different from that of
Y-pruning. 

In the case of pruning, the effective subjet radius is set dynamically
based on the jet mass. This means that at LO, when the jet consists of
just two partons, the subjet radius is always chosen such that the two
partons end up in different subjets. I.e., at LO, pruning and Y-pruning
are identical, and can probe arbitrarily small values of $\rho$.

In the case of trimming, the subjet radius is a fixed, user-chosen
parameter. Therefore, for sufficiently small values of $\rho$,
two-prong configurations are either entirely contained inside a single 
subjet, or else one of the prongs falls below the $\zcut$ requirement.
In other words for Y-trimming there will be a minimal value of $\rho$
that can be probed, which, in the small $\zcut$ approximation is
$\zcut r^2$, where we recall $r = \Rsub/R$. In effect the situation is
similar to that for normal jet finding with a fixed jet
radius.\footnote{Normal jet finding tends to be carried out with a
  fixed jet $p_{t,\min}$ cut, which leads to a different relation
  between minimum accessible mass and boosted-object $p_t$, $m^2 >
  p_{t,\min} p_t R^2$ for $p_{t,\min}\ll p_t$.}
This means that, unlike the other taggers we have considered,
Y-trimming is not ideally suited to probing a broad range of boosts.
It is for this reason that we have not included it in as part of our
main discussion of taggers.

In this context, it is interesting to note that a cut on the subjet
separation was used in early ATLAS work on MDT~\cite{ATLAS:2012am},
$\Delta_{j_1 j_2} > R_{\min}$ with $R_{\min} = 0.3$.
This cut has the same effect as the two-subjet requirement in
trimming, i.e.\ it leads to a minimal accessible value of $\rho$ of $\ycut r^2$,
where now $r =  R_{\min} / R$. 
The cut was imposed so as to reduce sensitivity to detector and
reconstruction granularity.
It is to be hoped that ongoing and future work by the ATLAS
collaboration will eliminate the need for such a cut in substructure
studies.

For completeness we provide here the exact LO result for
Y-trimming. We work in the small-$R$ limit, but relax the small
$\zcut$ and small $\rho$ approximations, because of the presence of
multiple transition points that are quite close to each other in $\ln
\rho$.
Defining
\begin{subequations}
  \begin{align}
    \label{eq:symp}
    \Pi_q(x) &= \int^{\frac12}_x dz \left[p_{qq}(z) +  p_{qq}(1-z)\right]\,,\\
    &= \ln\left(\frac1x -1 \right) - \frac34 + \frac{3x}{2}\,,
  \end{align}
\end{subequations}
the LO Y-trimming distribution is
\begin{equation}
  \label{eq:Y-trimming}
  \frac{\rho}{\sigma} \frac{d\sigma}{d \rho}^\text{(Y-trim, LO)}  
  = \frac{\as C_F}{\pi}
  \left\{
      \begin{array}{ll}
        \Pi_q\left(\frac12 - \sqrt{\frac14 - \rho}\right)\,, & 
        \displaystyle \zcut(1-\zcut) < \rho < \frac14\,,
        \\[10pt]
        \Pi_q\left(\zcut\right)\,, & 
        \displaystyle \frac{r^2}{4} < \rho < \zcut (1-\zcut)\,,
        \\[10pt]
        \Pi_q\left(\zcut\right) - \Pi_q\left(\frac12 - \sqrt{\frac14 -
            \frac{\rho}{r^2}}\right)\,, \quad
        &  \displaystyle  \zcut (1-\zcut) r^2  < \rho < \frac{r^2}{4}\,,
      \end{array}
    \right.
\end{equation}
and zero elsewhere. 
The LO result for (m)MDT with an $R_{\min}$ requirement is identical,
modulo the replacement $\zcut \to \frac{\ycut}{1+\ycut}$.

One can understand the structure of Eq.~(\ref{eq:Y-trimming}) as
follows: as we decrease $\rho$, the distribution starts to grow from
$\rho = \frac14$; it then saturates at $\rho = \zcut(1-\zcut)$ when
the $\zcut$ condition kicks in; when we reach $\rho = r^2/4$ then for
symmetric 2-prong configurations the two prongs are separated by
$\Rsub$ and give a single subjet, so the distribution starts to
decrease; finally for $\rho < r^2 \zcut(1-\zcut)$,
configurations where the two prongs are separated by an angle
greater than $\Rsub$ have one of the prongs carrying a momentum
fraction smaller than $\zcut$, i.e.\ the Y-trimmed distribution is
zero.

Eq.~(\ref{eq:Y-trimming}) is valid if $r^2 \le 4 \zcut(1-\zcut)$. 
For $r^2 = 4 \zcut(1-\zcut)$, the plateau region between $r^2/4$ and
$\zcut(1-\zcut)$ is replaced with a single peak transition point at
$\rho = r^2/4$, and a minimal $\rho$ of $r^4/4$.
For larger values of $r$, the result is left as an exercise for the reader.

\begin{figure}
  \centering
  \includegraphics[width=0.49\textwidth]{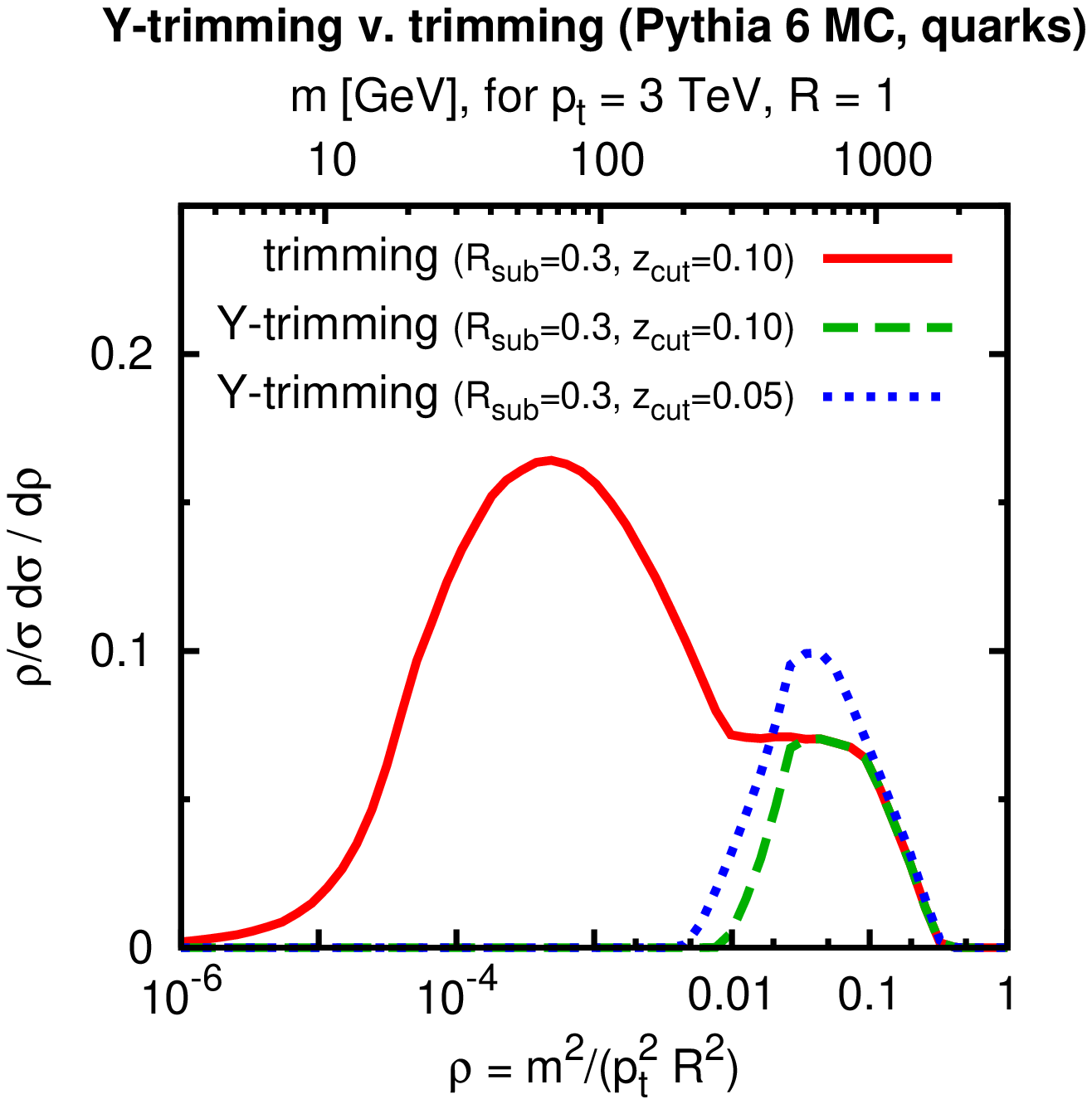}\hfill
  \includegraphics[width=0.49\textwidth]{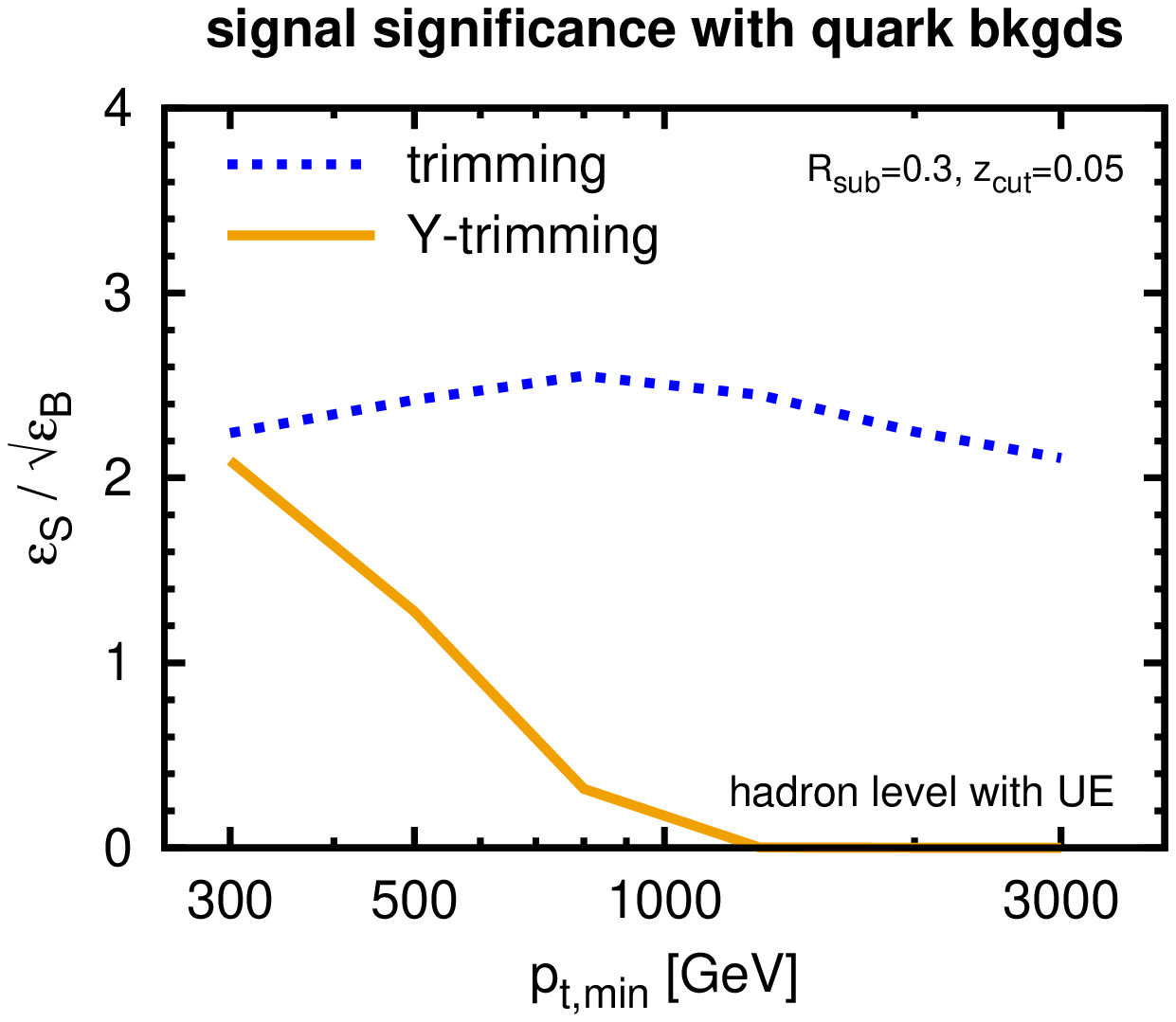}
  \caption{%
    Comparisons of trimming and Y-trimming.
    Left: the $\rho$ distributions. 
    Right: the signal significance for tagging $W$'s in the presence
    of quark backgrounds.
    The details of the MC event generation and cuts are as for
    Figs.~\ref{fig:tagged-mass-MC} and \ref{fig:SB-summary}
    respectively.  }
  \label{fig:trimming-v-Ytrimming}
\end{figure}

In figure~\ref{fig:trimming-v-Ytrimming} (left) we show the
$\rho$ distribution for Y-trimming and normal trimming, where the
transition points are clearly visible.
Finally, in the right-hand plot, we show the signal significances
versus minimum jet $p_t$ in the presence of quark jet backgrounds,
confirming that Y-trimming is not an adequate boosted-object tagger at
high transverse momenta.

%

\end{document}